%% file: main.tex
\documentclass[uplatex, prx,twocolumn,superscriptaddress]{revtex4-2}
\usepackage[utf8]{inputenc}

\usepackage{lmodern}
\usepackage{qcircuit}
\usepackage{braket}
\usepackage[version=3]{mhchem} 
\usepackage{here}
\usepackage{multirow}
\usepackage{longtable}
\usepackage{natbib}
\usepackage{graphicx}
\usepackage{amsfonts,amsmath,amssymb,amsthm}
\usepackage{todonotes}
\usepackage{bm}
\usepackage{algorithm}
\usepackage[noend]{algpseudocode}
\usepackage{comment}
\usepackage[flushleft]{threeparttable}

\algblock{Input}{EndInput}
\algnotext{EndInput}
\algblock{Output}{EndOutput}
\algnotext{EndOutput}

\newcommand{\lstickx}[1]{\lstick{\makebox[1.5em][l]{$#1$}}}
\newcommand{\arrep}[1]{\ar @<4pt> @/^/[#1]|-{\mbox{ $\times L$ }}}
\newcommand{\measo}[1]{\ar @<4pt> @/^/[#1]|-{\mbox{~$\left<\hat{O}\right>$~}}}
\newcommand{\measoz}[1]{\ar @<4pt> @/^/[#1]|-{\mbox{~$\left<\hat{Z}_a \otimes \hat{O}\right>$~}}}

\begin{document}
\title{Simulating Time Evolution with Fully Optimized Single-Qubit Gates on Parameterized Quantum Circuits}
\author{Kaito Wada}
\email{wkai1013keio840@keio.jp}
\affiliation{Department of Applied Physics and Physico-Informatics, Keio University, 3-14-1 Hiyoshi, Kohoku-ku, Yokohama, Kanagawa, 223-8522, Japan}
\author{Rudy Raymond}
\affiliation{IBM Quantum, IBM Japan 19-21 Nihonbashi Hakozaki-cho, Chuo-ku, Tokyo, 103-8510, Japan}
\affiliation{Quantum Computing Center, Keio University, 3-14-1 Hiyoshi, Kohoku-ku, Yokohama, Kanagawa, 223-8522, Japan}
\author{Yu-ya Ohnishi}
\affiliation{Materials Informatics Initiative, RD Technology \& Digital Transformation Center, JSR Corporation, 3-103-9, Tonomachi, Kawasaki-ku, Kawasaki, Yokohama, Kanagawa, 210-0821, Japan}
\affiliation{Quantum Computing Center, Keio University, 3-14-1 Hiyoshi, Kohoku-ku, Yokohama, Kanagawa, 223-8522, Japan}
\author{Eriko Kaminishi}
\affiliation{Quantum Computing Center, Keio University, 3-14-1 Hiyoshi, Kohoku-ku, Yokohama, Kanagawa, 223-8522, Japan}
\affiliation{JST PRESTO, 4-1-8 Honcho, Kawaguchi, Saitama 332-0012, Japan}
\author{Michihiko Sugawara}
\affiliation{Quantum Computing Center, Keio University, 3-14-1 Hiyoshi, Kohoku-ku, Yokohama, Kanagawa, 223-8522, Japan}
\author{Naoki Yamamoto}
\affiliation{Quantum Computing Center, Keio University, 3-14-1 Hiyoshi, Kohoku-ku, Yokohama, Kanagawa, 223-8522, Japan}
\affiliation{Department of Applied Physics and Physico-Informatics, Keio University, 3-14-1 Hiyoshi, Kohoku-ku, Yokohama, Kanagawa, 223-8522, Japan}
\author{Hiroshi C. Watanabe}
\email{hcwatanabe@keio.jp}
\affiliation{Quantum Computing Center, Keio University, 3-14-1 Hiyoshi, Kohoku-ku, Yokohama, Kanagawa, 223-8522, Japan}

\begin{abstract}
    We propose a novel method to sequentially optimize arbitrary single-qubit gates in parameterized quantum circuits for simulating real and imaginary time evolution.
    The method utilizes full degrees of freedom of single-qubit gates and therefore can potentially obtain better performance.
    Specifically, it simultaneously optimizes both the axis and the angle of a single-qubit gate, while the known methods either optimize the angle with the axis fixed, or vice versa. 
    It generalizes the known methods and utilizes sinusoidal cost functions parameterized by the axis and angle of rotation. 
    Furthermore, we demonstrate how it can be extended to optimize a set of parameterized two-qubit gates with excitation-conservation constraints, which includes the \textrm{Hop} and the \textrm{Reconfigurable Beam Splitter} gates.
    We perform numerical experiments showing the power of the proposed method to find ground states of typical Hamiltonians with quantum imaginary time evolution using parameterized quantum circuits.  
      In addition, we show the method can be applied to real time evolution and discuss the tradeoff between its simulation accuracy and hardware efficiency.

\end{abstract}

\maketitle

 \section{Introduction}
 \input{introduction}

\section{Methods}\label{sec:methods}
\input{methods}

\section{RESULTS AND DISCUSSION}\label{sec:results}
\input{results}

 \section{Conclusion}\label{sec:conclusion}
 \input{conclusion}

\section{Acknowledgements}
H.C.W. was supported by JSPS Grant Numbers 20K03885 and 20H05518, and JST PRESTO Grant number JPMJPR17GC. 
E.K was supported by JSPS Grant Number 20K14388 and JST PRESTO Grant number JPMJPR2011. 
In addition, H.C.W., E.K., M.S. and N.Y. were supported by the MEXT Quantum Leap Flagship Program Grant Number JPMXS0118067285 and JPMXS0120319794. We would like to thank  Dr.~Shumpei~Uno, Dr.~Yohichi~Suzuki, and Dr.~Antonio~Mezzacapo, for technical discussion, as well as Dr. Michael Lubasch for pointing out details in~\cite{benedetti2021}.

\bibliography{references}
\newpage
\onecolumngrid
 \section{Appendix}\label{sec:appendix}
 \input{appendix}

\end{document}

%% file: introduction.tex
Quantum simulations of materials, which are becoming popular as promising applications of quantum computing, are practically useful in designing functional materials.
In fact, there have been significant number of quantum and quantum-classical hybrid algorithms developed for such simulations~\cite{kassal2011simulating,cao2019quantum,mcardle2020quantum,bauer2020quantum,motta2021}.
One can track the time evolution of a quantum system by the real time evolution (RTE) algorithm, which is important to investigate, for example, the quantum dynamics under the irradiation of laser~\cite{charron1998femtosecond,yonehara2012fundamental}.
In addition to the quantum phase estimation (QPE) algorithm~\cite{kitaev1995quantum,lanyon2010towards}, 
and the variational quantum eigensolver (VQE)~\cite{peruzzo2014variational,mcclean2016theory,tilly2021variational}, the imaginary time evolution (ITE) algorithm~\cite{mcardle2019,motta2020} might be applied to obtain the ground state energy and wavefunction.
Although it is hard to run the QPE on current noisy intermediate-scale quantum devices, the VQE and ITE may be more realizable~\cite{peruzzo2014variational,kandala2017hardware,hempel2018quantum,kandala2019error,nam2020ground,tilly2020computation,eddins2021doubling,gao2021computational,gao2021applications, yeter2020practical}; actually several experimental studies on small systems have been reported~\cite{ma2020quantum, sun2021quantum}. 

The basic ingredients for running quantum-classical hybrid algorithms are first to set a parameterized quantum circuit (PQC), sometimes called the trial wave function or simply the ansatz, and then to iteratively update its parameters by classical optimizers so that its final output state approximates the target state.
The approximation accuracy of the target state achieved via quantum-classical hybrid algorithms heavily depends on the choice of PQC and the classical optimization strategy. 
Recent great efforts have revealed several essential properties of the components of quantum-classical hybrid algorithms~\cite{sim2019expressibility,cerezo2021variational,bharti2021noisy}. 
Among those literature works, we find interesting gradient-free optimizers that make full use of the specific parameterization of standard PQCs.

More precisely, those optimizers can analytically optimize a subset of the parameters at each iteration by exploiting the special type of analytic form of the cost function. 
That is, the parameters are locally optimized in a coordinate-wise manner and updated deterministically. 
Specifically, Nakanishi~\textit{et al}.~\cite{nakanishi2020sequential} and Ostaszewski~\textit{et al}.~\cite{ostaszewski2021structure} showed that the cost of VQE becomes a sinusoidal function of a single-qubit rotation, and thus we can determine the optimal rotational angle. 
Ostaszewski~\textit{et al}. also proposed a sequential optimization method for selecting the best rotational axes of qubits from the $x$, $y$, or $z$ rotations.
Going further to relax the ansatz-dependency of VQE, a generalization of such gradient-free optimizers called the free-axis selection (or, Fraxis)~\cite{watanabe2021} was proposed.
The Fraxis algorithm analytically determines the best rotational axis (not limited to the discrete set consisting of $x$, $y$, or $z$ rotations) for each single-qubit gate when the rotational angles are fixed.
Note that these gradient-free optimizers directly determine the local optima and are different from the generic ones such as the Nelder-Mead~\cite{nelder1965simplex} or SPSA~\cite{spall1992multivariate}. 

In this article, with a particular attention to the task of simulating real and imaginary time evolution on a PQC, we make a further progress of this gradient-free optimization method.
That is, for simulating time evolution we prove that both the rotational axis and angle of each single-qubit gate in the PQC can be analytically optimized in a coordinate-wise manner. 
The set of rotational axes and angles of single-qubit gates constitutes the parameters of the PQC, and hence we have derived a fully-optimized gradient-free optimizer for simulating the time evolution.
This means that the PQC now acquires the ability for searching the quantum state in the Hilbert space that best approximates the time evolution in the coordinate-wise sense, and hence opening the new path to efficiently approximate the target state being driven via the time evolving propagators.
Moreover, we show that this method can be applied to optimize some multi-qubit gates, including the excitation-conserving two-qubit gates which play important roles in the quantum chemistry applications.
We conducted numerical simulations of ITE for 1-dimensional (1D) Heisenberg model and \ce{H2} molecule, and showed that the proposed algorithm could approximate the target ground state faster and better than some previous methods.
In addition, we also conducted numerical simulations of RTE for 1D Ising model using the proposed algorithms, and confirmed the reproduction of more accurate dynamics in our methods. 

Here we provide a summary of some related works on the quantum-computing approach for quantum simulations.
Note that some existing methods for RTE and ITE~\cite{yuan2019theory, mcardle2019variational, yao2021adaptive} utilize McLachlan's variational principle~\cite{mclachlan1964variational} to update the parameters of PQC.
The resultant updating rule requires, however, evaluating an inverse matrix of the size of parameters dimension at each iteration step, which is computationally demanding.
Several approaches circumvent this issue.
The method proposed by Motta \textit{et al}.\ \cite{motta2020} is to use a length-varying quantum circuit where a local Hamiltonian approximating the ITE can be efficiently optimized at each depth (corresponding to the time step).
The resultant quantum circuit achieves a controllable accuracy, while the length inevitably becomes large.
Nishi \textit{et al}.~\cite{nishi2021implementation} extended this result and showed that a non-local approximation of Hamiltonian could result in a shallower circuit and applied the method to the MaxCut problem.
As for RTE, the algorithm proposed by Barison \textit{et al}.~\cite{barison2021efficient} utilize a global optimization for parameters without matrix inversion. However, it requires the direct implementation of time evolving propagators, and then the application is limited to small quantum systems.
Benedetti \textit{et al}.~\cite{benedetti2021} reported a new algorithm that optimizes the parameters of a fixed-length PQC in a coordinate-wise manner to approximate the RTE and ITE, which is based on parametererization of each single-qubit gate with one parameter as in~\cite{ostaszewski2021structure,nakanishi2020sequential,Parrish2019arXiv,vidal2018calculus} and therefore not fully optimized.
Although there was an attempt to optimize a PQC whose some of single-qubit gates are constrained to have the same rotational angle~\cite{nakanishi2020sequential}, we are not aware of previous methods that explicitly optimize parameterized multi-qubit gates similarly as ours.

This article is organized as follows. 
The problem formulation and the proposed method are described Section~II. 
Some extension to optimize multi-qubit gates is also shown there. 
Section~III is devoted to show the numerical simulations. 
Finally we conclude the paper in Section~IV. 
Detailed derivation of equations are given in Appendix.

%% file: methods.tex
We first review the hardware-efficient
objective function for time evolving simulation proposed by Benedetti~\textit{et al}.~\cite{benedetti2021}, which focuses on the overlap between a target state and a trial state,
and introduce a measure of hardware-efficiency of objective function.
Then, we introduce \textit{Free Quaternion Selection for Quantum Simulation} (abbreviated as \textbf{FQS}, where \textbf{QS} has two meanings: \textit{Quaternion Selection} and \textit{Quantum Simulation}), which can fully optimize an arbitrary single-qubit gate with hardware efficiency.
Next, we show that the FQS can be extended to special multi-qubit gates such as the excitation-conserving gates.

\section*{A. Hardware-Efficient Quantum simulation of time evolution}
Suppose a time evolution simulation based on the Hamiltonian given as $\hat{H}=\sum_{k=1}^K h_k \hat{O}_k$, where $\hat{O}_k$ denotes a tensor product of Pauli operators for $m$-qubit $\hat{O}_k\in \{\sigma_0 = I,\sigma_x,\sigma_y,\sigma_z\}^{\otimes m}$, $h_k$ is a real valued coefficient, and $K\sim\mathcal{O}({\rm poly}(m))$.
The time evolving propagator $e^{-i\hat{H}t}$ is approximated with the first-order Trotter decomposition as
\begin{align}\label{eqn:TrotterDecomposition}
    &e^{-i \hat{H}t} = \left(e^{-i\hat{H}\Delta t}\right)^N \notag \\[4pt]
    &~~~~~\simeq\left(e^{-i h_K \hat{O}_{K} \Delta t}\cdots e^{- i h_k \hat{O}_{k} \Delta t} \cdots e^{- i h_1 \hat{O}_1 \Delta t } \right)^N,
\end{align}
where $N$ is the number of time steps and $\Delta t \equiv  t/N$.
Now a propagator $e^{-i h_k \hat{O}_k \Delta t}$ is applied to an arbitrary state $\ket{\psi_{k-1}}$, which results in a state $\ket{\psi_k}$ as
\begin{equation}\label{eqn:propagation}
    \ket{\psi_k}=e^{-ih_k \hat{O}_k \Delta t} \ket{\psi_{k-1}}.
\end{equation}
Suppose an initial state $\ket{\psi_{k-1}}$ is approximated by a PQC as $U(\vec{\vartheta}^*_{[k-1]})\ket{\bm{0}}$, where $\vec{\vartheta}^*_{[k-1]}$ is an optimal parameter set, and $\ket{\bm{0}}$ is the computational basis state. 
Provided that a PQC has sufficient expressibility, there exists an optimal parameter set $\vec{\vartheta}_{[k]}^*$ such that $\ket{\psi_k}\simeq U(\vec{\vartheta}_{[k]}^*)\ket{\bm{0}}$.
Therefore, a time evolution can be simulated if parameter sets that reproduce the time evolving propagator are somehow found.
Hereafter, we focus on a series of processes to determine the $k$th optimal parameter set $\vec{\vartheta}^*_{[k]}$ from $\vec{\vartheta}^*_{[k-1]}$.
For readability, we write $\vec{\vartheta}^*_{[k]}$ and $\vec{\vartheta}^*_{[k-1]}$ as $\vec{\vartheta}^*,\vec{\vartheta}'$, respectively, when it is obvious from the context.
To determine the optimal parameter set $\vec{\vartheta}^*$, Benedetti~\textit{et al}.~\cite{benedetti2021} proposed a recursive optimization using the objective function $\mathcal{M}(\vec{\vartheta})$ based on the Euclidean distance as
\begin{eqnarray}\label{eqn:Euclidean_distance}
  \mathcal{M}(\vec{\vartheta}) 
  &\equiv& \left\|U(\vec{\vartheta})\ket{\bm{0}}-e^{-i h \hat{O} \Delta t}U(\vec{\vartheta}')\ket{\bm{0}} \right\|^2 \notag\\[4pt]
  &=& 2-2\mathrm{Re}\left[ \bra{\bm{0}}U^\dagger(\vec{\vartheta}') e^{i h \hat{O}\Delta t} U(\vec{\vartheta})\ket{\bm{0}}\right],
\end{eqnarray}
and
\begin{equation} \label{eqn:opt_vartheta_all}
  \vec{\vartheta}^*  \equiv ~\underset{\vec{\vartheta}\in\Theta}{\mathrm{argmin}}~\mathcal{M}(\vec{\vartheta})
  = ~\underset{\vec{\vartheta}\in\Theta}{\mathrm{argmax}}~\mathcal{F}(\vec{\vartheta}),
\end{equation}
where $\Theta$ is the parameter space, and $\mathcal{F}(\vec{\vartheta})$ is defined as
\begin{equation}\label{eqn:primitive_objective_function}
\mathcal{F}(\vec{\vartheta}) \equiv \mathrm{Re}\left[\bra{\bm{0}}U^\dagger(\vec{\vartheta}') e^{i h \hat{O}\Delta t}U(\vec{\vartheta})\ket{\bm{0}}\right].
\end{equation}

Now we suppose a PQC $U(\vec{\vartheta})$ consists of $D$ parameterized single-qubit gates and arbitrary unitary gates without parameters such as CNOT gate.
An element of the parameter space $\Theta$ is given as $\vec{\vartheta}=(\vartheta_1,\cdots,\vartheta_D)$, where $\vartheta_d$ represents the parameters for the $d$th parameterized gate.
Conventionally, $\vartheta_d$ is a scalar value, e.g., rotational angle $\vartheta=\theta\in \mathbb{R}$ for a single-qubit axis-fixed rotation gate such as $R_x$, but in this paper to deal with a general single-qubit gate $\vartheta_d$ can be vector-valued as detailed in the next section.
To evaluate $\mathcal{M}(\vec{\vartheta})$, it is common to apply Hadamard tests with controlled gates for different parameters between $\vec{\vartheta}$ and $\vec{\vartheta}'$.
Because All $D$ elements in $\vec{\vartheta}$ usually differ from those in $\vec{\vartheta}'$,
the Hadamard test requires additional $\mathcal{O}(D)$ controlled gates and ancilla qubits, it may be difficult for near-term quantum devices when $D$ is large.

The Hadamard test can be replaced by direct measurements as proposed by Mitarai~\textit{et al}.~\cite{mitarai2019}, which are often called hardware-efficient because they need less controlled gates and ancilla qubits.
However, the direct measurements require more types of circuits, i.e., replacing $D$ controlled gates with direct measurements incur the cost of evaluating $\mathcal{O}(4^D)$ types of circuits.
Here we see the trade-off among different quantum resources.
To reduce the required quantum resources, it is possible to restrict the number of the gates to be updated in the objective function Eq.~(\ref{eqn:primitive_objective_function}).
On the other hand, such restricted updates may not be sufficient to minimize the simulation error $\mathcal{M}(\vec{\vartheta})$ for $e^{{-i h \hat{O}} \Delta t}$.
To balance between the hardware efficiency and simulation accuracy,
we employ \textit{$P \in\mathbb{N}$ series of updates} procedure where parameterized gates are grouped into $P$ sets, and the respective sets are represented by $\Lambda_{p},~(p=1,2,\cdots,P)$.
Here, $\Lambda_{p}$ consists of the gate indices as $\Lambda_{p}\subseteq \{1,2,\cdots,D\}$.
In the $P$ series update, we divide the propagator into $P$ terms maintaining total $\Delta t$ time evolution.
Although the division of $\Delta t$ is not unique, we uniformly assign $\Delta t/P$ 
similarly as \cite{benedetti2021}.
Then, the original objective function in Eq.~(\ref{eqn:primitive_objective_function}) is replaced by a series of the following objective functions
\begin{widetext}
\begin{align}\label{eqn:generalized_objective}
  \mathcal{F}^{(p)}(\{\vartheta_d\}) 
  \equiv {\rm Re}\left[\bra{\bm{0}} U^\dagger(\vec{\vartheta}^{(p-1)})
  e^{i h\hat{O}\Delta t /P} U(\{\vartheta_d\};\vec{\vartheta}^{(p-1)})\ket{\bm{0}}\right],~~~d\in{\Lambda_{p}},
\end{align}
\end{widetext}
where $\vec{\vartheta}^{(p-1)}$ denotes a parameter set whose $\vartheta_d,~(d\in \cup_{q=1}^{p-1}\Lambda_{q})$ have been updated from those in $\vec{\vartheta}^{(0)}\equiv\vec{\vartheta}'$.
Here, $U(\{\vartheta_d\};\vec{\vartheta}^{(p-1)}),~d\in\Lambda_{p}$ denotes a PQC in which the elements of $\Lambda_{p}$ are variable.
$\vec{\vartheta}^{(p)}$ is recursively obtained from $\vec{\vartheta}^{(p-1)}$ 
by solving the following problems
\begin{align}\label{eqn:generalized_problem}
  &\{\vartheta^*_d\}=\underset{\{\vartheta_d\}}{\mathrm{argmax}}~\mathcal{F}^{(p)} (\{\vartheta_d\}),~~~d\in{\Lambda_{p}}, \notag\\[4pt]
  &~~~{\rm for}~~~p=1,2,\cdots,P,
\end{align}
where $\vec{\vartheta}^{(p)}$ is defined by substituting $\{\vartheta^*_d\},~d\in\Lambda_{p}$ into $\vec{\vartheta}^{(p-1)}$.
Eventually, we obtain the solution $\vec{\vartheta}^*=\vec{\vartheta}^{(P)}$, which approximates the state evolved by the propagator $e^{-ih\hat{O}\Delta t}$ from the state with $\vec{\vartheta}'$.
This update procedure are repeated for the total $K$ Trotterized time propagators to simulate time evolution of $\Delta t$ as in Eq.~(\ref{eqn:TrotterDecomposition}) and (\ref{eqn:propagation}).
The optimization method for Eq.~(\ref{eqn:generalized_objective}) is not unique, and classical optimizers are conventionally employed, most of which update simultaneously multiple parameters with the cost of Hadamard tests consisting of multiple controlled gates and ancilla qubits.
In the present study, instead, we employ coordinate-wise update, where parameters are sequentially updated for respective single-qubit gates.
The procedure is summarized in Algorithm 1.
\renewcommand{\baselinestretch}{1.3}
\begin{figure*}[!t]
 \begin{minipage}{\linewidth}
 \begin{algorithm}[H]
	\caption{Time evolution with sequentially update of gate parameters for divided time propagators}\label{alg:FQS_general}
	\begin{algorithmic}
	\Input~{(1) Trotter-decomposed time propagators $\{e^{-ih_k\hat{O}_k\Delta t}\}_{k=1}^K$, (2) a PQC $U(\vec{\vartheta})$ with $D$ single-qubit gates, $\vec{\vartheta}=(\vartheta_1,\cdots,\vartheta_D)$, (3)~the number of time step $N$, (4) gate sets $\Lambda_{p}\subseteq\{1,2,\cdots,D\},~(p=1,2,\cdots,P)$}
	\EndInput
	\State \textbf{Initialize}~Choose initial gate parameters $\vec{\vartheta}$. 
	\Repeat
	   \For{$k$ = 1, 2, \ldots, $K$}
		  \For{$p$ = 1, 2, \ldots, $P$}
		  
		  \State{Set}
		  $\vec{\vartheta}_{\rm temporal}\leftarrow\vec{\vartheta}$

		  \For{$d$ in $\Lambda_{p}$}
		      \State{Set} a PQC from $U(\vartheta_d;\vec{\vartheta}_{\rm temporal})$.
              \vspace{3mm}
              \State Compute 
              $\vartheta^*_d = {\mathrm{argmax}}~{\rm Re}\left[\bra{\bm{0}} U^\dagger(\vec{\vartheta})e^{i h_k\hat{O}_k\Delta t /P} U(\vartheta_d;\vec{\vartheta}_{\rm  temporal})\ket{\bm{0}}\right]$
              \vspace{3mm}
              
              \State{Set} 
              $\left(\vec{\vartheta}_{\rm temporal}\right)_d\leftarrow \vartheta^*_d$
	       \EndFor
	       \State {Set} $\vec{\vartheta}\leftarrow\vec{\vartheta}_{\rm temporal}$

	       \EndFor
	   \EndFor
	\Until{$N\Delta\tau$ time evolution}
	\Output~Gate parameters $\vec{\vartheta}$
	\EndOutput
	\end{algorithmic}
 \end{algorithm}
 \end{minipage}
\end{figure*}
\renewcommand{\baselinestretch}{1.0}

Because there are at most $|\Lambda_{p}|$ different parameterized gates  between $U^\dagger(\vec{\vartheta}^{(p-1)})$ and $U(\{\vartheta_d\};\vec{\vartheta}^{(p-1)})$, the evaluation of the objective function Eq.~(\ref{eqn:generalized_objective}) requires additional $\mathcal{O}(|\Lambda_{p}|)$ controlled gates and one ancilla qubit.
We introduce a hardware-efficiency measure $\eta$ for an algorithm to simulate time evolution with Eq.~(\ref{eqn:generalized_objective}) defined as
\begin{equation}
    \eta\equiv \underset{p}{\rm min}~\frac{1}{|\Lambda_{p}|}.
\end{equation}
The measure varies in the range of $1/D\leq\eta\leq 1$; in case of the lowest value $\eta=1/D$ the objective function in Eq.~(\ref{eqn:generalized_objective})  is regressed to the original form in Eq.~(\ref{eqn:primitive_objective_function}).
On the other hand, in the highest value $\eta = 1$, namely, the most hardware-efficient case, the variables in each objective function are parameters of one single-qubit gate.

Note that the objective function in Eq.~(\ref{eqn:generalized_objective}) does not prescribe any optimization methods and PQC structures.
In general, it is important to employ a PQC with sufficient expressibility to describe the state of interest.
While it is common to extend a quantum circuit by adding parameterized gates, the correlation among parameters emerges as a new obstacle for optimization upon increase of parameters.
To circumvent this problem, it is required to both (1) simultaneously update  correlating parameters and (2) construct a high-expressibility PQC with as fewer number of parameters as possible.
To this end, we propose a new optimization method for time evolution simulation by generalizing the free-axis selection (Fraxis) algorithm~\cite{watanabe2021}, which makes full use of degree of freedom with respect to a single-qubit gate.
It is different from the previous work~\cite{benedetti2021}, where the objective function is analytically optimized for axis-fixed rotation gates by the analog of the NFT(Rotosolve) method.
Since a general single-qubit gate with three parameters is decomposed into $R_z$-$R_y$-$R_z$ gates, the NFT can be applied to a general single-qubit gates by sequential update of the three gates.
However, it may fail to consider the correlation among parameters of the $R_z$-$R_y$-$R_z$ gates.

In contrast, the new optimization method can incorporate correlation between all parameters of a single-qubit gate.
The three parameters of a single-qubit gate corresponds to a three-dimensional rotation which is best captured by the selection of \textit{quaternion} system, and hence the name \textit{free quaternion selection} of our proposed method.
It is also worth noting, when $\eta=1~(|\Lambda_p|=1,~\forall p$), the full optimization of single-qubit gate by FQS can be conducted with only seven types of direct measurements without Hadamard test, which is smaller than the types required for sequential optimization of a decomposed generalized single-qubit gate proposed in the previous work~\cite{benedetti2021}.

In the next subsections, we firstly introduce the FQS formulation based on imaginary time evolution using the objective function for the most hardware-efficient case $\eta=1$, where real time $t$ is replaced by imaginary time $\tau$ as $t\to -i\tau$.
More specifically, in imaginary time evolution, a target state as in Eq.~(\ref{eqn:propagation}) becomes
\begin{equation}\label{eqn:imaginary_time_propargation}
    \frac{1}{\sqrt{\mathcal{N}}}e^{- h \hat{O} \Delta \tau}U(\vec{\vartheta}') \ket{\bm{0}},
\end{equation}
where $\Delta \tau \equiv \tau/N $, and $\mathcal{N}$ is a normalization factor as $\mathcal{N} = \|e^{-h \hat{O} \Delta \tau}U(\vec{\vartheta}') \ket{\bm{0}}\|^2$, which can be ignored in optimization problem of the objective function Eq.~(\ref{eqn:primitive_objective_function}).
In the Result section, we demonstrate the applications of imaginary time evolution for finding the ground state of the 1D Heisenberg model and \ce{H2} molecule. 

Although we supposed $\eta=1$ and imaginary time evolution for simplicity in the following derivation,
we emphasize that the FQS algorithm is neither limited to the most hardware-efficient objective functions of $\eta=1$ nor imaginary time evolution.
Note that the FQS algorithm with $\eta\neq1$ is a simple extension of $\eta=1$ by using a coordinate-wise update for each general single-qubit gate in the same $\Lambda_p$ as in Algorithm 1.
In particular, we also demonstrate the FQS application to simultaneous optimization for an excitation-conserving gate consisting of two parameterized single-qubit gates, and thus $\eta=1/2$.

The appropriate hardware efficiency $\eta$ should be determined from the trade-off between the performance of quantum devices and the required simulation accuracy.
To demonstrate this point, we also applied FQS to real time evolution of the 1D Ising model with several hardware-efficiency levels.

\section*{B. The proposed method}
\subsection*{Free Quaternion Selection for Quantum Simulation}
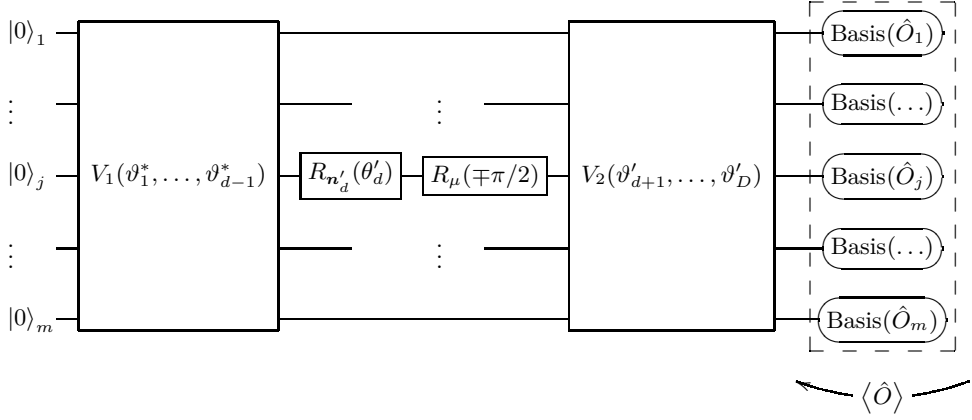
\begin{figure*}[t]
\centering
\begin{tabular}{c}
\Qcircuit @C=0.9em @R=1.2em {
  &\lstickx{\ket{0}_1} & \multigate{4}{V_1(\vartheta^*_1,\ldots,\vartheta^*_{d-1})} & \qw & \qw& \multigate{4}{V_2(\vartheta'_{d+1},\ldots,\vartheta'_{D})} & \qw & \measure{\mbox{Basis}(\hat{O}_1)} \\ 
  &\lstickx{\vdots}     & \ghost{V_1(\vartheta^*_1,\ldots,\vartheta^*_{d-1})} & \qw & \lstickx{\vdots} & \ghost{V_2(\vartheta'_{d+1},\ldots,\vartheta'_{D})} & \qw &  \measure{\mbox{Basis}(\ldots)}\\
  &\lstickx{\ket{0}_j} & \ghost{V_1(\vartheta^*_1,\ldots,\vartheta^*_{d-1})} & \gate{R_{\bm{n}'_d}(\theta'_d)} & \gate{R_{\mu}(\mp\pi/2)}& \ghost{V_2(\vartheta'_{d+1},\ldots,\vartheta'_{D})}  & \qw &  \measure{\mbox{Basis}(\hat{O}_j)}\\
  &\lstickx{\vdots}     & \ghost{V_1(\vartheta^*_1,\ldots,\vartheta^*_{d-1})} & \qw & \lstickx{\vdots} & \ghost{V_2(\vartheta'_{d+1},\ldots,\vartheta'_{D})} & \qw &  \measure{\mbox{Basis}(\ldots)}\\
  &\lstickx{\ket{0}_m} & \ghost{V_1(\vartheta^*_1,\ldots,\vartheta^*_{d-1})} & \qw & \qw& \ghost{V_2(\vartheta'_{d+1},\ldots,\vartheta'_{D})} & \qw &  \measure{\mbox{Basis}(\hat{O}_m)}\\ 
  & & & & & & & &\measo{ll}\gategroup{1}{8}{5}{8}{.7em}{--}}
\end{tabular}
\caption{The quantum circuits for computing the expected values $\mathcal{Q}_{\pm,\mu},(\mu=0,x,y,z)$ as in Eq.~(\ref{eqn:observable}) to update the parameter of the $d$th single-qubit gate. 
Here, $R_{\bm{n}'_d}(\theta'_d)$ is the $d$th single-qubit gate with a parameter $\vartheta'_d=(\theta'_d, \bm{n}'_d)$ obtained in the previous processes. 
$R_{\mu}(\pm \pi/2)$, which is defined as $e^{\mp i\sigma_{\mu}\pi/4}$, is inserted after the gate of interest. 
$V_1$ is the sub-circuit containing all parameterized single-qubit gates that have been updated, $V_2$ is that containing the rest of single-qubit gates to be updated, and $\hat{O}$ is that containing measurements related to the Trotter-decomposition term (namely, the $j$-th qubit is rotated by $H$ or $HS^\dagger$ before measured in the computational basis when $\hat{O}_j$ is $\sigma_x$ or $\sigma_y$, respectively).}
  \label{fgr:PQC_measureQ}
\end{figure*}
A general single-qubit gate with parameters of a rotational angle $\theta$ and a rotational axis $\bm{n}$ is written as
\begin{equation}\label{eqn:general}
  R_{\bm{n}}(\theta)\equiv e^{-{i}\frac{\theta}{2}\bm{n}\cdot\bm{\sigma}}=\cos{\left(\frac{\theta}{2}\right)}\sigma_0-{i}\sin{\left(\frac{\theta}2\right)}\bm{n}\cdot\bm{\sigma},
\end{equation}
where $\sigma_0$ is identity and $\bm{\sigma}=(\sigma_x, \sigma_y, \sigma_z)$.
Here, $\bm{n}=(n_x,n_y,n_z)$ denotes a normalized vector corresponding to the rotational axis. 
Suppose a PQC $U(\vec{\vartheta})$ consisting of $D$ general single-qubit gates.
In the PQC, a parameter set $\vartheta_d$ of the $d$th parameterized gate denotes
\begin{equation}\label{eqn:parameterset}
  \vartheta_d=(\theta_d,\bm{n}_d)=(\theta_d,{n}_{dx},{n}_{dy},{n}_{dz}),
\end{equation}
where $\theta_d\in \mathbb{R}$ and $\|\bm{n}_d\|=1$.
For simplicity, we suppose the $d$th gate set $\Lambda_d$ contains only the $d$th parameterized gate (i.e., the most hardware-efficient case $\eta=1$) and the total number of the gate sets is $D$.
Then, the unitary operator with $\Lambda_d$ in Eq.~(\ref{eqn:generalized_objective}) is written as
\begin{equation}\label{U_tilde}
    {U}(\vartheta_d;\vec{\vartheta}^{(d-1)})=V_2R_{\bm{n}_d}(\theta_d)V_1,
\end{equation}
where $V_1$ and $V_2$ denotes the unitary operators corresponding to the partial circuits prior and posterior to the $R_{\bm{n}_d}$ in the PQC, respectively.
Substituting Eq.~(\ref{U_tilde}) into Eq.~(\ref{eqn:generalized_objective}) we obtain 
\begin{align}\label{gate_optimization}
  \mathcal{F}^{(d)}(\theta_d,\bm{n}_d)&= \sqrt{g^2_0+\left(\bm{{n}}_d\cdot \bm{g}\right)^2} \nonumber \\[4pt]
  &~~~\times \sin{\left[\frac{\theta_d}{2}+{\rm arctan2}\left(g_0,\bm{n}_d\cdot \bm{g}\right)\right]},
\end{align}
where $g_0$ and $\bm{g}=(g_x,g_y,g_z)$ are defined as
\begin{align}\label{eqn:g_mu}
  g_\mu &\equiv {\rm Re}\left[\bra{\bm{0}} {U}^\dagger(\vec{\vartheta}^{(d-1)}) e^{-h \hat{O}{\Delta\tau}/{D}} V_2 \varsigma_\mu V_1\ket{\bm{0}}\right]\notag \\[6pt]
  &=\cosh{\left(\frac{\Delta\tau}{D}h\right)} {\rm Re}\left[{\rm tr} \left(\varsigma_{\mu}R^\dagger_{\bm{n}'_d}(\theta'_d)
  \rho'\right)\right]\notag\\[4pt]
  &~~~-\sinh{\left(\frac{\Delta\tau}{D}h\right)} {\rm Re}\left[{\rm tr} \left(\hat{O}'\varsigma_{\mu}R^\dagger_{\bm{n}'_d}(\theta'_d) \rho'\right)\right],
\end{align}
where $\vartheta'_d=(\theta_d', \bm{n}'_d)$ is the $d$th component of $\vec{\vartheta}^{(d-1)}$, and $\varsigma_\mu \in \{\sigma_0, -i \sigma_x,-i \sigma_y,-i \sigma_z \}$ (See Appendix~A, B for detailed derivation).
Here, we used the following notations,  
\begin{equation}
    \hat{O}'\equiv V_2^{\dagger }\hat{O} V_2,~~~\rho' \equiv  R_{\bm{n}'_d}(\theta'_d)V_1\ket{\bm{0}} \bra{\bm{0}} V_1^{\dagger} R^\dagger_{\bm{n}'_d}(\theta'_d).
\end{equation}
Because the objective function Eq.~(\ref{gate_optimization}) has sinusoidal form, the optimal parameter $\vartheta^*_d=(\theta^*_d, \bm{n}^*_d)$ are trivially obtained as
\begin{equation}\label{eqn:optimal_parameter}
  \bm{{n}}^*_d=\frac{\bm{g}}{\|\bm{g}\|},~\theta^*_d=\pi-2\mathrm{arctan2}\left(g_0,\|\bm{g}\|\right)+4l\pi,
\end{equation}
where we choose $l \in \mathbb{Z}$ satisfying $\theta_d^*\in[0,4\pi]$.
Thus, the optimal value of $\vartheta^*_d$ can be determined from $g_\mu$.

The first term in Eq.~(\ref{eqn:g_mu}) can be determined by a parameter $\vartheta'_d=(\theta'_d, \bm{n}'_d)$ without quantum computation as (See Appendix B)
\begin{align}\label{eqn:first_term}
  {\rm Re}\left[{\rm tr}\left(\varsigma_{\mu}
  R^\dagger_{\bm{n}'_d}(\theta'_d)\rho'\right)\right]&=\begin{cases}
  \cos{\left({\theta'_d}/{2}\right)},&\mu=0\\[4pt]
  n'_{d\mu}\sin{\left({\theta'_d}/{2}\right)},&\mu\neq 0
  \end{cases}.
\end{align}
On the other hand, quantum computations are required for the second term of Eq.~(\ref{eqn:g_mu}).
In general, the evaluation of the second terms requires four types of measurements in total using the Hadamard test with controlled operation on $\varsigma_\mu R^{\dagger}_{\bm{n}'_d}(\theta'_d)$.
For $\eta=1$, however, we can replace the Hadamard test with modest number of direct measurements. 
\renewcommand{\baselinestretch}{1.3}
\begin{figure*}[!t]
 \begin{minipage}{\linewidth}
 \begin{algorithm}[H]
	\caption{Imaginary Time Evolution with Free Quaternion Selection for Quantum Simulation}\label{alg:FQS}
	\begin{algorithmic}
	\Input~{(1) Trotter-decomposed time propagators $\{e^{-h_k\hat{O}_k\Delta\tau }\}_{k=1}^K$, (2) a PQC $U(\vec{\vartheta})$ with $D$ general single-qubit gates, $\vec{\vartheta}=(\vartheta_1,\cdots,\vartheta_D),~\vartheta_d=(\theta_d,\bm{n}_d)\in{\mathbb{R}^4},~\|\bm{n}_d\|=1$, and (3)~the number of time step $N$.}
	\EndInput
	\State \textbf{Initialize}~Choose initial gate parameters $\vec{\vartheta}$. 
	\Repeat
	   \For{$k$ = 1, 2, \ldots, $K$}
		  \For{$d$ = 1, 2, \ldots, $D$}
		        \State{Determine} $\mathcal{Q}_{+,0},\mathcal{Q}_{\pm,x},\mathcal{Q}_{\pm,y}$, $\mathcal{Q}_{\pm,z}$ with  \textit{quantum computers}~~~[Eq.~(\ref{eqn:observable})] 
		        \State Set a function $\mathcal{G}_d(\theta,\bm{n})$ by $\mathcal{Q}_{+,0},\mathcal{Q}_{\pm,x},\mathcal{Q}_{\pm,y}$, $\mathcal{Q}_{\pm,z}$
		        \For{$\mu = 0, x, y, z$}
		           \State Compute $\varphi_\mu,  \bm{\bar{n}}_\mu$~~~ [Eq.~(\ref{eqn:varphi_0})--(\ref{eqn:varphi_mu})] 
		          \State Compute $g_\mu=(g_0,\bm{g})$ from $\vartheta_d, \mathcal{G}_d(\varphi_\mu,\bm{\bar{n}}_\mu)$~~~[Eq.~(\ref{eqn:g_mu})]
	            \EndFor
	            \State {Set} $\vartheta_d\leftarrow(\theta^*_d,\bm{n}^*_d)=({\pi}-2\mathrm{arctan2}\left(g_0,\|\bm{g}\|\right),~{\bm{g}}/{\|\bm{g}\|})$
	       \EndFor
	   \EndFor
	\Until{$N\Delta\tau$ time evolution}
	\Output~Gate parameters $\vec{\vartheta}$
	\EndOutput
	\end{algorithmic}
 \end{algorithm}
 \end{minipage}
\end{figure*}
\renewcommand{\baselinestretch}{1.0}

In the following part of this section, we describe the details of the direct measurement protocol.
First, we note $\varsigma_\mu R^{\dagger}_{\bm{n}'_d}(\theta'_d)$ in the second term of Eq.~(\ref{eqn:g_mu}) can be transformed into single-qubit gate as $R_{\bm{\bar{n}}_\mu}(\varphi_\mu)$, where $\varphi_\mu$ and $\bm{\bar{n}}_\mu$ are determined with $\theta'$ and $\bm{n}'_d$ that are obtained in the previous processes as for $\mu = 0$ 
\begin{align} \label{eqn:varphi_0}
   \varphi_0 = -\theta'_d,~~~\bm{\bar{n}}_0=\bm{n}'_d,
\end{align}
and for $\mu \neq 0 $ 
\begin{align} \label{eqn:varphi_mu}
   \varphi_\mu &=
   2\mathrm{arctan2}\left(\sqrt{1-n'_{d\mu}\sin^2{\left(\frac{\theta'_d}{2}\right)}}, ~n'_{d\mu} \sin{\left(\frac{\theta'_d}{2}\right)} \right),\notag\\[4pt]
   \bar{n}_{\mu t} &=
   \frac{1}{\sqrt{1-{n'}^2_{d\mu}\sin^2{\left(\frac{\theta'_d}{2}\right)}}} \notag\\
   &~~~\times {\left(\cos{\left(\frac{\theta'_d}{2}\right)}\delta_{\mu t}-\sin{\left(\frac{\theta'_d}2\right)} \sum_{s=x,y,z}\epsilon_{\mu s t}{{n}'_{ds}}\right)},\notag\\[4pt]
   &~~~t=x,y,z,
\end{align}
where $\delta_{\mu t}$ and $\epsilon_{\mu st}$ denotes the Kronecker delta and the three-dimensional Levi-Civita symbol, respectively (See Appendix B for details).
Next, we define a generator $\mathcal{G}_d$ as a function of rotation angle $\theta$ and axis $\bm{n}$
\begin{align}\label{eqn:exp_generator}
 \mathcal{G}_d(\theta,\bm{n}) 
 &\equiv{\rm Re}\left[{\rm tr}\left(\hat{O}' R_{\bm{n}}({\theta})\rho'\right)\right]
\end{align}
Furthermore, the generator is transformed as 
\begin{align}
 \mathcal{G}_d(\theta,\bm{n}) 
 &=\cos{\left(\frac{\theta}{2}\right)}\mathcal{Q}_{+,0} \notag\\[4pt]
 &~~~+\sin{\left(\frac{\theta}2\right)}
 \sum_{s=x,y,z} n_{s}\frac{\left(\mathcal{Q}_{+,s}-\mathcal{Q}_{-,s}\right)}{2},
\end{align}
where 
\begin{equation}\label{eqn:observable}
  \mathcal{Q}_{\pm,\nu}\equiv{\rm tr}\left(\hat{O}'e^{\mp i\sigma_{\nu}\pi/4}\rho'e^{\pm i\sigma_{\nu}\pi/4}\right),~~~\nu=0,x,y,z.
\end{equation}
Note that $\mathcal{Q}_{\pm,\nu}$ is independent of $(\theta, \bm{n})$ and can be evaluated by direct measurement with a PQC in Fig.~\ref{fgr:PQC_measureQ}, where a single-qubit gate $e^{\mp i\sigma_{\mu}\pi/4}$ is inserted after the gate of interest.
The generator agrees with the second term in Eq.~(\ref{eqn:g_mu}) when ($\varphi_\mu,\bm{\bar{n}}_\mu)$ is substituted into $(\theta, \bm{n})$. 
Hence, once $\mathcal{Q}_{+,0},\mathcal{Q}_{\pm,x},\mathcal{Q}_{\pm,y},\mathcal{Q}_{\pm,z}$ are obtained for the $d$th gate, we can evaluate the second term in Eq.~(\ref{eqn:g_mu}) without additional quantum computation as shown in Algorithm~\ref{alg:FQS}.
Therefore, the optimal values of single-qubit gate parameters can be analytically determined with only seven types of expectation values in the direct measurement scheme: no ancilla qubits, no additional CNOT gates, which is rather advantageous on present real devices with limited qubit connectivity and significant error from control operation.  
Therefore, we believe the direct measurement scheme of FQS as the one of the most hardware-efficient protocol for time evolving simulation.

In the following, for simplicity we refer to the algorithm described in this section as FQS(1q,~3p), where the 1q denotes its targeting parameterized single-qubit gates, and the 3p denotes the full parameterization of each gate: the (\textit{unit}) \textit{quaternion} which can be identified with a set of rotational angle and axis or direct parameters of a single-qubit gate \cite{wharton2015unit}. 
We emphasize the fact that all parameters of a single-qubit gate are simultaneously optimized in FQS(1q,~3p).
Thus, the time evolution is more accurately simulated by making full use of expressibility of a target gate.
Obviously, FQS can be used to optimize the specific single-qubit gates such as rotation gates with fixed axis (i.e., NFT~\cite{nakanishi2020sequential} or Rotosolve~\cite{ostaszewski2021structure}) and the Fraxis~\cite{watanabe2021} gate, which is equivalent to $R_{\bm{n}}(\pi)$.
PQCs consisting of fixed-axis rotation gates each with 1 parameter, or the Fraxis gates each with 2 parameters can be optimized with FQS(1q,~1p) or FQS(1q,~2p), respectively.
In particular, the objective functions tailored to the specialized FQS are shown in Eq.~(\ref{eqn:apdx_Fraxis}) and Eq.~(\ref{eqn:NFT}) at the Appendix. 
Although NFT~\cite{nakanishi2020sequential} and Fraxis~\cite{watanabe2021} are mainly used for VQE, there are many similarities with the specialized FQS in optimization procedure.
For simplicity, we also refer to FQS(1q,~1p) and FQS(1q,~2p) as NFT and Fraxis, respectively.

\begin{table*}
\begin{threeparttable}[t]
  \centering
  \caption{Examples of the special two-qubit gates in the form of $AR_{\bm{n}}(\theta)BR^\dagger_{\bm{n}}(\theta)C$.
  The parameterized single-qubit gate $R_{\bm{n}}(\theta)$ is supposed to act on the second qubit (more precisely, $R_{\bm{n}}(\theta)$ denotes $I_1\otimes R_{\bm{n}}(\theta)$).
  The arbitrary gates $A$, $B$, and $C$ without parameters are represented by $Z_2$ (Pauli Z), $X^{\rm{c}}_{i,j}$ (CNOT), $Z^{\rm{c}}_{i,j}$ (CZ) gates, and their products, where the superscript c represents a controlled-gate with the first (control) and second (target) subscripts.
  In the right-most column, the direct sum of matrices $1 \oplus \mathbf{M} \oplus \pm 1$ denotes the block diagonal matrix $\mbox{Diag}\left(1, \mathbf{M}, \pm 1\right)$.
  }
  \label{tbl:multi-qubit gates}
  \begin{tabular}{cccccc}
    \hline
    Gate Type&$A$&$B$&$C$&$R_{\bm{n}}(\theta)$&$AR_{\bm{n}}(\theta)BR^\dagger_{\bm{n}}(\theta)C$\\
    \hline
    \hline
    excitation-$\mathrm{conserving}\tnote{a}$
    &{$X^{\rm c}_{2,1}$}
    &{$Z^{\rm c}_{1,2}$}
    &{$X^{\rm c}_{2,1}$}
    &\begin{tabular}{c}
    $\theta=\pi$\\
    $\bm{n}(\psi,\phi)$
    \end{tabular}
    &{$1\oplus\begin{pmatrix}
    -\cos{\psi}&e^{i\phi}\sin{\psi}\\
    e^{-i\phi}\sin{\psi}&\cos{\psi}\\
    \end{pmatrix}\oplus1$} \rule[-6mm]{0mm}{14mm} \\
    swap
    &{$X^{\rm c}_{2,1}$}
    &{$X^{\rm c}_{1,2}$}
    &{$X^{\rm c}_{2,1}$}&
    \begin{tabular}{c}
    $\theta$\\
    $\bm{n}=(0,0,1)$
    \end{tabular}
    &{$1\oplus\begin{pmatrix}
    0 & e^{i\theta}\\
    e^{-i\theta} & 0\\
    \end{pmatrix}\oplus 1$}
    \rule[-6mm]{0mm}{14mm} \\
     Hop
    &$Z_{2} X^{\mathrm{c}}_{2,1}$
    &$Z^{\rm c}_{1,2}$
    &$X^{\rm c}_{2,1}$
    &\begin{tabular}{c}
    $\theta=\pi$\\
    $\bm{n}(\psi,\phi=0)$
    \end{tabular}
    &$1\oplus\begin{pmatrix}
    \cos{\psi} & -\sin{\psi}\\
    \sin{\psi} & \cos{\psi}\\
    \end{pmatrix}\oplus -1$
    \rule[-6mm]{0mm}{14mm} \\
   RBS
    &$X^{\rm c}_{2,1}$
    &$Z^{\rm c}_{1,2}$
    &{$X^{\rm c}_{2,1}Z_2Z^{\rm c}_{1,2}$}&
    \begin{tabular}{c}
    $\theta=\pi$\\
    $\bm{n}(\psi,\phi=0)$
    \end{tabular}
    &{$1\oplus\begin{pmatrix}
    \cos{\psi} & \sin{\psi}\\
    -\sin{\psi} & \cos{\psi}\\
    \end{pmatrix}\oplus 1$}
    \rule[-6mm]{0mm}{14mm} \\
    \hline
  \end{tabular}
  \begin{tablenotes}
    \item[a] excitation-conserving gate with 2 parameters ($\psi,\phi$) are represented by the polar coordinate of the rotational axis $\bm{n}$ such as $\bm{n}(\psi,\phi)=(\sin{\left(\psi/{2}\right)}\cos{(\phi)},\sin{\left(\psi/{2}\right)}\sin{(\phi)},  \cos{\left(\psi/{2}\right)})$.
    
  \end{tablenotes}
\end{threeparttable}
\end{table*}

\subsection*{Free quaternion selection for multi-qubit gate}
Here, we extend the FQS to special multi-qubit gates that can be decomposed as, 
\begin{equation}\label{eqn:gate_decomposition}
  A\cdot R_{\bm{n}}(\theta)\cdot B\cdot R^\dagger_{\bm{n}}(\theta)\cdot C
\end{equation}
where $A$, $B$, and $C$ denote arbitrary unitary gates without parameters (such as the CNOT and Pauli gates), and $R$ and $ R^{\dagger}$ share the same parameter $\vartheta=(\theta, \bm{n})$.
Table \ref{tbl:multi-qubit gates} lists some examples of the well-known gates in this class.
They are excitation-conserving, swap, Hop, and Reconfigurable Beam Splitter (RBS) gates. 
Strictly speaking, they are all instances of the excitation-conserving (or, particle-number-preserving) gate set which can be implemented by the native gate set of the Fermionic Simulation as in~\cite{Foxen2020}, but for convenience, we name the gates as in Table \ref{tbl:multi-qubit gates}.
A product of Hop gates is known to be universal with respect to real-valued probability amplitudes of quantum states on fixed particle number, making the gates attractive for quantum chemistry applications~\cite{eddins2021doubling}.
The RBS gates have been proposed as building blocks of quantum neural network~\cite{kerenidis2021}.

Because $R_{\bm{n}}(\theta)$ and $R^{\dagger}_{\bm{n}}(\theta)$ gates share the same parameters in those multi-qubit gates, these gates are simultaneously updated with an optimization scheme similar to the FQS method.
In this case, each $\Lambda$ includes two single-qubit gates making the hardware-efficiency measure $\eta \le 1/2$.
Although in the following, we suppose the simple case that each $\Lambda_p$ consists of 
only two single-qubit gates, namely $\eta=1/2~(|\Lambda_p| = 2,~\forall p=1,2,\cdots,P)$, with shared parameters which are written as $\vartheta_p=(\theta_p,\bm{n}_p)$, it is straightforward to generalize it to $|\Lambda|>2$ by using a coordinate-wise update with a multi-qubit gate in Eq.~(\ref{eqn:gate_decomposition}).

Given the excitation-conserving gate in Table \ref{tbl:multi-qubit gates}, the objective function for the $p$th gate set is rewritten by substituting Eq.~(\ref{eqn:gate_decomposition}) and $\theta=\pi$ into 
Eq.~(\ref{eqn:generalized_objective}) as
\begin{widetext}

\begin{align}\label{eqn:2q_object_func}
  \mathcal{F}^{(p)}(\pi,\bm{n}_p)
  &=\mathrm{Re}\left[\bra{\bm{0}}U^\dagger(\vec{\vartheta}^{(p-1)})e^{-h \hat{O}{\Delta\tau}/{P}}V_2AR_{\bm{n}_p}(\pi)BR^\dagger_{\bm{n}_p}(\pi)CV_1\ket{\bm{0}}\right]\notag\\[4pt]
  &=\sum_{s,t=x,y,z}n_{ps}n_{pt}{\rm Re}\left[\bra{\bm{0}}U^\dagger(\vec{\vartheta}^{(p-1)})e^{-h \hat{O}{\Delta\tau}/{P}}V_2 A {\sigma}_{t} B {\sigma}_{s} C V_1\ket{\bm{0}}\right],
\end{align}
\end{widetext}
where $V_1$ and $V_2$ denotes the parts of the PQC as in the previous subsection.
By definition of a $3\times 3$ asymmetric matrix $G^{(p)}$ as
\begin{align}\label{eqn:G_matrix}
  &G^{(p)}_{st}\equiv{\rm Re}\left[\bra{\bm{0}}U^\dagger(\vec{\vartheta}^{(p-1)})e^{-h \hat{O}{\Delta\tau}/{P}}V_2 A {\sigma}_{t} B {\sigma}_{s} C V_1\ket{\bm{0}}\right],\notag \\[4pt]
  &s,t \in \{x,y,z\},
\end{align}
Eq.~(\ref{eqn:2q_object_func}) is transformed in a quadratic form as
\begin{align}\label{eqn:2q_gate_opt}
  &\mathcal{F}^{(p)}(\pi,\bm{n}_p)=\bm{n}_p^T G^{(p)}\bm{n}_p,
\end{align}
where superscript $T$ denotes a transpose operation.
The optimal value can be computed from the symmetric matrix 
$S^{(p)}$, defined as
\begin{equation}
  S^{(p)}\equiv\frac{G^{(p)}+\left(G^{(p)}\right)^T}{2},
\end{equation}
so that the optimal parameter becomes
\begin{equation} \label{eqn:2q_optimal_value}
  \bm{n}_p^*
  =\underset{\bm{n}_p}{\mathrm{argmax}}~\bm{n}_p^T S^{(p)} \bm{n}_p,
\end{equation}
where the eigenvector corresponding to the largest eigenvalue of $S^{(p)}$ is the analytical solution.
As for the objective function Eq.~(\ref{eqn:generalized_objective}) with respect to Swap, Hop, or RBS gates, we also derive the analytical optimization form as describe in Appendix~D.

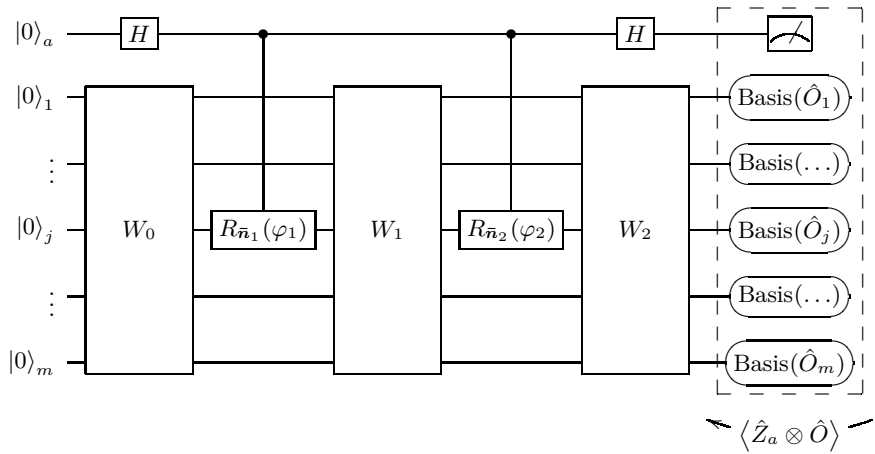
\begin{figure*}[tb]
\centering
\begin{tabular}{c}
\Qcircuit @C=0.75em @R=1.0em {
   &\lstick{\ket{0}_a}             &\gate{H}           &\ctrl{3} & \qw                 &\ctrl{3}& \gate{H}           & \qw  & \meter \\ 
   &\lstick{\ket{0}_1}             & \multigate{4}{~~~W_0~~~}&\qw      & \multigate{4}{~~~W_1~~~}  &\qw     & \multigate{4}{~~~W_2~~~} & \qw & \measure{\mbox{Basis}(\hat{O}_1)}   \\
   &\lstick{\vdots}             & \ghost{~~~W_0~~~}       &\qw      & \ghost{~~~W_1~~~}  &\qw     & \ghost{~~~W_2~~~} & \qw & \measure{\mbox{Basis}(\ldots)}   \\
   &\lstick{\ket{0}_j}             & \ghost{~~~W_0~~~}       &\gate{R_{\bm{\bar{n}}_1}(\varphi_1)}      & \ghost{~~~W_1~~~}  &\gate{R_{\bm{\bar{n}}_2}(\varphi_2)}     & \ghost{~~~W_2~~~} & \qw & \measure{\mbox{Basis}(\hat{O}_j)}   \\
   &\lstick{\vdots}             & \ghost{~~~W_0~~~}       &\qw      & \ghost{~~~W_1~~~}  &\qw     & \ghost{~~~W_2~~~} & \qw & \measure{\mbox{Basis}(\ldots)}   \\
   &\lstick{\ket{0}_m}             & \ghost{~~~W_0~~~}       &\qw      & \ghost{~~~W_1~~~}  &\qw     & \ghost{~~~W_2~~~} & \qw & \measure{\mbox{Basis}(\hat{O}_m)}  \\
   & & & & & & &  & &  \measoz{ll}\gategroup{1}{9}{6}{9}{.7em}{--}
}
\end{tabular}
\caption{A quantum circuit for Hadamard test to evaluate Eq.~(\ref{eqn:expectationsfor2qopt}) with respect to the optimization of the $p$th gate set. $\ket{0}_a$ is the ancilla qubit. $W_0$, $W_1$, and $W_2$ are the components of the PQC defined in Eq.~(\ref{eqn:components_expectationsfor2qopt}), where $W_0$ includes the gates that have been updated, and $W_2$ includes the rest of gates to be updated, and $R_{\bm{\bar{n}}_1}(\varphi_1), R_{\bm{\bar{n}}_2}(\varphi_2)$ act on the same qubit as the $d$th gate. 
Using the circuits, we can evaluate the expectations $\langle \hat{Z}_a\otimes \hat{O}\rangle$ and $\langle \hat{Z}_a\otimes \hat{I}\rangle$, and the linear combination of them with their coefficients $-\sinh{(\Delta\tau h/P)}$ and $\cosh{(\Delta\tau h/P)}$ yields the quantity ${\rm Re}\left[\bra{\bm{0}}U^\dagger(\vec{\vartheta}^{(p-1)})e^{-h \hat{O}{\Delta\tau}/P}W_2 R_{\bm{\bar{n}}_2}(\varphi_2)W_1R_{\bm{\bar{n}}_1}(\varphi_1)W_0 \ket{\bm{0}} \right]$.}
\label{fgr:circ_Gh}
\end{figure*}

To obtain a solution of Eq.~(\ref{eqn:2q_optimal_value}), we need to evaluate the elements of $G^{(p)}$ in Eq.~(\ref{eqn:G_matrix}), 
which can be written in an unified expression
similarly to the previous subsection, as
\begin{align}\label{eqn:expectationsfor2qopt}
    &{\rm Re}\Bigl[
    \bra{\bm{0}}U^\dagger(\vec{\vartheta}^{(p-1)})e^{-h \hat{O}{\Delta\tau}/P}\notag\\
    &~~~~~~~~~~~~~~~\times W_A R_{\bm{\bar{n}}_2}(\varphi_2) W_B R_{\bm{\bar{n}}_1}(\varphi_1)W_C \ket{\bm{0}}\Bigr],
\end{align}
where $W_A$, $W_B$, and $W_C$ are defined as
\begin{align}\label{eqn:components_expectationsfor2qopt}
    W_A=V_2A,~~~
    W_B=R_{\bm{n}'_p}(\theta'_p)BR^\dagger_{\bm{n}'_p}(\theta'_p),~~~
    W_C=CV_1.
\end{align}
Note that $(\varphi_1, \bm{\bar{n}}_1)$, $(\varphi_2, \bm{\bar{n}}_2)$ are determined by each element of $G^{(p)}$ as in the previous subsection.
In addition, an analytical solution of the objective function for Swap, Hop, or RBS gates derived in Appendix~D also requires quantities in the same form as Eq.~(\ref{eqn:expectationsfor2qopt}).
In principle, these quantities in Eq.~(\ref{eqn:expectationsfor2qopt}) can be evaluated with Hadamard test with two control operations on $R_{\bm{\bar{n}}_2}(\varphi_2)$ and $R_{\bm{\bar{n}}_1}(\varphi_1)$
as shown in Fig.~\ref{fgr:circ_Gh}.
However, direct measurements without ancilla qubits and CNOT gates are available similarly as proposed in literature \cite{mitarai2019}. It should be also noted that in the case of excitation-conserving gate, eight Hadamard tests are required because $G_{xy}=-G_{yx}$, while four measurements for the Hop and RBS gates (See Appendix~D).

Here, we denote the FQS method for optimizing Eq.~(\ref{eqn:2q_optimal_value}) as \textit{FQS with $u$-qubit gates of 2 parameters}; FQS($u$q,~2p), where $u$ is the number of qubits subject to nontrivial action of $A$, $B$, and $C$. 
On the other hand, the FQS method to optimize only one degree of freedom out of three in $\vartheta_p=(\theta_p,\bm{n}(\psi_p,\phi_p))$ is termed \textit{FQS with $u$-qubit gates of 1 parameter}; FQS($u$q,~1p), which can be applied to Hop, RBS, swap, and the excitation-conserving gate with one fixed degree of freedom.

In particular, as shown in Table~\ref{tbl:multi-qubit gates}, FQS(2q,~2p) generalizes the excitation-conserving gate, the Hop gate, and the RBS gate.
In contrast to the conventional excitation-conserving gate where only rotational angle $\psi$ is the optimization target, the FQS(2q,~2p) can simultaneously update not only $\psi$ but also $\phi$, which seems to be advantageous to exhibit higher expressibility and to avoid local minimum and saddle points.
To verify this feature, we also carry out controlled experiments where the two parameters $\psi,\phi$ of an excitation-conserving gate are sequentially and separately optimized.
We also note that because the total number of electrons has to be preserved in quantum chemistry calculation, limiting the parameter search space by taking into account the preservation is essential to reduce the computational cost.
The FQS(2q,~*p) methods are useful for such preservation constraints.

%% file: results.tex
In this section, we verify the performance of the proposed FQS methods in real and imaginary time simulations of typical Hamiltonians.
The ITE simulations were executed with $\eta=1$ to find the ground state of the 1D Heisenberg model and \ce{H2} molecule.
As for real time evolution, we simulated the 1D Ising model with various hardware-efficiency. All simulations presented in the paper were carried out using statevector simulator of Qiskit~\cite{Qiskit}.
The settings of each experiments are detailed in Appendix~E.

\section*{A. 1D Heisenberg model with FQS(1q,~3p)}
Here, we consider a 5-qubit 1D Heisenberg model under the periodic boundary condition.
The Hamiltonian is given as
\begin{equation}\label{eqn:Heisenberg}
  H=J\sum_{(i,j)\in E}(X_iX_j+Y_iY_j+Z_iZ_j)+h\sum_{i\in V} Z_i,
\end{equation}
where the coupling constant $J$ and the external fields $h$ satisfy $J=h=1$, and $G=\{{V,E}\}$ is the undirected graph of the lattice with 5 nodes.
The imaginary time propagators were prepared under the first-order Trotter decomposition with fixed time step $\Delta\tau=0.50$.
We carried out 21 independent ITE simulations obtained with FQS in comparison with NFT and Fraxis, where we consistently employed an ansatz with ladder-like entangler shown in Fig.~\ref{fgr:PQC_Heisenberg}(a).

Since it is not possible to employ the identical state as the initial conditions consistently for NFT, Fraxis, and FQS, we separately compared FQS with Fraxis and NFT by preparing different initial conditions as shown in Fig.~\ref{fgr:vs_NFT_Fraxis}.
In comparison to NFT with $R_y$ ansatz, the initial rotational axes in FQS were slightly perturbed from the y-axis while identical values were assigned to the initial rotational angles, which were randomly selected.
Otherwise the axes were not updated from the initial direction keeping the state vector in real space.
Although unvarying axis is not trivial from Eq.~(\ref{eqn:g_mu}), it seems to be reasonable, because a state expressed by $R_y$ ansatz corresponds to a real vector, and a resulting state by exact ITE based on the stationary Hamiltonian initiating from a state in real vector space trivially remains in real.
Note that the axis rearrangement from the y-axis is critical for the Heisenberg model, because its ground state cannot be expressed by the $R_y$ ansatz.
Indeed, NFT based on the $R_y$ ansatz did not sufficiently reach to the ground state.
Although the effect of the initial perturbation was not pronounced in the beginning of the simulation, the difference of ITE paths of FQS and NFT became distinct around $\tau=100$ as seen in Fig.~\ref{fgr:vs_NFT_Fraxis}, which implies the complex component gradually increased in the course of the simulation and became dominant at this point.
In this case, obviously, FQS is advantageous with respect to expressibility for the ground state.

As confirmed in the previous study~\cite{watanabe2021}, the ansatz shown in Fig.~\ref{fgr:PQC_Heisenberg}(a) with the Fraxis gates can express states with lower energy than the $R_y$ ansatz.
FQS, however, made distinct difference in comparison with Fraxis as in Fig.~\ref{fgr:vs_NFT_Fraxis}, where we employed the identical initial states whose rotational axes were randomly selected. 
Since in conventional optimization rotational axes of single-qubit gates are fixed, the initial choice of the axes, namely ansatz design, plays critical roles.
In contrast, because the rotational axes are adaptively switched during simulation in circuit structure optimization such as Rotoselect \cite{ostaszewski2021structure} and Fraxis \cite{watanabe2021}, they alleviate circuit design dependency achieving higher expressibility. 
While the axis selection are limited to x-, y-, and z-axes in Rotoselect, Fraxis can find better axes under the condition of fixed rotational angle $\theta=\pi$. Nevertheless, due to the fixed rotational angles, Fraxis may not be sufficient to reproduce accurate states on ITE path.
Indeed, Fig.~\ref{fgr:vs_NFT_Fraxis} shows that the ITE path reproduced by FQS is more accurate than the one by Fraxis.
This shows that the full expressibility of single-qubit gates in FQS brings large improvement in ITE simulations.

\begin{figure}[tb]
    \centering
    \begin{tabular}{c}
    (a) A 5-qubit PQC for 1D Heisenberg model in result A\\
    \\
    \Qcircuit @C=1.2em @R=1.2em {
  \lstickx{\ket{0}_0} & \qw & \gate{U_{5l-4}} & \ctrl{1}          & \qw              & \qw& \gate{U_{5L+1}}  & \meter   & \qw     & \qw \\
  \lstickx{\ket{0}_1} & \qw & \gate{U_{5l-3}} & \control \qw & \ctrl{1}          & \qw& \gate{U_{5L+2}}  & \meter    & \qw      & \qw \\
  \lstickx{\ket{0}_2} & \qw & \gate{U_{5l-2}}& \ctrl{1}         & \control \qw  & \qw& \gate{U_{5L+3}}  & \meter    & \qw     & \qw \\
  \lstickx{\ket{0}_3} & \qw & \gate{U_{5l-1}} & \control \qw & \ctrl{1}          & \qw& \gate{U_{5L+4}} & \meter   & \qw      & \qw \\
  \lstickx{\ket{0}_4} & \qw & \gate{U_{5l}} & \qw              & \control \qw  & \qw& \gate{U_{5L+5}}  & \meter   & \qw      & \qw \\
  & & & & &    \arrep{llll}
  \gategroup{1}{3}{5}{5}{.5em}{--}}\\
  \\
  (b) A 4-qubit PQC for 1D Ising model in result C\\
  \\
  \Qcircuit @C=1.2em @R=1.2em {
  \lstickx{\ket{0}_0} & \qw & \gate{U_{4l-3}} & \ctrl{1}    & \qw           & \qw        & \qw& \gate{U_{4L+1}}  & \meter   & \qw     & \qw \\
  \lstickx{\ket{0}_1} & \qw & \gate{U_{4l-2}} & \control \qw& \ctrl{1}      & \qw        & \qw& \gate{U_{4L+2}}  & \meter   & \qw     & \qw \\
  \lstickx{\ket{0}_2} & \qw & \gate{U_{4l-1}} & \qw         & \control \qw  & \ctrl{1}   & \qw& \gate{U_{4L+3}}  & \meter   & \qw     & \qw \\
  \lstickx{\ket{0}_3} & \qw & \gate{U_{4l}}   & \qw         & \qw           & \control \qw   & \qw& \gate{U_{4L+4}}  & \meter   & \qw     & \qw \\
  & & & & & &    \arrep{lllll}
  \gategroup{1}{3}{4}{6}{.5em}{--}}
    \end{tabular}
    \caption{PQCs for result A and result C. The PQC (a) and (b) are used for simulating imaginary time evolution with a 5-qubit 1D Heisenberg model and real time evolution with a 4-qubit 1D Ising model, respectively. The number of layers is represented by $L~(l=1,\cdots,L)$, and $U_d$ is the $d$th single-qubit gate whose type is determined corresponding to the optimization method, e.g. $U_d=R_{y}(\theta)$ for NFT, $U_d=R_{\bm{n}}(\pi)$ for Fraxis, and $U_d=R_{\bm{n}}(\theta)$ for FQS in the first part of result A.} \label{fgr:PQC_Heisenberg}
\end{figure}

\begin{figure}[tb]
  \centering
  \includegraphics[scale=0.7]{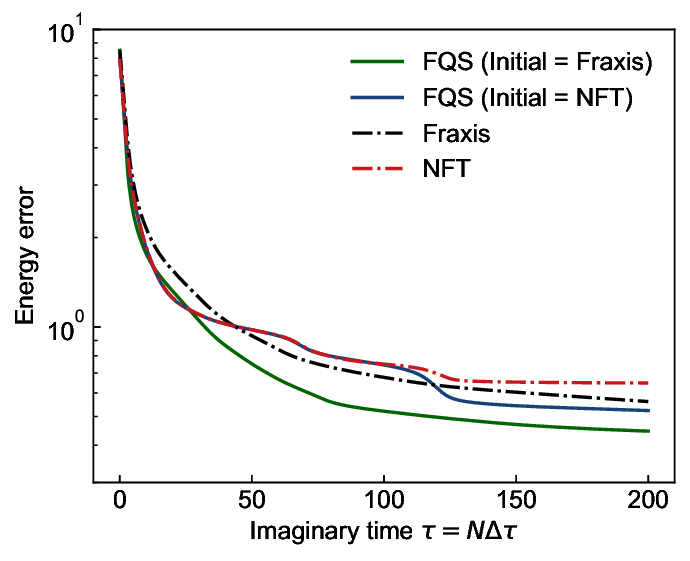}
  \caption{Averaged trajectories of 21 independent simulations for 1D Heisenberg model using the PQC shown in Fig.~\ref{fgr:PQC_Heisenberg}(a) with two layers. Black and red lines represent the results obtained with NFT and Fraxis, where initial rotational angles and axes are randomly selected, respectively. Blue and Green lines represent the result with FQS, where the initial states are identical to those of NFT and Fraxis, respectively. }\label{fgr:vs_NFT_Fraxis}
\end{figure}
In addition to higher expressibility, the advantage of FQS is its simultaneous optimization of multi parameters. 
Similar to NFT and Fraxis, FQS can analytically obtain the exact landscape of cost function.
While the number of parameters to be updated at each iteration is limited to one in NFT and two in Fraxis, FQS can update at most three parameters which fully parameterize a single-qubit gate. 
To confirm it, we carried out additional ITE simulations for the 1D Heisenberg model in Eq.~(\ref{eqn:Heisenberg}) with the ansatz in Fig.~\ref{fgr:PQC_Heisenberg}(a).

For fair comparison, we prepared two settings such that two simulations were performed on PQC with equivalent expressibility.
In the first condition (Setting-A), all $D=15$ single-qubit gates $U_d(\vartheta)$ in the ansatz with two layers were treated by FQS(1q,~3p) as shown in Table \ref{tbl:simulation_settings}.
On the other hand, in the second condition (Setting-B), these 15 gates were decomposed into 15 $R_y$ and 30 $R_z$ gates ($D=45$) as $U(\vartheta) = R_{z}(\phi)R_{y}(\psi)R_{z}(\lambda)$, and the 45 gates were sequentially optimized with NFT.

\begin{figure}[tb]
    \centering
    \includegraphics[scale=0.52]{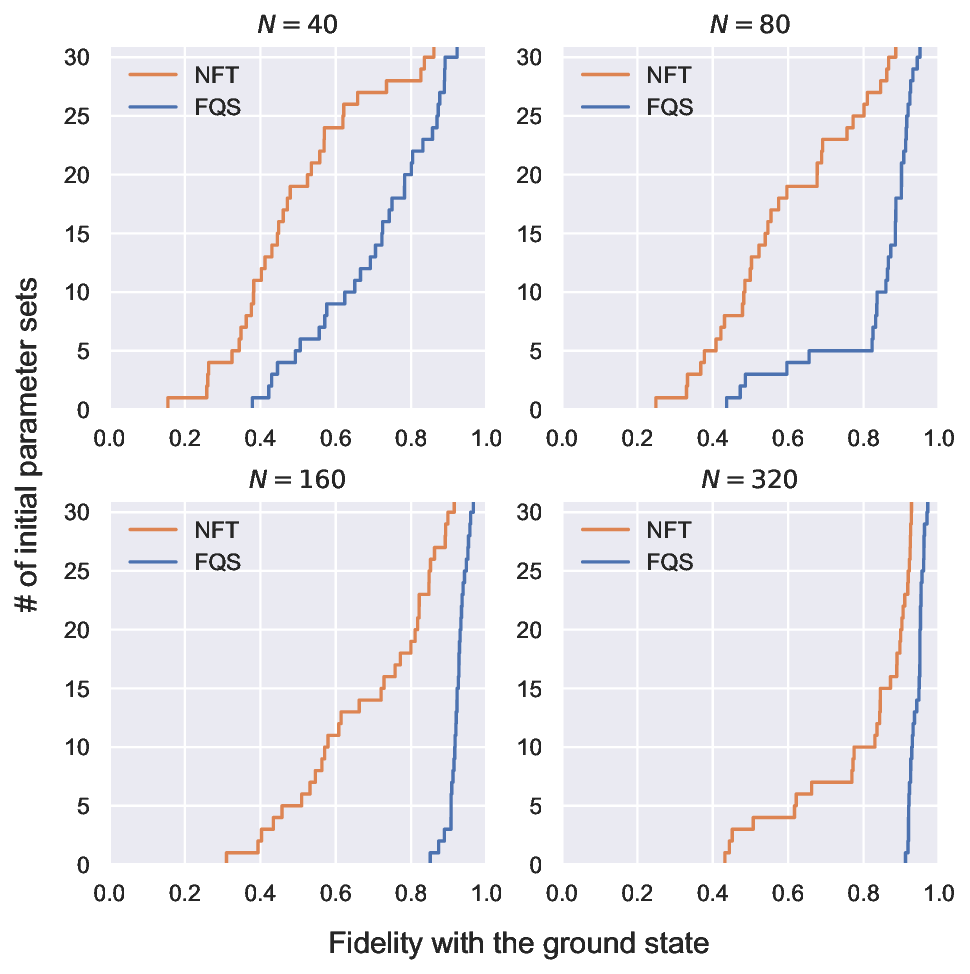}
    \caption{Cumulative distributions of the fidelity between the ground state and the imaginary time evolved state in 1D Heisenberg model using the PQC shown in Fig.~\ref{fgr:PQC_Heisenberg}(a) with 2 layers. Orange and blue lines represent the cumulative distributions of NFT and FQS(1q,~3p), respectively.}
    \label{fgr:Heisenberg_Fidelity}
\end{figure}

We carried out 30 independent simulations with different initial parameters that were shared by the two settings, where the initial rotational axes were determined randomly, and the initial rotational angles were fixed to $\pi$.
Figure \ref{fgr:Heisenberg_Fidelity} shows the cumulative distributions of the fidelity between the ground state and the resulting state at 40, 80, 160, and 320 time step.
Since it is less likely to be trapped at local minimum for ITE, the fidelity in both simulations became gradually larger with the time steps.
Note that because the total number of gates in NFT simulation is larger than that in FQS, actions of the respective ITE propagators in NFT are scaled down more than FQS for better reproduction even with the coordinate-wise update.
Nevertheless it is obvious that FQS reproduced ITE paths more accurately than NFT, which shows the importance to simultaneously optimize multiple parameters.
\renewcommand{\arraystretch}{1.5}
\begin{center}
\begin{table*}
  \caption{Simulation settings of imaginary time evolution in Section III-A and III-B}
  \label{tbl:simulation_settings}
  \centering
  \begin{tabular}{c||p{6em}p{6em}|p{6em}p{6em}p{6em}p{6em}}
    \hline
    Simulation setting
    &\multicolumn{1}{|p{6em}|}{\centering A}
    &\multicolumn{1}{ p{6em}|}{\centering B}
    &\multicolumn{1}{ p{6em}|}{\centering C}
    &\multicolumn{1}{ p{6em}|}{\centering D}
    &\multicolumn{1}{ p{6em}|}{\centering E}
    &\multicolumn{1}{ p{6em} }{\centering F}\\
    \hline
    Hamiltonian $\hat{H}$&\multicolumn{2}{|c|}{1D Heisenberg model}&\multicolumn{4}{c}{H$_2$ molecule}\\
    \hline
    \# of qubits&\multicolumn{2}{|c|}{5}
    &\multicolumn{4}{c}{4}\\
    \hline
    \# of Trotter terms $K$&\multicolumn{2}{|c|}{$20$}
    &\multicolumn{4}{c}{$15$}\\
    \hline
    Time step $\varDelta\tau$
    &\multicolumn{2}{c|}{0.50}
    &\multicolumn{4}{c }{1.0}\\  
    \hline
    Ansatz structure
    &\multicolumn{2}{|c|}{Fig.~\ref{fgr:PQC_Heisenberg}(a)}
    &\multicolumn{4}{ c }{Fig.~\ref{fgr:PQC_H2}}\\    
    \hline
    Simulation method&
     \multicolumn{1}{ c|}{FQS(1q,~3p)}&
     \multicolumn{1}{ c}{FQS(1q,~1p)}
    &\multicolumn{1}{|c|}{FQS(2q,~1p)}
    &\multicolumn{2}{|c|}{FQS(2q,~2p)}
    &\multicolumn{1}{ c }{FQS(2q,~1p)}\\
    \hline
    Gate type&
    \multicolumn{1}{c|}{~General 1-qubit~~}
    &\centering$R_z R_y R_z$\:
    &\multicolumn{1}{|c|}{$\phi=\pi$}
    &\multicolumn{2}{|c|}{Excitation-conserving}
    &\multicolumn{1}{c}{Fig.~\ref{fgr:gate_decomposition}}\\ 
    \hline
    \# of parameterized gates $D$ 
    &\multicolumn{1}{c|}{15}
    &\multicolumn{1}{c|}{45}
    &\multicolumn{3}{c|}{10}
    &\multicolumn{1}{c}{20}\\
    \hline
    \# of gate sets $P$ &
    \multicolumn{1}{c|}{15}
    &\multicolumn{1}{c|}{45}
    &\multicolumn{2}{c|}{5}
    &\multicolumn{2}{c}{10}\\
    \hline
    Hardware efficiency $\eta$ &
    \multicolumn{2}{c|}{1}
    &\multicolumn{4}{c}{1/2}\\    
    \hline
  \end{tabular}
\end{table*}
\end{center}
\renewcommand{\arraystretch}{1.0}
It should be also noted that NFT and FQS(1q,~3p) require three and seven types measurements per gate update, respectively (see Section~II-B).
NFT was originally proposed for Variational Quantum Eigensolver (VQE), where the number of required measurement types can be reduced to two except for the first optimization step, because the cost function is equivalent to the observable (expectation value of energy), and thus the estimated value of the cost function in the prior optimization can be reused in the following optimization.
In contrast, because in this time evolution simulation the objective function in Eq.~(\ref{eqn:primitive_objective_function}) differs from the observable in Eq.~(\ref{eqn:observable}), it is not possible to reduce the number of measurement types in each optimization step.
As a result, the NFT optimization of a general single-qubit gate requires nine measurement types because the gate is decomposed into three fixed-axis rotation gates, e.g., one $R_y$ and two $R_z$ gates.
Therefore, the computational cost  for a general single-qubit gate by FQS(1q,~3p), which requires seven measurement types, is actually smaller than that of NFT while achieving higher accuracy due to taking into account correlation among parameters.

\section*{B. \ce{H2} molecule with FQS(2q,~2p)}
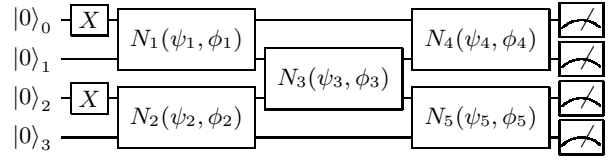
\begin{figure}[t] 
  \centering
  \begin{tabular}{c}
  \Qcircuit @C=0.4em @R=0.2em {
      \lstickx{\ket{0}_0} & \gate{X} & \multigate{1}{N_1(\psi_1,\phi_1)} & \qw & \multigate{1}{N_4(\psi_4,\phi_4)} &  \meter\\
      \lstickx{\ket{0}_1} & \qw & \ghost{N_1(\psi_1,\phi_1)} & \multigate{1}{N_3(\psi_3,\phi_3)} & \ghost{N_1(\psi_4,\phi_4)}  & \meter  \\
      \lstickx{\ket{0}_2} & \gate{X} & \multigate{1}{N_2(\psi_2,\phi_2)}& \ghost{N_3(\psi_3,\phi_3)} & \multigate{1}{N_5(\psi_5,\phi_5)} & \meter \\
      \lstickx{\ket{0}_3} & \qw & \ghost{N_2(\psi_2,\phi_2)} & \qw & \ghost{N_5(\psi_5,\phi_5)} & \meter 
      }
      \vspace{3mm}
  \end{tabular}
  \caption{
    The excitation-conserving PQC for \ce{H2} molecule with Jordan-Wigner mapping.}
  \label{fgr:PQC_H2}
\end{figure}

As described in Method Section, the FQS can be applied to the set of two-qubit gate decomposable as in Eq.~(\ref{eqn:gate_decomposition}).
In this section, we confirm performance of the FQS for excitation-conserving gates useful for quantum chemical calculation.
We chose \ce{H2} molecule with the atomic distance of 0.74 {\AA} as a benchmark system, where the molecular Hamiltonian obtained by Hartree-Fock method with STO-3G basis was mapped to 4-qubit Hamiltonian by Jordan-Wigner transformation.
The imaginary time propagators were prepared as well as in the previous section with fixed time step $\Delta\tau=1.0$. 
In this section, we employed an ansatz with the structure shown in Fig.~\ref{fgr:PQC_H2}.

When compared to the Hop and RBS gates, the excitation-conserving gate has an additional degree of freedom, which allows to express relative phase in complex space.
Although the advantage of FQS(2q,~2p) is the simultaneous update of two parameters taking the correlation into account, the advantage is not unveiled trivially in treatment of molecular Hamiltonian.
This is because the eigenstates can be represented by vectors in real space.
To confirm this point, we compared two ITE simulations, where $\psi$ and $\phi$ are variable for FQS(2q,~2p), while $\phi=\pi$ for FQS(2q,~1p).
Here, we refer the simulation conditions for FQS(2q,~1p) and FQS(2q,~2p) as Setting-C and -D, respectively, (See Table~\ref{tbl:multi-qubit gates}).
In both setting, the gate set $\Lambda_p$ consists of the single-qubit gates in $p$th excitation-conserving gate.

\begin{figure}[t]
 \centering
 \begin{tabular}{c}
   \Qcircuit @C=.3em @R=.4em {
       \lstickx{}&\qw&\multigate{1}{N(\psi,\phi)}
       &\qw \\
       \lstickx{} &\qw&\ghost{N(\psi,\phi)}
       &\qw   
       }
   \\
   \Qcircuit @C=.24em @R=.32em 
   {\Updownarrow }
   \\
   \Qcircuit @C=0.24em @R=.32em {
       & \qw &\targ &\qw &\qw &\ctrl{1}  
       &\qw &\qw &\targ &\qw\\ 
       &\qw &\ctrl{-1}  &\gate{R^{\dagger}_{z}(\phi)} &\gate{R^{\dagger}_{y}(\psi+\frac{3}{2}\pi)}&\targ &\gate{R_{y}(\psi+\frac{3}{2}\pi)}&\gate{R_{z}(\phi)} &\ctrl{-1} &\qw}
      \\
   \end{tabular}
   \vspace{3mm}
      \caption{Decomposition of the excitation-conserving gate $N(\psi, \phi)$ using $R_y$ and $R_z$ gates.
      }\label{fgr:gate_decomposition}
\end{figure}
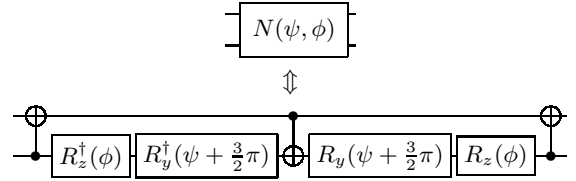

\begin{figure}[h] 
 \centering
 \begin{tabular}{c}
     (a) Applications to excitation-conserving gate\\
     \\
     \includegraphics[scale=0.3]{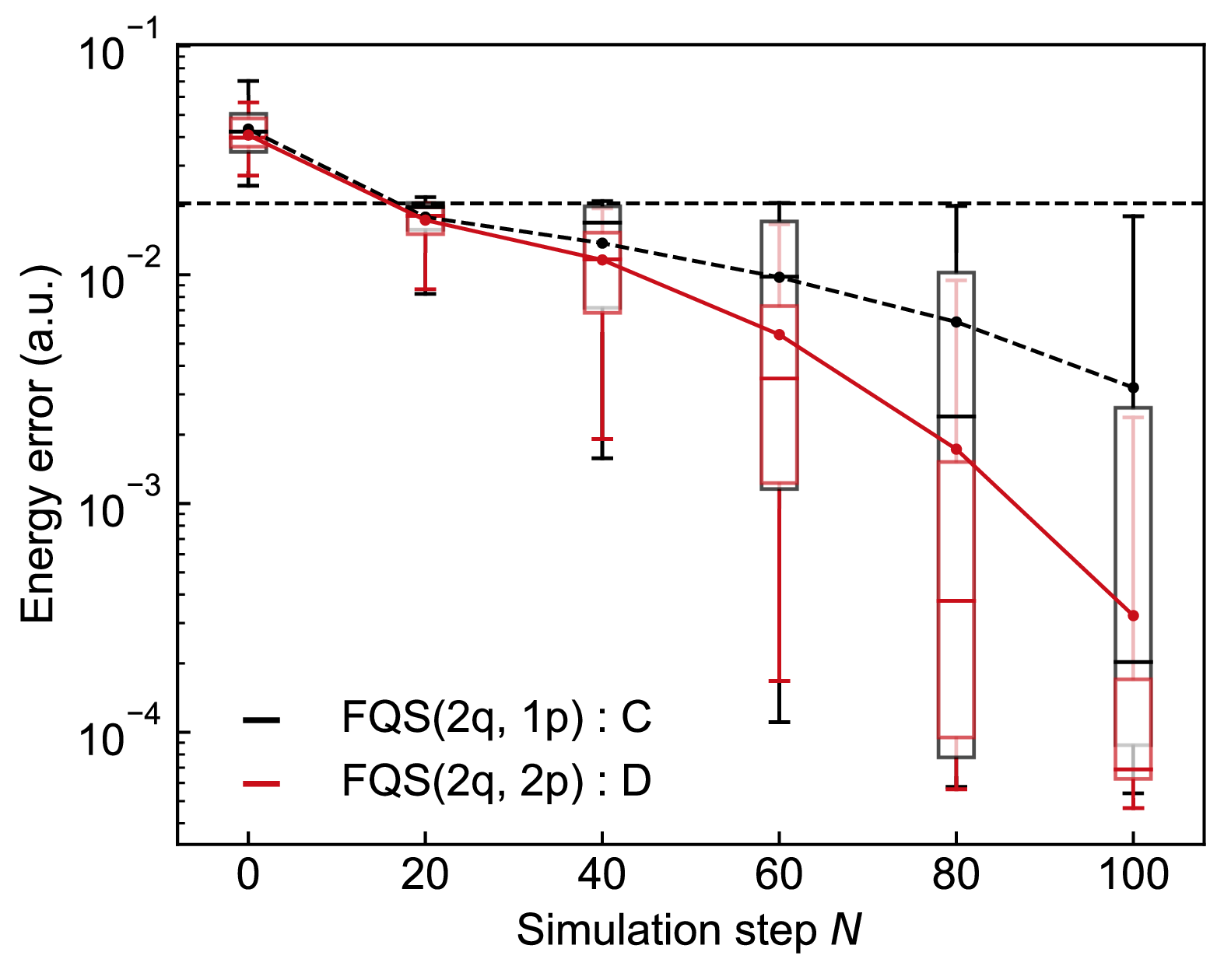}\\
     \\
     (b) Consistent number of optimizations per one propagator\\
     \\
     \includegraphics[scale=0.3]{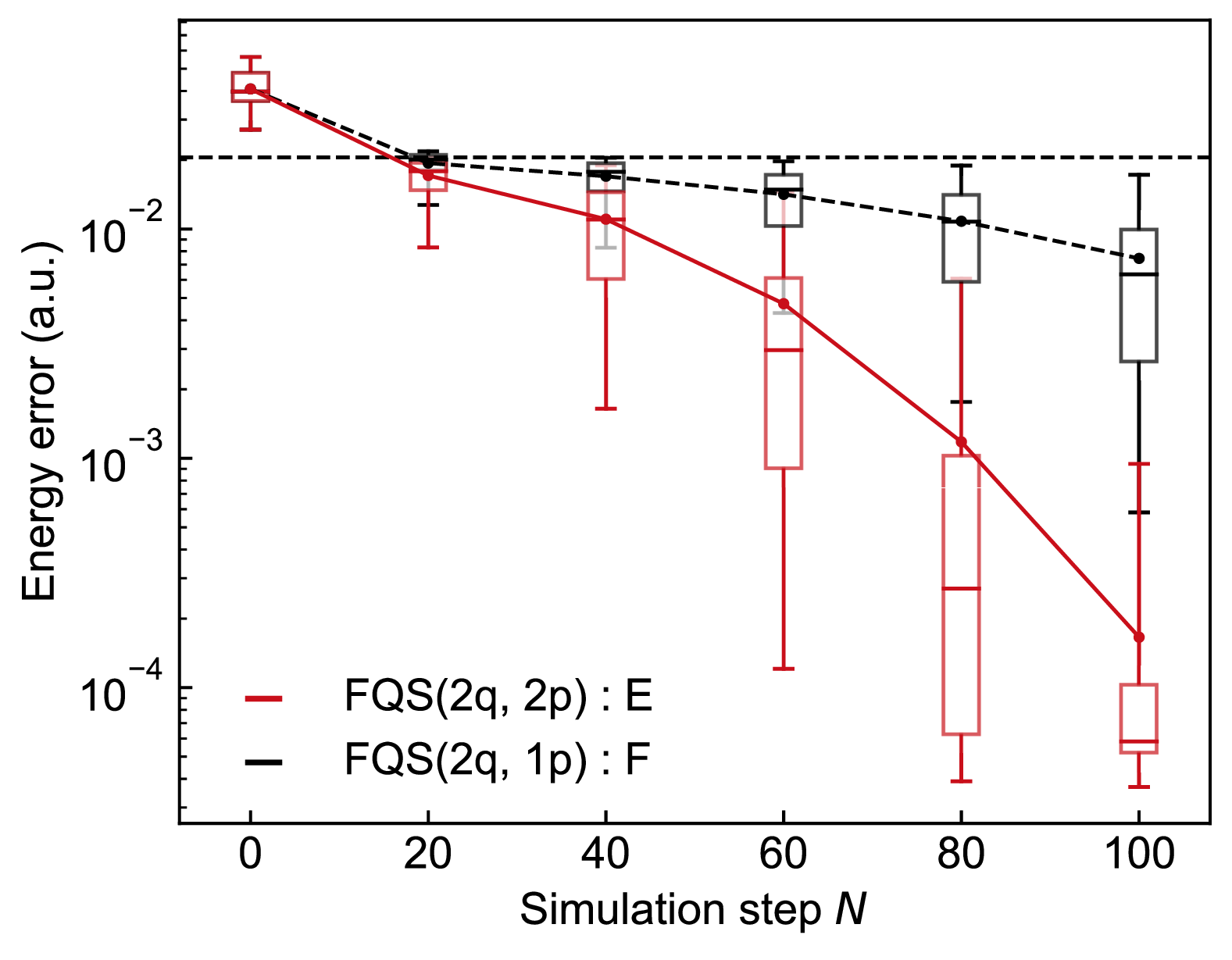}
 \end{tabular}
 \caption{Energy in the course of simulation steps $N$ with $\Delta \tau = 1$. 
 Black and red lines correspond to the results of FQS(2q, 1p) and FQS(2q, 2p), respectively. 
 Box and whisker plot denote quantiles obtained from the ITE simulation results with 20 different initial parameter sets. 
 Dash line represents the HF energy level.} 
 \label{fgr:H2_energy}
\end{figure}

Given $\phi=\pi$ as an initial condition, FQS(2q,~2p) yielded the same trajectory as FQS(2q,~1p) with $\phi=\pi$ (Data is not shown).
The unvarying behavior of $\phi$ is not trivial, but seems to be reasonable because a state represented by a real vector remains in real space by ITE with molecular Hamiltonian.
Hence, for FQS(2q,~2p) simulations we randomly chose the initial value of $\phi$. 
On the other hand, FQS(2q,~1p) and FQS(2q,~2p) shared the same initial values of $\psi$, which were randomly generated.

Figure \ref{fgr:H2_energy}(a) shows all simulations started from the similar energy level, which implies the initial states contain the ground state with amplitude in the same scale.
Note that FQS(2q,~1p) with $\phi=\pi$ reached to the chemical accuracy ($\Delta E=10^{-3}$ a.u.) in the best case, which implies the excitation-conserving ansatz restricted to $\phi=\pi$ can sufficiently cover the ground state.
However, in the worst case of FQS(2q,~1p), the energy was not improved from the HF level, which implies that the expressibility for the ground state does not necessarily guarantee sufficient expressibility to reproduce accurate ITE path.
In contrast, FQS(2q,~2p) initiating from a state in complex space outperformed FQS(2q,~1p) with $\phi=\pi$.
This is presumably because FQS initiating from complex vector can make use of larger Hilbert space which includes not only complex space represented by $\phi \ne \pi$ but also extended real space caused by the phase cancellation among $\phi$'s in excitation-conserving gates.
Thus, the time evolution in FQS can possibly be described better with $\phi \neq \pi$.

Next, we evaluated the correlation of two parameters $(\psi, \phi)$ in each excitation-conserving gate.
To this end, we decomposed an excitation-conserving gate into $R_y$ and $R_z$ gates as Fig.~\ref{fgr:gate_decomposition} according to \cite{gard2020}.
Then we sequentially updated $\psi$ and $\phi$ with FQS(2q,~1p) in different gate sets (Setting-F).
Since these multiple $R_y$ and $R_z$ gates share the parameters $\psi$ and $\phi$, respectively, required controlled operations still remain two for one excitation-conserving gate, namely $\eta=0.5$, even after the gate decomposition.
On the other hand, the number of optimization is doubled as $P=10$, that is effective time step is scale down, according to increase of the total number of parameterized gate $D$, which can lead to overestimation of its performance through scaling effect of the propagator $e^{-ih\hat{O}\Delta t /P}$ in Eq.~(\ref{eqn:generalized_objective}).
Hence, for fair comparison with FQS(2q,~1p) and FQS(2q,~2p), we carried out \textit{twice sweep update} for each optimization of Eq.~(\ref{eqn:generalized_objective}) in the FQS(2q,~2p) method, where we employed $P=10$ allowing overlap of $\Lambda$ (see Appendix E).
We refer this simulation condition as Setting-E in Table \ref{tbl:simulation_settings}.

For statistical accuracy, we independently conducted 20 ITE simulations by using randomly-generated common initial states for both FQS(2p,~1p) and FQS(2q,~2p) methods. 
Figure~\ref{fgr:H2_energy} showed a boxplot of the energy in course of simulation time, which exhibits distinct difference where FQS(2q,~2p) reached to lower energy states when compared to FQS(2q,~1p).
This discrepancy implies the importance of taking into account correlation between $\psi$ and $\phi$.

FQS(2q,~2p) requires eight-type measurements to evaluate Eq.~(\ref{eqn:expectationsfor2qopt}), because $G_{xy}=-G_{yx}$ holds for the excitation-conserving gate.
On the other hand, four-type measurements are required in FQS(2q,~1p) for the respective gates.
Therefore, the number of required measurements for a single excitation-conserving gate in Eq.~(\ref{eqn:gate_decomposition}) with FQS(2q,~2p) is equivalent to that in separate optimization with FQS(2q,~1p).

Considering the twice sweep, FQS(2q,~2p) simulation in Fig.~\ref{fgr:H2_energy}(b) are twice as expensive with respect to the measurement cost.
To evaluate the simulation accuracy with consistent measurement cost,  FQS(2p,~1p) with Setting-F should be compared with FQS(2q,~2p) with Setting-D where all parameters are updated once per a propagator. Eight types of measurements are required in both simulations.

As shown in Fig.~\ref{fgr:H2_energy}, notably, the performance of FQS(2q,~2p) (Setting-D) was almost retained in this case when compared to the twice-sweep simulation (Setting-E) in Fig.~\ref{fgr:H2_energy}(b), although the worst case in single sweep resulted in slightly larger energy. Altogether, the FQS(2q,~2p) application to the excitation-conserving gate realizes incorporation of the parameter correlation without any additional cost when compared to FQS(2q,~1p), which remarkably outweighs the time scaling of the propagator according to the number of the parameterized gates.

\section*{C. Real time evolution for 1D Ising model with FQS(1q,~3p)}

We applied FQS algorithm to real time evolution of a 4-qubit 1D Ising model with transverse-field under the open boundary condition.
The Hamiltonian is given as 
\begin{equation}
H=-J\sum_{i=0}^{N-2}Z_iZ_{i+1}-h\sum_{i=0}^{N-1} X_i,
\end{equation} 
where $J=h=1$.
We employed the ground state $\ket{0}^{\otimes 4}$ of the Hamiltonian without transverse-field as the initial state at $t=0$.
The time propagators were prepared using the first-order Trotter decomposition with fixed time step $\Delta t=0.01$.
We used the 2-layer 
PQC with a linear entangler as shown in Fig.~\ref{fgr:PQC_Heisenberg}(b), where the single-qubit gate $U_d$ represents a general single-qubit gate in FQS and $R_z$-$R_y$-$R_z$ gates in NFT so that FQS and NFT simulations were performed on PQC with equivalent expressibility as in the previous section.
\begin{figure}[h] 
  \centering
  \begin{tabular}{c}
       \includegraphics[scale=0.7]{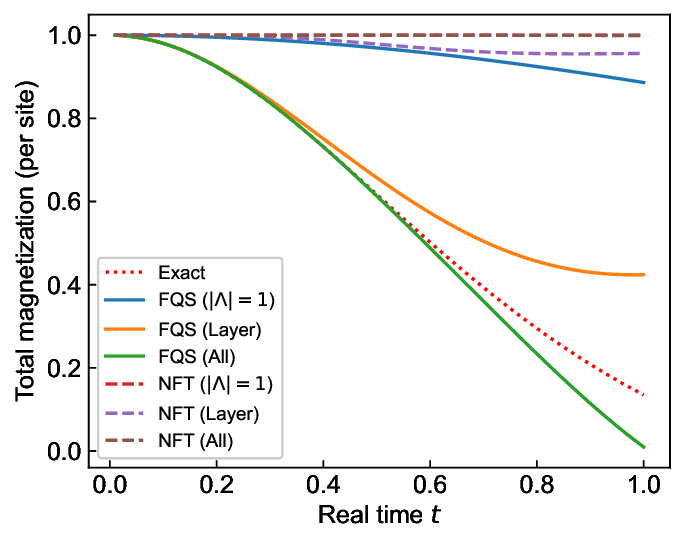}\\
       \includegraphics[scale=0.7]{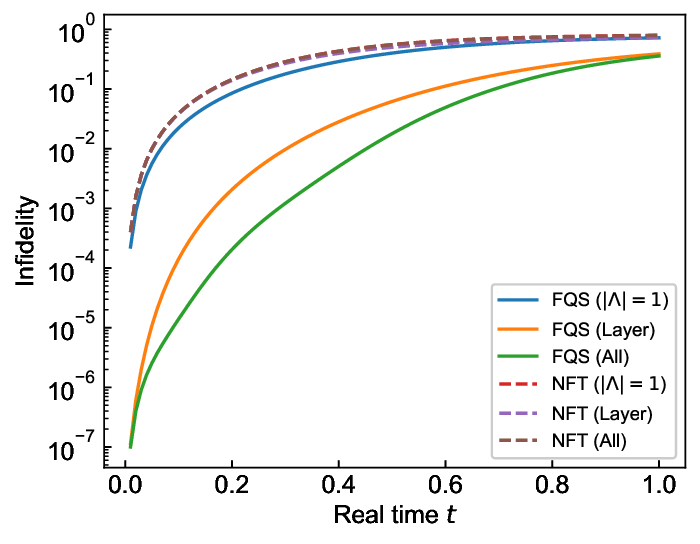}
  \end{tabular}
    \caption{Dynamics of (Top) total magnetization per site and (Bottom) the infidelity of simulation for 4-qubit 1D Ising model, where the initial states were $\ket{0}^{\otimes 4}$. 
    The solid lines and dashed lines represent the results of FQS and NFT, respectively. The red dotted line represents the exact simulation. "$|\Lambda|=1$", "Layer", and "All" denote the level of hardware-efficiency determining the gate set in one optimization problem in Eq.~(\ref{eqn:generalized_problem}).}
    \label{fgr:real_time}
\end{figure}

In order to demonstrate the trade-off between simulation accuracy and hardware efficiency, we prepared several simulation settings with different values of hardware efficiency $\eta$.
In the most hardware-efficient case ($\eta=1$), a propagator is divided by the total number of single-qubit gates $D$, and the respective single-qubit gates are updated according to Eq.~(\ref{eqn:generalized_objective}) with respect to a divided propagator $e^{-ih_k\hat{O}_k\Delta t /D}$, where $D=12$ in FQS and $D=36$ in NFT.

As the next hardware-efficient case, we employed a group update for a propagator of $e^{-ih_k\hat{O}_k\Delta t/P}$ with the objective function in Eq.~(\ref{eqn:generalized_objective}).
Here, the respective gate sets $\Lambda_p,~(p=1,\dots,P)$ consist of single-qubit gates in identical layers as in Fig.~\ref{fgr:PQC_Heisenberg}(b) ($|\Lambda|=m$ and Hardware efficiency $\eta = 1/m$, where $m$ is the number of qubits). 
Then, in the respective optimizations, we conducted FQS algorithm $|\Lambda|$ times in coordinate-wise fashion.
Because in the case of $\eta<1$, more than one parameterized gate is different between two PQCs $U^{\dagger}(\vec{\vartheta}^{(p-1)})$ and $U(\{\vartheta_d\};\vec{\vartheta}^{(p-1)}),~d\in\Lambda_p$, we employed the Hadamard test with an ancilla qubit and at most $\mathcal{O}(|\Lambda_{p}|)$ controlled gates instead of direct measurements. 

For comparison, we also carried out the simulations of $\eta=1/D$ using either NFT or FQS(1q,~3p), where all parameterized gates are sequentially updated to approximate the action of one propagator $e^{-i h_k\hat{O}_k\Delta t}$. We refer to such simulations as NFT(All) and FQS(All). 

Figure~\ref{fgr:real_time} shows total magnetization per site $\langle N^{-1}\sum_{i} Z_i\rangle$ and infidelity at each time step with different levels of hardware-efficiency, where the infidelity is defined as
\begin{equation}
  1-\|\bra{\psi_{\rm FQS/NFT}(t)}e^{-iHt}\ket{0}^{\otimes 4}\|^2.
\end{equation}
The NFT method failed to reproduce dynamics of the total magnetization for all three cases, in particular with $\eta=1/D$ and 1. Although NFT improved with the use of different initial state (See Supplementary materials), FQS still outperformed NFT for all cases.
Considering equivalent expressibility of PQC between NFT and FQS, the results imply parameter correlations within a single gate ({\it intragate} correlation) are more important in real time dynamics than in imaginary time evolution. 
It seems to be inline with the fact that the required accuracy with respect to reproduction of propagators in real-time evolution is much higher than that in imaginary time evolution to find the ground state. 
In addition, although the performance of FQS with $\eta=1$ was not necessarily satisfactory, it was drastically improved with group update in FQS with $\eta= 1/m$ and $1/D$.
The improvement was due to incorporation of the correlations between multiple gates in one group ({\it intergate} correlation) that are assigned to one divided propagator.
In FQS~$(\eta=1)$, the divided propagators are uniformly assigned to all parameterized gates.
However, although the error with respect to one divided propagator can be suppressed by fine-grained time steps, the action of each divided propagator may not be sufficiently represented by a single gate update.
On the other hands, in FQS with $\eta=1/m$ and $1/D$, the description of action of one propagator reproduced better by multi-gate update in one optimization.
Indeed, higher accuracy was achieved in FQS with $\eta<1$ where the respective propagators were assigned to more parameterized gates in one optimization.

%% file: conclusion.tex
In this paper, we proposed a new method called FQS for time evolution simulation with full optimization of a single-qubit gate with respect to its rotational angle and axis.
The time evolution is reproduced by sequential optimization of Euclidean norm between target and trial states, instead of the conventional gradient based approach.
Because FQS can incorporate correlation among parameters into optimization, it can achieve more accurate simulation.
We extended FQS to the excitation-conserving gates that have been widely employed in quantum chemical applications.
To verify the performance of the proposed method, we applied it to quantum imaginary time evolution for 1D Heisenberg model and \ce{H2} molecule and confirmed that it effectively led to quantum states that were closer to the true ground states.
We also applied FQS to real time evolution of 1D Ising model. 
Unlike imaginary time evolution, the most hardware-efficient setting with FQS did not reproduce the dynamics with satisfactory accuracy in real time evolution, although the advantage of FQS over other methods was significant. However, its dynamics accuracy was drastically improved when the hardware efficient condition was relaxed with the use of $\mathcal{O}(m)$ controlled gates, where $m$ is the number of qubits.

Although in the present work the gate updating order is fixed from left to right (from the one closest to the input qubits to that closest to the output qubits) of the quantum circuit assuming generality of ansatz and Hamiltonian, the order may not be optimal and there remain rooms for improvement. One way to determine the order is based on the support of the time propagator.
These technical improvement may allow the use of FQS in real applications implemented 
on a bigger size circuit; for instance, calculation of broad vibrational absorption spectra of floppy molecules~\cite{engel1990time}, 
simulation of short time molecular dynamics observed by femto-second time-resolved spectroscopy~\cite{zewail2000femtochemistry}.
Lastly note that, other than time evolution simulation, FQS is applicable to general 
optimization problems whose objective functions are given by Euclidean norm between 
a target state and a trial state of a PQC. 
Therefore, an improved FQS may also be potentially applicable to such optimization 
problems in a practical level. 

%% file: appendix.tex
\renewcommand{\theequation}{A.\arabic{equation}}
\setcounter{equation}{0}
\section*{A. Derivation of Free Quaternion Selection for Quantum Simulation}\label{sec:appendix_A}
In this section, we derive the objective function of $\eta=1$ ($|\Lambda_p|=1,~\forall p$) tailored to a general single-qubit gate $R_{\bm{n}}(\theta)$.
Let the parameters of the $d$th parameterized gate be $\vartheta_d = (\theta_d, \bm{n}_d) = (\theta_d, n_{dx}, n_{dy}, n_{dz})$.
Using a parameter set $\vec{\vartheta}=(\vartheta_1,\vartheta_2,\cdots,\vartheta_D)$, an arbitrary PQC is written as $U(\vec{\vartheta})$.
For simplicity, we suppose the $d$th gate set $\Lambda_d$ contains only the $d$th parameterized gate and the total number of the gate sets is $D$.
Focusing on $\Lambda_d$, the PQC is decomposed as
\begin{equation}\label{eqn:apdxA_U=VRV}
  {U}(\vartheta_d;\vec{\vartheta}^{(d-1)})=V_2R_{\bm{n}_d}(\theta_d)V_1,
\end{equation}
where $V_1$ and $V_2$ are the partial circuits prior and posterior to the $d$th gate $R_{\bm{n}_d}(\theta_d)$, respectively.
By substituting Eq.~(\ref{eqn:apdxA_U=VRV}) into the objective function $\mathcal{F}^{(d)}(\vartheta_d)$,
\begin{align}\label{eqn:apdxA_prmitive_objective}
  \mathcal{F}^{(d)}(\vartheta_d)
  &\equiv\mathrm{Re}\left[\bra{\psi^{(d-1)}_{\Delta \tau/D}}{U}(\vartheta_{d};\vec{\vartheta}^{(d-1)})\ket{\bm{0}}\right]\notag\\[4pt]
  &=\mathrm{Re}\left[\bra{\psi^{(d-1)}_{\Delta \tau/D}}V_2 V_1\ket{\bm{0}}\right]\cos{\left(\frac{\theta_d}{2}\right)}
  +{\rm Im}\left[\bra{\psi^{(d-1)}_{\Delta \tau/D}} V_2 \bm{n}_d \cdot \bm{\sigma} V_1 \ket{\bm{0}}\right]\sin{\left(\frac{\theta_d}{2}\right)}\notag \\[4pt]
  &={\rm Re}\left[\bra{\psi^{(d-1)}_{\Delta \tau/D}}V_2V_1\ket{\bm{0}}\right]\cos{\left(\frac{\theta_d}{2}\right)}
  +\sum_{s = x,y,z}n_{ds}{\rm Im}\left[\bra{\psi^{(d-1)}_{\Delta \tau/D}}V_2\sigma_{s}V_1\ket{\bm{0}}\right]\sin{\left(\frac{\theta_d}{2}\right)},
\end{align}
where $\ket{\bm{0}}\equiv\ket{0}^{\otimes m}$ and
\begin{equation}
    \ket{\psi^{(d-1)}_{\xi}}\equiv e^{-h\hat{O}\xi}{U}(\vec{\vartheta}^{(d-1)})\ket{\bm{0}}.
\end{equation}
Here, we introduce a convenient notation $a_{\mu}\equiv(a_0,\bm{a})$, where $a_{\mu},~(\mu=0,x,y,z)$ denotes the $\mu$th component of a set $(a_0, \bm{a})=(a_0, a_x,a_y,a_z)$.
Using this notation, we define $\varsigma_{\mu}$ and $g_{\mu}$ as follows,
\begin{equation}\label{eqn:apdx_sigma}
    \varsigma_{\mu}\in
    \{\sigma_0,-i\bm{\sigma}\}
    =\{\sigma_0,-i\sigma_x,-i\sigma_y,-i\sigma_z\},
\end{equation}
\begin{equation}\label{eqn:apdx_g_mu_tensor}
    g_\mu\in\{g_0,\bm{g}\}=\{g_0,g_x,g_y,g_z\},
\end{equation}
where $\sigma_0$ denotes identity, and each component of $g_\mu$ is
\begin{equation}\label{eqn:apdx_g_mu}
  g_\mu={\rm Re}\left[\bra{\psi^{(d-1)}_{\Delta \tau/D}}V_2\varsigma_{\mu}V_1\ket{\bm{0}}\right],~~~\mu=0,x,y,z.
\end{equation}
Substituting Eq.~(\ref{eqn:apdx_g_mu}) into Eq.~(\ref{eqn:apdxA_prmitive_objective}), we obtain
\begin{align}\label{eqn:apdx_gate_optimization}
  \mathcal{F}^{(d)}(\vartheta_d)
  &=g_0\cos{\left(\frac{\theta_d}{2}\right)}+\bm{n}_d\cdot \bm{g}\sin{\left(\frac{\theta_d}{2}\right)}=\sqrt{g^2_0+\left(\bm{n}_d\cdot \bm{g}\right)^2}\sin{\left[\frac{\theta_d}{2}+{\rm arctan2}\left(g_0,\bm{n}_d\cdot \bm{g}\right)\right]}.
\end{align}
The exact solution for maximization of $\mathcal{F}^{(d)}(\vartheta_d)$ is trivially given as
\begin{equation}\label{eqn:exact_solution_appendixs}
  \bm{{n}}^*_d=\frac{\bm{g}}{\|\bm{g}\|},~\theta^*_d=\pi-2\mathrm{arctan2}\left(g_0,\|\bm{g}\| \right)+4l\pi,
\end{equation}
where $l\in\mathbb{Z}$ is chosen such that  $\theta^*_d\in[0,4\pi]$.
The sign of $\bm{n}^*_d$ does not affect the expectation values while it changes the global phase.

\renewcommand{\theequation}{B.\arabic{equation}}
\setcounter{equation}{0}
\section*{B. Evaluation of $g_\mu$ without Hadamard test}\label{sec:appendix_B}
Here, we clarify that $g_\mu$ in Eq.~(\ref{eqn:apdx_g_mu}) can be estimated by evaluation of several expectation values.
To this end, we transform $g_\mu$ as
\begin{align}\label{eqn:apdx_g_mu_trans}
  g_\mu
  &= {\rm Re}\left[\bra{\psi^{(d-1)}_{\Delta \tau/D}}V_2\varsigma_{\mu}V_1\ket{\bm{0}}\right]\notag \\[4pt]
  &=\cosh{\left(\frac{\Delta\tau}{D}h\right)}{\rm Re}\left[\bra{\psi^{(d-1)}_0}V_2\varsigma_{\mu}V_1\ket{\bm{0}}\right]
  -\sinh{\left(\frac{\Delta\tau}{D}h\right)}{\rm Re}\left[\bra{\psi^{(d-1)}_0}\hat{O} V_2\varsigma_{\mu}V_1\ket{\bm{0}}\right]\notag \\[4pt]
  &=\cosh{\left(\frac{\Delta\tau}{D}h\right)}{\rm Re}\left[\bra{\bm{0}}V_1^\dagger R^\dagger_{\bm{n}'_d}(\theta'_d)\varsigma_{\mu}V_1\ket{\bm{0}}\right]
  -\sinh{\left(\frac{\Delta\tau}{D}h\right)}{\rm Re}\left[\bra{\bm{0}}V_1^\dagger R^\dagger_{\bm{n}'_d}(\theta'_d)V_2^\dagger \hat{O} V_2\varsigma_{\mu}V_1\ket{\bm{0}}\right]\notag \\[4pt]
  &=\cosh{\left(\frac{\Delta\tau}{D}h\right)}{\rm Re}\left[{\rm tr}\left(\varsigma_{\mu}R^\dagger_{\bm{n}'_d}(\theta'_d)
  \rho'\right)\right]
  -\sinh{\left(\frac{\Delta\tau}{D}h\right)}{\rm Re}\left[{\rm tr}\left(\hat{O}'\varsigma_{\mu}R^\dagger_{\bm{n}'_d}(\theta'_d)\rho' \right)\right],
\end{align}
where we employed the condition that $U(\vartheta_d;\vec{\vartheta}^{(d-1)})$ differs from $U(\vec{\vartheta}^{(d-1)})$ with respect to only the $d$th gate in the third equality.
$\vartheta'_d=(\theta'_d,\bm{n}'_d)$ denotes the $d$th component of $\vec{\vartheta}^{(d-1)}$, and $\rho', \hat{O}'$ are defined as
\begin{equation}
    \rho' \equiv R_{\bm{n}'_d}(\theta'_d) V_1 \ket{\bm{0}}\bra{\bm{0}}V_1^\dagger R^{\dagger}_{\bm{n}'_d}(\theta'_d),
\end{equation}
\begin{equation}
    \hat{O}'\equiv V_2^\dagger \hat{O} V_2.
\end{equation}
Note that the first term in Eq.~(\ref{eqn:apdx_g_mu_trans}) is transformed as
\begin{align}
  {\rm Re}\left[{\rm tr}\left(\varsigma_{\mu}
  R^\dagger_{\bm{n}'_d}(\theta'_d)\rho'\right)\right]&=\begin{cases}
  \cos{\left({\theta'_d}/{2}\right)},&\mu=0\\[4pt]
  n'_{d\mu}\sin{\left({\theta'_d}/{2}\right)},&\mu\neq 0
  \end{cases}.
\end{align}
\begin{proof}
For $\mu=0$,
\begin{align}
  {\rm Re}\left[{\rm tr}\left(\varsigma_{\mu}
  R^\dagger_{\bm{n}'_d}(\theta'_d)\rho'\right)\right]\notag
  &={\rm Re}\left[{\rm tr}\left(R^\dagger_{\bm{n}'_d}(\theta'_d)\rho'\right)\right]\notag\\[4pt]
  &={\rm Re}\left[{\rm tr}\left(\cos{\left(\frac{\theta'_d}{2}\right)}\rho'
  +i\sin{\left(\frac{\theta'_d}2\right)}\bm{n}'_d\cdot\bm{\sigma}~\rho'\right)\right]\notag\\[4pt]
  &=\cos{\left(\frac{\theta'_d}{2}\right)}.\notag
\end{align}
For $\mu\neq0$,
\begin{align}
  {\rm Re}\left[{\rm tr}\left(\varsigma_{\mu}
  R^\dagger_{\bm{n}'_d}(\theta'_d)\rho'\right)\right]
  &={\rm Re}\left[{\rm tr}\left((-i\sigma_{\mu})
  R^\dagger_{\bm{n}'_d}(\theta'_d)\rho'\right)\right]\notag\\[4pt]
  &={\rm Im}\left[{\rm tr}\left(\cos{\left(\frac{\theta'_d}{2}\right)}\sigma_{\mu}\rho+i\sin{\left(\frac{\theta'_d}2\right)}\sigma_{\mu} \bm{n}'_d\cdot\bm{\sigma}~
  \rho'\right)\right]\notag\\[4pt]
  &=\sin{\left(\frac{\theta'_d}2\right)}\sum_{p=x,y,z}\mathrm{Im}\left[{\rm tr}\left(i{n}'_{dp}\sigma_{\mu}{\sigma}_p
  \rho'\right)\right]\notag\\[4pt]
  &=\sin{\left(\frac{\theta'_d}2\right)}\sum_{p=x,y,z}{\rm Im}\left[{\rm tr}
  \left(i{n}'_{dp}
  \left\{\delta_{\mu p}I
  +\sum_{q=x,y,z}i\epsilon_{\mu p q}\sigma_{q}\right\}
  \rho'\right)\right]\notag\\[4pt]
  &=n'_{d\mu}\sin{\left(\frac{\theta'_d}{2}\right)}\notag.
\end{align}
To obtain the above, we used a relation $\sigma_{a}\sigma_{b}=\delta_{ab}I+\sum_{c=x,y,z}i\epsilon_{abc}\sigma_c$, where $\delta_{ab}$ and  $\epsilon_{abc}$ denotes the Kronecker delta and the  three-dimensional Levi-Civita symbol, respectively.
\end{proof}

Because a parameter $\vartheta'_d=(\theta'_d,\bm{n}_d')$ is already known, the first term in Eq.~(\ref{eqn:apdx_g_mu_trans}) can be calculated without quantum circuits.
Thus, to obtain $g_\mu$, only the following quantities in the second term in Eq.~(\ref{eqn:apdx_g_mu_trans}) are required 
\begin{equation}\label{eqn:Appendix3}
  {\rm Re}\left[{\rm tr}\left(\hat{O}'\varsigma_{\mu}R^\dagger_{\bm{n}'_d}(\theta'_d)\rho'  \right)\right],~~~\mu=0,x,y,z.
\end{equation}
Here, we introduce a real value $\alpha_\mu$ and a real unit vector $\bm{\beta}_{\mu} = (\beta_{\mu x},\beta_{\mu y},\beta_{\mu z})$ such that
\begin{equation}\label{eqn:alpha_beta_Origin}
    R_{\bm{\beta}_\mu}(\alpha_\mu) =\varsigma_{\mu} R^\dagger_{\bm{n}'_d}(\theta'_d).
\end{equation}
It is straightforward to prove the existence of the $\alpha_\mu$ and $\bm{\beta}_\mu$ as follows.
\begin{proof}
For $\mu=0$, $\alpha_0$ and $\bm{\beta}_0$ are trivially given as
\begin{equation}
  \alpha_0 = -\theta'_d,~~~\bm{\beta}_0=\bm{n}'_d.
\end{equation}
For $\mu\neq 0$, $\varsigma_{\mu} R^\dagger_{\bm{n}'_d}(\theta'_d)$ can be transformed as
\begin{align}
  \varsigma_{\mu} R^\dagger_{\bm{n}'_d}(\theta'_d)
  &=-i\cos{\left(\frac{\theta'_d}{2}\right)}\sigma_\mu+\sin{\left(\frac{\theta'_d}2\right)}\sum_{p=x,y,z}{{n}'_{dp}}\sigma_\mu{\sigma}_{p}\notag\\[4pt]
  &=-i\cos{\left(\frac{\theta'_d}{2}\right)}\sigma_\mu+\sin{\left(\frac{\theta'_d}2\right)}\sum_{p=x,y,z}{{n}'_{dp}}\left(\delta_{\mu p}I+\sum_{q=x,y,z}i\epsilon_{\mu p q}\sigma_q\right)\notag\\[4pt]
  &=-i\cos{\left(\frac{\theta'_d}{2}\right)}\sigma_\mu+\sin{\left(\frac{\theta'_d}2\right)}\sum_{p=x,y,z}\left({{n}'_{dp}}\delta_{\mu p}I+\sum_{q=x,y,z}i\epsilon_{\mu p q}{{n}'_{dp}}\sigma_q\right)\notag\\[4pt]
  &=-i\cos{\left(\frac{\theta'_d}{2}\right)}\sigma_\mu+{{n}'_{d\mu}}\sin{\left(\frac{\theta'_d}{2}\right)}I+i\sin{\left(\frac{\theta'_d}{2}\right)}\sum_{p, q=x,y,z}\epsilon_{\mu p q}n'_{d p}\sigma_ q \notag\\[4pt]
  &={{n}'_{d\mu}}\sin{\left(\frac{\theta'_d}2\right)}I-i\cos{\left(\frac{\theta'_d}{2}\right)}\sum_{ q=x,y,z}\delta_{\mu q}\sigma_ q+i\sin{\left(\frac{\theta'_d}2\right)}\sum_{p, q=x,y,z}\epsilon_{\mu p q}{{n}'_{dp}}\sigma_ q\notag\\[4pt]
  &={{n}'_{d\mu}}\sin{\left(\frac{\theta'_d}2\right)}I-i\sum_{ q=x,y,z}\left(\cos{\left(\frac{\theta'_d}{2}\right)}\delta_{\mu q}-\sin{\left(\frac{\theta'_d}{2}\right)}\sum_{p=x,y,z}\epsilon_{\mu p q}{{n}'_{dp}}\right)\sigma_ q\notag\\[4pt]
  &={{n}'_{d\mu}}\sin{\left(\frac{\theta'_d}2\right)}I-i\sqrt{1-{n'}^2_{d\mu}\sin^2{\left(\frac{\theta'_d}{2}\right)}}\sum_{ q=x,y,z}\frac{\left(\cos{\left(\frac{\theta'_d}{2}\right)}\delta_{\mu q}-\sin{\left(\frac{\theta'_d}2\right)}\sum_{p=x,y,z}\epsilon_{\mu p q}{{n}'_{dp}}\right)}{\sqrt{1-{n'}^2_{d\mu}\sin^2{\left(\frac{\theta'_d}{2}\right)}}}\sigma_ q.
\end{align}
Thus, one can always find $\alpha_\mu$ and $\bm{\beta}_\mu,~(\mu\neq0)$ as follows,
\begin{equation}
    \alpha_\mu
    = 2\mathrm{arctan2}\left(\sqrt{1-n'_{d\mu}\sin^2{\left(\frac{\theta'_d}{2}\right)}}, ~n'_{d\mu} \sin{\left(\frac{\theta'_d}{2}\right)} \right)
\end{equation}
\begin{equation}
    \beta_{\mu q} = \frac{1}{\sqrt{1-{n'}^2_{d\mu}\sin^2{\left(\frac{\theta'_d}{2}\right)}}}{\left(\cos{\left(\frac{\theta'_d}{2}\right)}\delta_{\mu q}-\sin{\left(\frac{\theta'_d}2\right)}\sum_{p=x,y,z}\epsilon_{\mu p q}{{n}'_{dp}}\right)},~~~q=x,y,z.
\end{equation}
\end{proof}

We can evaluate all the required quantities, i.e., $g_{\mu}$ via the generator $\mathcal{G}(\theta,\bm{n})$ defined as
\begin{align}\label{eqn:apdx_generator}
\mathcal{G}(\theta,\bm{n})&\equiv{\rm Re}\left[{\rm tr}\left(\hat{O}'R_{\bm{n}}({{\theta}})\rho'\right)\right]\notag\\[4pt]
&=\cos{\left(\frac{\theta}{2}\right)}
\mathrm{tr}\left(\hat{O}'\rho'\right)
+\sin{\left(\frac{\theta}{2}\right)}\sum_{p=x,y,z}n_p{\rm Im}
\left[{\rm tr}\left(\hat{O}'{\sigma}_p\rho'\right)\right].
\end{align}
Here, let $\mathcal{Q}_{\pm,\nu}$ be an expectation value of $\hat{O}'$ defined as
\begin{equation}
  \mathcal{Q}_{\pm,\nu}\equiv{\rm tr}\left(\hat{O}'e^{\mp i\sigma_{\nu}\pi/4}\rho'e^{\pm i\sigma_{\nu}\pi/4}\right),~~~\nu=0,x,y,z,
\end{equation}
where $\sigma_\nu\equiv(I,\bm{\sigma})$. Then,  simply Eq.~(\ref{eqn:apdx_generator}) is written as
\begin{align}
  \mathcal{G}(\theta,\bm{n})
  &=\cos{\left(\frac{\theta}{2}\right)}\mathcal{Q}_{+,0}
  +\sin{\left(\frac{\theta}2\right)}\sum_{p=x,y,z}\frac{n_p}{2}\left(\mathcal{Q}_{+,p}-\mathcal{Q}_{-,p}\right).
\end{align}
Therefore, Eq.~(\ref{eqn:Appendix3}) is evaluated with seven types of measurements $\mathcal{Q}_{+,0},\mathcal{Q}_{\pm,x},\mathcal{Q}_{\pm,y},\mathcal{Q}_{\pm,z}$ and classical computation
for Eq.~(\ref{eqn:alpha_beta_Origin}).

\renewcommand{\theequation}{C.\arabic{equation}}
\setcounter{equation}{0}
\section*{C. The relationship of FQS with NFT and Fraxis}\label{sec:appendix_C}
In this section, we derive FQS(1q,~2p) (Fraxis) and FQS(1q,~1p) (NFT) from FQS(1q,~3p).
First, in a Fraxis gate the rotational angle $\theta$ in a general single-qubit expression $R_{\bm{n}}(\theta)$ is fixed to $\pi$ while the rotation axis $\bm{n}$ is arbitrary. 
Substituting $\theta_d=\pi$ and $\vartheta_d=\bm{n}_d$ into Eq.~(\ref{eqn:apdxA_prmitive_objective}), we obtain the objective function for Fraxis
\begin{align}\label{eqn:apdx_Fraxis}
  \mathcal{F}^{(d)}_{\rm Fraxis}(\bm{n}_d)
  &=\sum_{s = x,y,z}n_{ds}{\rm Im}\left[\bra{\psi^{(d-1)}_{\Delta \tau/D}}V_2\sigma_{s}V_1\ket{\bm{0}}\right]\notag\\[4pt]
  &=\bm{n}_d\cdot \bm{g}.
\end{align}
Note that $g_0$ in Eq.~(\ref{eqn:apdx_g_mu}) does not appear in this objective function.
Thus, the optimal $\bm{n}^*_d $ that maximizes the objective function is trivially $\bm{n}_d^* = \bm{g}/\|\bm{g}\|$ as in Eq.~(\ref{eqn:exact_solution_appendixs}).
The same procedure as for FQS(1q,~3p) can be used for evaluation of $\bm{g}$, and thus FQS(1q,~2p) requires only six types of measurements $\mathcal{Q}_{\pm,x},\mathcal{Q}_{\pm,y},\mathcal{Q}_{\pm,z}$.

In NFT method a single-qubit gate $R_{\bm{n}}(\theta)$ has a fixed axis $\bm{\tilde{n}}$ such as $R_{x}(\theta)$. 
From Eq.~(\ref{eqn:apdxA_prmitive_objective}), the objective function for NFT is given as
\begin{align}\label{eqn:NFT}
  \mathcal{F}^{(d)}_{\rm NFT}(\theta_d)&=\mathrm{Re}\left[\bra{\psi^{(d-1)}_{\Delta \tau/D}}V_2 V_1\ket{\bm{0}}\right]\cos{\left(\frac{\theta_d}{2}\right)}
  +{\rm Im}\left[\bra{\psi^{(d-1)}_{\Delta \tau/D}} V_2 \bm{\tilde{n}}_d \cdot \bm{\sigma} V_1 \ket{\bm{0}}\right]\sin{\left(\frac{\theta_d}{2}\right)}\notag\\[4pt]
  &=g_0 \cos{\left(\frac{\theta_d}{2}\right)}   +g_d \sin{\left(\frac{\theta_d}{2}\right)}\notag\\[4pt]
  &=\sqrt{g_0^2+g_d^2}\sin{\left[\frac{\theta}{2}+\arctan2\left(g_0, g_d \right)\right]}
\end{align}
where $g_d \equiv \bm{\tilde{n}}_d\cdot\bm{g}$.
Then, we only evaluate $g_0$ and $g_d$ for executing FQS(1q,~1p).
More specifically, the following quantities corresponding to the second term in Eq.~(\ref{eqn:apdx_g_mu_trans}) are required,
\begin{align}\label{eqn:NFT_0}
  {\rm Re}\left[{\rm tr}\left(\hat{O}'R^\dagger_{\bm{\tilde{n}}_d}(\theta'_d)\rho'\right)\right],
\end{align}
\begin{align}\label{eqn:NFT_d}
  \sum_{p=x,y,z} \tilde{n}_{dp}{\rm Re}\left[{\rm tr}\left(\hat{O}'\varsigma_{p}R^\dagger_{\bm{\tilde{n}}_d}(\theta'_d)\rho' \right)\right]
  &={\rm Re}\left[{\rm tr}\left(\hat{O}'\bm{\tilde{n}}_d\cdot(-i\bm{\sigma}) R^\dagger_{\bm{\tilde{n}}_d}(\theta'_d)\rho'\right)\right]\notag\\[4pt]
  &={\rm Re}\left[{\rm tr}\left(\hat{O}'R^\dagger_{\bm{\tilde{n}}_d}\left(-{\pi}\right)R^\dagger_{\bm{\tilde{n}}_d}(\theta'_d)\rho'\right)\right]\notag\\[4pt]
  &={\rm Re}\left[{\rm tr}\left(\hat{O}'R^\dagger_{\bm{\tilde{n}}_d}\left(\theta'_d-{\pi}\right)\rho'\right)\right].
\end{align}
Defining $\alpha_0$ and $\alpha_d$ as $-\theta_d'$ and $-\theta_d'+\pi$, respectively, we rewrite Eq. (\ref{eqn:NFT_0}) and Eq. (\ref{eqn:NFT_d}) all together in the form of a generator $\mathcal{G}_{\rm NFT}(\theta)$ definded as
\begin{align}
  \mathcal{G}_{\rm NFT}(\theta)&\equiv{\rm Re}\left[{\rm tr}\left(\hat{O}'R_{\bm{\tilde{n}}_d}(\theta)\rho'\right)\right]\notag\\[4pt]
  &=\cos{\left(\frac{\theta}{2}\right)}{\rm Re}\left[{\rm tr}\left(\hat{O}'\rho'\right)\right]+\sin{\left(\frac{\theta}2\right)}{\rm Im}\left[{\rm tr}\left(\hat{O}'\bm{\tilde{n}}_d\cdot\bm{\sigma}\rho'\right)\right]\notag\\[4pt]
  &=\cos{\left(\frac{\theta}{2}\right)}\mathcal{Q}_{+,0}+\sin{\left(\frac{\theta}2\right)}\frac{\left(\mathcal{Q}_{+,d}-\mathcal{Q}_{-,d}\right)}{2},
\end{align}
where the expectations $\mathcal{Q}_{+,0},\mathcal{Q}_{\pm,d}$ are defined as
\begin{align}
  \mathcal{Q}_{+,0}&\equiv{\rm tr}\left(\hat{O}'\rho'\right),~~~\mathcal{Q}_{\pm,d}\equiv{\rm tr}\left(\hat{O}'e^{\mp i\bm{\bar{n}}_d\cdot\bm{\sigma}\pi/4}\rho'e^{\pm i\bm{\bar{n}}_d\cdot\bm{\sigma}\pi/4}\right).
\end{align}
Then, we can evaluate Eq.~(\ref{eqn:NFT_0}) and Eq.~(\ref{eqn:NFT_d}) in the same way as FQS(1q,~3p), and the number of types of required measurements is three for FQS(1q,~1p).

\renewcommand{\theequation}{D.\arabic{equation}}
\setcounter{equation}{0}
\section*{D. FQS for Hop, RBS, and Swap gates}\label{sec:apdx_opt_hopgate}
Considering the case $\eta=1/2~(|\Lambda_p|=2,~\forall p=1,2,\cdots,P)$, we derive the analytical optimization for the objective function in Eq.~(\ref{eqn:generalized_objective}) with specific multi-qubit gates.
Here we suppose the $p$th gate set includes only two single-qubit gates with shared parameters $\vartheta_p=(\theta_p,\bm{n}_p)$ as in Eq.~(\ref{eqn:gate_decomposition}).
Given the swap gate in Table~\ref{tbl:multi-qubit gates} where the rotational axis is fixed to $\bm{\tilde{n}}=(0,0,1)$, the objective function $\mathcal{F}^{(p)}(\vartheta_p)$ is rewritten as
\begin{align}\label{eqn:FQS(2q,1p)_objective}
  \mathcal{F}^{(p)}(\theta_p,\bm{\tilde{n}}_p)
  &=\mathrm{Re}\left[\bra{\psi^{(p-1)}_{{\Delta\tau}/{P}}}V_2AR_{\bm{\tilde{n}}_p}(\theta_p)BR^\dagger_{\bm{\tilde{n}}_p}(\theta_p)CV_1\ket{\bm{0}}\right]\notag\\[4pt]
  &=\mathrm{Re}\left[\bra{\psi^{(p-1)}_{{\Delta\tau}/{P}}}V_2 A B C V_1\ket{\bm{0}}\right]\cos^2{\frac{\theta_p}{2}}+\mathrm{Im}\left[\bra{\psi^{(p-1)}_{{\Delta\tau}/{P}}}V_2 A\bm{\tilde{n}}_p\cdot \bm{\sigma} B C V_1\ket{\bm{0}}\right]\cos{\frac{\theta_p}{2}}\sin{\frac{\theta_p}{2}}\notag\\[4pt]
  &~~~-\mathrm{ Im}\left[\bra{\psi^{(p-1)}_{{\Delta\tau}/{P}}}V_2 A B \bm{\tilde{n}}_p\cdot \bm{\sigma} C V_1\ket{\bm{0}}\right]\sin{\frac{\theta_p}{2}}\cos{\frac{\theta_p}{2}}+{\rm Re}\left[\bra{\psi^{(p-1)}_{{\Delta\tau}/{P}}}V_2 A \bm{\tilde{n}}_p\cdot \bm{\sigma} B \bm{\tilde{n}}_p\cdot \bm{\sigma} C V_1\ket{\bm{0}}\right]\sin^2{\frac{\theta_p}{2}}\notag\\[4pt]
  &=h_0\cos^2{\frac{\theta_p}{2}}+h_1\cos{\frac{\theta_p}{2}}\sin{\frac{\theta_p}{2}} \notag+h_2\sin{\frac{\theta_p}{2}}\cos{\frac{\theta_p}{2}}+h_3\sin^2{\frac{\theta_p}{2}}\notag\\[4pt]
  &=\sqrt{\left(\frac{h_1+h_2}{2}\right)^2+\left(\frac{h_0-h_3}{2}\right)^2}\sin{\left[\theta_p+{\rm arctan2}\left(h_0-h_3,h_1+h_2\right)\right]}+\frac{h_0+h_3}{2},
\end{align}
where $h_0,h_1,h_2$, and $h_3$ are defined as 
\begin{align}
    h_0&\equiv\mathrm{Re}\left[\bra{\psi^{(p-1)}_{{\Delta\tau}/{P}}}V_2 A B C V_1\ket{\bm{0}}\right],\label{eqn:h_element}\\[4pt]
    h_1&\equiv\mathrm{Im}\left[\bra{\psi^{(p-1)}_{{\Delta\tau}/{P}}}V_2 A\bm{\tilde{n}}_p\cdot \bm{\sigma} B C V_1\ket{\bm{0}}\right],\\[4pt]
    h_2&\equiv-\mathrm{ Im}\left[\bra{\psi^{(p-1)}_{{\Delta\tau}/{P}}}V_2 A B \bm{\tilde{n}}_p\cdot \bm{\sigma} C V_1\ket{\bm{0}}\right],\\[4pt]
    h_3&\equiv{\rm Re}\left[\bra{\psi^{(p-1)}_{{\Delta\tau}/{P}}}V_2 A \bm{\tilde{n}}_p\cdot \bm{\sigma} B \bm{\tilde{n}}_p\cdot \bm{\sigma} C V_1\ket{\bm{0}}\right].\label{eqn:h_element2}
\end{align}
Then, the optimal value of $\theta^*_p$ for the optimization problem Eq.~(\ref{eqn:FQS(2q,1p)_objective}) is given as
\begin{equation}\label{eqn:2q_optimal_value2}
    \theta_p^*=\frac{\pi}{2}-{\rm arctan2}\left(h_0-h_3,h_1+h_2\right)+2l\pi,
\end{equation}
where we choose $l\in\mathbb{Z}$ satisfying $\theta_p^*\in[0,2\pi]$.

For the cases of the Hop and RBS gates in Table~\ref{tbl:multi-qubit gates}, the rotational angle $\theta$ is fixed to $\pi$ and the rotational axis $\bm{n}$ is restricted in the XZ-plane. Thus, $\bm{n}$ can be expressed with one parameter $\psi$ as $\bm{n}(\psi)=(\sin{\left(\psi/{2}\right)},0,\cos{\left(\psi/{2}\right)})$.
Expanding Eq.~(\ref{eqn:2q_gate_opt}) with $\bm{n}_p(\psi_p)$ as a function of $\psi_p$, we obtain the the objective function that has the same form as Eq.~(\ref{eqn:FQS(2q,1p)_objective}), where $h_0,h_1,h_2,h_3$, and $\theta_p$ are replaced by $G_{33},G_{31},G_{13},G_{11}$, and $\psi_p$, respectively.
Altogether, for any gate types in Table \ref{tbl:multi-qubit gates} only Eq.~(\ref{eqn:expectationsfor2qopt}) are required for estimation of the optimal parameter.

\section*{E. Experimental details}\label{sec:appendix_E}

\subsection*{1D Heisenberg model}

Figure~\ref{fgr:vs_NFT_Fraxis}:
In the case of NFT and Fraxis, the single-qubit gates in Fig.~\ref{fgr:PQC_Heisenberg}(a) were replaced by $R_y$ and $R_n(\pi)$ gates, respectively, where indices of the parameterized gates are also shown in Fig.~\ref{fgr:PQC_Heisenberg}(a).
We consistently employed the following conditions \textit{Setting-A} for NFT, Fraxis, and FQS.

\textit{Setting-A}: 
The PQC employed contains 15 parameterized single-qubit gates and the gates in an identical layer are indexed from ascending order of corresponding qubits ($D=15$).
We used 15 gate sets in total ($P=15$) where $\Lambda_p=\{p\},~(p=1,\cdots,15)$.
In the optimization for the respective gate sets $\Lambda_p$, the gate parameters of each gate set were updated once in ascending order of gate indices.

\textit{Setting-B}: 
The 15 general single-qubit gates in the FQS simulation in Setting-A were decomposed into 15 Ry and 30 Rz gates as $U(\vartheta) = R_{z}(\phi)R_{y}(\psi)R_{z}(\lambda)$ ($D = 45$).
Then, the 45 gate sets are prepared as $\Lambda_p=\{p\},~(p=1,\cdots,45)$ ($P=45$).
The three gates obtained by decomposition are assigned a series of indices from the front of the circuit and separately updated with the NFT method in ascending order of the gate set index.

\subsection*{\ce{H2} molecules}
For imaginary time evolution of \ce{H2} molecular Hamiltonian, we employed a PQC shown in Fig.~\ref{fgr:PQC_H2}, which contains five excitation-conserving gates.
An index of single-qubit gates in $p$th excitation-conserving gate is assigned from the front of the circuit such as $2p-1,2p$.

\textit{Setting-C and D}: 
Since the excitation-conserving gate consists of two parameterized single-qubit gates as in Eq.~(\ref{eqn:gate_decomposition}), the optimization gate set in the FQS(2q, *p) simulation are $|\Lambda|=2M, (M\in \mathbb{N}$).
Here, we employed the minimum gate set as $|\Lambda|=2$ and prepared five gate sets ($P=5,D=10$), and thus $\Lambda_{p} =\{2p-1, 2p\}, (p=1,\cdots,5)$.
The simulations were conducted with use of two controlled operations.

\textit{Setting-E}: 
While we use the same ansatz as Setting-C and D,
we carried out \textit{twice sweep update} for each optimization of Eq.~(\ref{eqn:generalized_objective}) in the FQS(2q, 2p) method, where 
$\Lambda_{p} = \Lambda_{p+5} =\{2p-1, 2p\}, (p=1,\cdots,5)$ and thus the total number of gate sets $P=10$.

\textit{Setting-F}: 
The $d$th excitation-conserving gates are decomposed into multiple $R_y$ and $R_z$ gates as in Fig.~\ref{fgr:gate_decomposition}, 
where the parameters $(\psi, \phi)$ in excitation-conserving gates are shared by two $R_z$ gates and two $R_y$ gates, respectively.
Then $\psi$ and $\phi$ are separately and sequentially updated once for through ten optimizations in the order of $\phi_1\to\psi_1\to\phi_2\to\psi_2\cdots\phi_5\to\psi_5$ by using FQS(2q, 1p).

\subsection*{Real time evolution for 1D Ising model}
Based on gate indices of the parameterized gates in the ansatz for four qubits shown in Fig.~\ref{fgr:PQC_Heisenberg}(b), we prepared three types of the gate sets, (1) $\Lambda_p=\{p\},~(p=1,\cdots,12)$, (2) $\Lambda_p=\{4p-3,\cdots,4p\},~(p=1,\cdots,3)$, (3) $\Lambda=\{1,\cdots,12\}$.
For convenience, we refer the second and the third setting as FQS(Layer) and FQS(All), respectively.
In the FQS(Layer) and FQS(All) simulations, gate parameters $\vartheta$ in a single gate set $\Lambda$ are sequentially updated for action of one propagator in ascending order of the gate index in $\Lambda$.

\section*{F. Additional Experiments}\label{sec:additional_realtime}

\begin{figure}[h]
    \centering
    \begin{tabular}{cc}
  \includegraphics[scale=0.7]{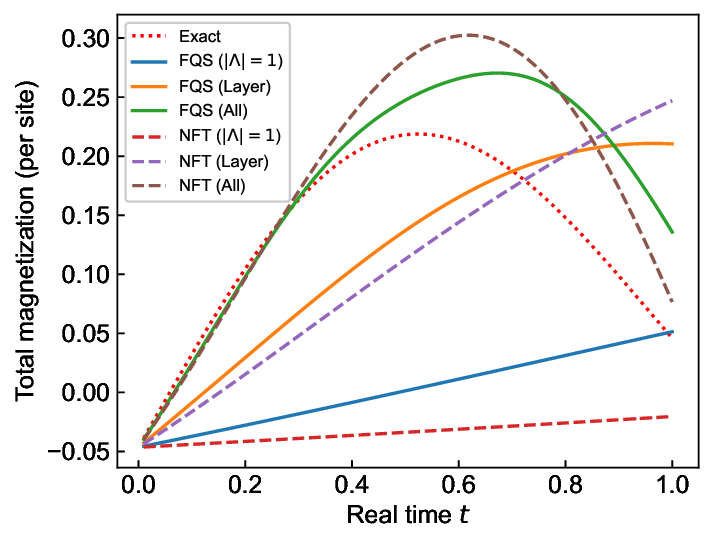}& 
  \includegraphics[scale=0.7]{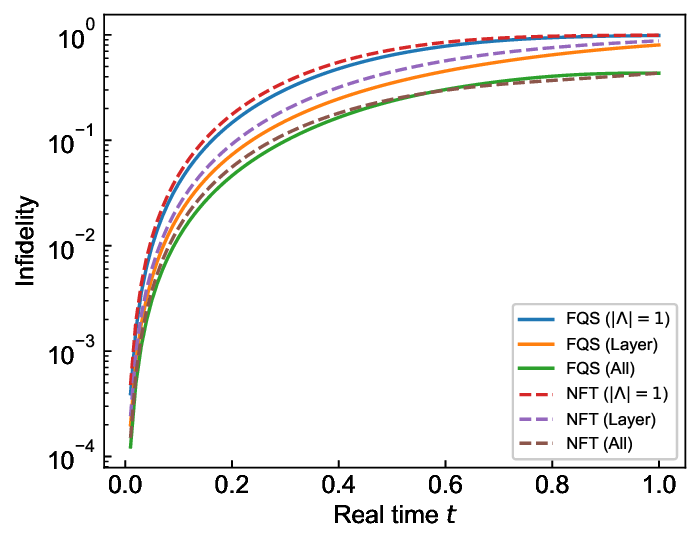}
    \end{tabular}
    \caption{Dynamics of (Left) total magnetization per site and (Right) the infidelity of simulation for 4-qubit 1D Ising model. The solid lines and dashed lines represent the results of FQS and NFT, respectively. The red dotted line represents the exact simulation. "$|\Lambda|=1$", "Layer", and "All" denote the level of hardware-efficiency.}
    \label{fgr:apdx_real_time}
\end{figure}
We provide additional results for real-time evolution with a 4-qubit 1D Ising model.
Figure~\ref{fgr:apdx_real_time} shows the results for simulating the dynamics from a randomly-generated initial state.
The simulation settings except for the initial state were the same as in the main article Section III-C.

%% file: main.bbl
\begin{thebibliography}{50}%
\makeatletter
\providecommand \@ifxundefined [1]{%
 \@ifx{#1\undefined}
}%
\providecommand \@ifnum [1]{%
 \ifnum #1\expandafter \@firstoftwo
 \else \expandafter \@secondoftwo
 \fi
}%
\providecommand \@ifx [1]{%
 \ifx #1\expandafter \@firstoftwo
 \else \expandafter \@secondoftwo
 \fi
}%
\providecommand \natexlab [1]{#1}%
\providecommand \enquote  [1]{``#1''}%
\providecommand \bibnamefont  [1]{#1}%
\providecommand \bibfnamefont [1]{#1}%
\providecommand \citenamefont [1]{#1}%
\providecommand \href@noop [0]{\@secondoftwo}%
\providecommand \href [0]{\begingroup \@sanitize@url \@href}%
\providecommand \@href[1]{\@@startlink{#1}\@@href}%
\providecommand \@@href[1]{\endgroup#1\@@endlink}%
\providecommand \@sanitize@url [0]{\catcode `\\12\catcode `\$12\catcode
  `\&12\catcode `\#12\catcode `\^12\catcode `\_12\catcode `\%12\relax}%
\providecommand \@@startlink[1]{}%
\providecommand \@@endlink[0]{}%
\providecommand \url  [0]{\begingroup\@sanitize@url \@url }%
\providecommand \@url [1]{\endgroup\@href {#1}{\urlprefix }}%
\providecommand \urlprefix  [0]{URL }%
\providecommand \Eprint [0]{\href }%
\providecommand \doibase [0]{https://doi.org/}%
\providecommand \selectlanguage [0]{\@gobble}%
\providecommand \bibinfo  [0]{\@secondoftwo}%
\providecommand \bibfield  [0]{\@secondoftwo}%
\providecommand \translation [1]{[#1]}%
\providecommand \BibitemOpen [0]{}%
\providecommand \bibitemStop [0]{}%
\providecommand \bibitemNoStop [0]{.\EOS\space}%
\providecommand \EOS [0]{\spacefactor3000\relax}%
\providecommand \BibitemShut  [1]{\csname bibitem#1\endcsname}%
\let\auto@bib@innerbib\@empty
\bibitem [{\citenamefont {Kassal}\ \emph {et~al.}(2011)\citenamefont {Kassal},
  \citenamefont {Whitfield}, \citenamefont {Perdomo-Ortiz}, \citenamefont
  {Yung},\ and\ \citenamefont {Aspuru-Guzik}}]{kassal2011simulating}%
  \BibitemOpen
  \bibfield  {author} {\bibinfo {author} {\bibfnamefont {I.}~\bibnamefont
  {Kassal}}, \bibinfo {author} {\bibfnamefont {J.~D.}\ \bibnamefont
  {Whitfield}}, \bibinfo {author} {\bibfnamefont {A.}~\bibnamefont
  {Perdomo-Ortiz}}, \bibinfo {author} {\bibfnamefont {M.-H.}\ \bibnamefont
  {Yung}},\ and\ \bibinfo {author} {\bibfnamefont {A.}~\bibnamefont
  {Aspuru-Guzik}},\ }\bibfield  {title} {\bibinfo {title} {Simulating chemistry
  using quantum computers},\ }\href@noop {} {\bibfield  {journal} {\bibinfo
  {journal} {Annu. Rev. Phys. Chem.}\ }\textbf {\bibinfo {volume} {62}},\
  \bibinfo {pages} {185} (\bibinfo {year} {2011})}\BibitemShut {NoStop}%
\bibitem [{\citenamefont {Cao}\ \emph {et~al.}(2019)\citenamefont {Cao},
  \citenamefont {Romero}, \citenamefont {Olson}, \citenamefont {Degroote},
  \citenamefont {Johnson}, \citenamefont {Kieferová}, \citenamefont
  {Kivlichan}, \citenamefont {Menke}, \citenamefont {Peropadre}, \citenamefont
  {Sawaya}, \citenamefont {Sim}, \citenamefont {Veis},\ and\ \citenamefont
  {Aspuru-Guzik}}]{cao2019quantum}%
  \BibitemOpen
  \bibfield  {author} {\bibinfo {author} {\bibfnamefont {Y.}~\bibnamefont
  {Cao}}, \bibinfo {author} {\bibfnamefont {J.}~\bibnamefont {Romero}},
  \bibinfo {author} {\bibfnamefont {J.~P.}\ \bibnamefont {Olson}}, \bibinfo
  {author} {\bibfnamefont {M.}~\bibnamefont {Degroote}}, \bibinfo {author}
  {\bibfnamefont {P.~D.}\ \bibnamefont {Johnson}}, \bibinfo {author}
  {\bibfnamefont {M.}~\bibnamefont {Kieferová}}, \bibinfo {author}
  {\bibfnamefont {I.~D.}\ \bibnamefont {Kivlichan}}, \bibinfo {author}
  {\bibfnamefont {T.}~\bibnamefont {Menke}}, \bibinfo {author} {\bibfnamefont
  {B.}~\bibnamefont {Peropadre}}, \bibinfo {author} {\bibfnamefont {N.~P.~D.}\
  \bibnamefont {Sawaya}}, \bibinfo {author} {\bibfnamefont {S.}~\bibnamefont
  {Sim}}, \bibinfo {author} {\bibfnamefont {L.}~\bibnamefont {Veis}},\ and\
  \bibinfo {author} {\bibfnamefont {A.}~\bibnamefont {Aspuru-Guzik}},\
  }\bibfield  {title} {\bibinfo {title} {Quantum chemistry in the age of
  quantum computing},\ }\href {https://doi.org/10.1021/acs.chemrev.8b00803}
  {\bibfield  {journal} {\bibinfo  {journal} {Chem. Rev.}\ }\textbf {\bibinfo
  {volume} {119}},\ \bibinfo {pages} {10856} (\bibinfo {year}
  {2019})}\BibitemShut {NoStop}%
\bibitem [{\citenamefont {McArdle}\ \emph {et~al.}(2020)\citenamefont
  {McArdle}, \citenamefont {Endo}, \citenamefont {Aspuru-Guzik}, \citenamefont
  {Benjamin},\ and\ \citenamefont {Yuan}}]{mcardle2020quantum}%
  \BibitemOpen
  \bibfield  {author} {\bibinfo {author} {\bibfnamefont {S.}~\bibnamefont
  {McArdle}}, \bibinfo {author} {\bibfnamefont {S.}~\bibnamefont {Endo}},
  \bibinfo {author} {\bibfnamefont {A.}~\bibnamefont {Aspuru-Guzik}}, \bibinfo
  {author} {\bibfnamefont {S.~C.}\ \bibnamefont {Benjamin}},\ and\ \bibinfo
  {author} {\bibfnamefont {X.}~\bibnamefont {Yuan}},\ }\bibfield  {title}
  {\bibinfo {title} {Quantum computational chemistry},\ }\href
  {https://doi.org/10.1103/RevModPhys.92.015003} {\bibfield  {journal}
  {\bibinfo  {journal} {Rev. Mod. Phys.}\ }\textbf {\bibinfo {volume} {92}},\
  \bibinfo {pages} {015003} (\bibinfo {year} {2020})}\BibitemShut {NoStop}%
\bibitem [{\citenamefont {Bauer}\ \emph {et~al.}(2020)\citenamefont {Bauer},
  \citenamefont {Bravyi}, \citenamefont {Motta},\ and\ \citenamefont
  {Chan}}]{bauer2020quantum}%
  \BibitemOpen
  \bibfield  {author} {\bibinfo {author} {\bibfnamefont {B.}~\bibnamefont
  {Bauer}}, \bibinfo {author} {\bibfnamefont {S.}~\bibnamefont {Bravyi}},
  \bibinfo {author} {\bibfnamefont {M.}~\bibnamefont {Motta}},\ and\ \bibinfo
  {author} {\bibfnamefont {G.~K.-L.}\ \bibnamefont {Chan}},\ }\bibfield
  {title} {\bibinfo {title} {Quantum algorithms for quantum chemistry and
  quantum materials science},\ }\href@noop {} {\bibfield  {journal} {\bibinfo
  {journal} {Chem. Rev.}\ }\textbf {\bibinfo {volume} {120}},\ \bibinfo {pages}
  {12685} (\bibinfo {year} {2020})}\BibitemShut {NoStop}%
\bibitem [{\citenamefont {Motta}\ and\ \citenamefont {Rice}(2021)}]{motta2021}%
  \BibitemOpen
  \bibfield  {author} {\bibinfo {author} {\bibfnamefont {M.}~\bibnamefont
  {Motta}}\ and\ \bibinfo {author} {\bibfnamefont {J.}~\bibnamefont {Rice}},\
  }\bibfield  {title} {\bibinfo {title} {Emerging quantum computing algorithms
  for quantum chemistry},\ }\href@noop {} {\bibfield  {journal} {\bibinfo
  {journal} {arXiv:2109.02873}\ } (\bibinfo {year} {2021})}\BibitemShut
  {NoStop}%
\bibitem [{\citenamefont {Charron}\ and\ \citenamefont
  {Suzor-Weiner}(1998)}]{charron1998femtosecond}%
  \BibitemOpen
  \bibfield  {author} {\bibinfo {author} {\bibfnamefont {E.}~\bibnamefont
  {Charron}}\ and\ \bibinfo {author} {\bibfnamefont {A.}~\bibnamefont
  {Suzor-Weiner}},\ }\bibfield  {title} {\bibinfo {title} {Femtosecond dynamics
  of nai ionization and dissociative ionization},\ }\href@noop {} {\bibfield
  {journal} {\bibinfo  {journal} {J. Chem. Phys.}\ }\textbf {\bibinfo {volume}
  {108}},\ \bibinfo {pages} {3922} (\bibinfo {year} {1998})}\BibitemShut
  {NoStop}%
\bibitem [{\citenamefont {Yonehara}\ \emph {et~al.}(2012)\citenamefont
  {Yonehara}, \citenamefont {Hanasaki},\ and\ \citenamefont
  {Takatsuka}}]{yonehara2012fundamental}%
  \BibitemOpen
  \bibfield  {author} {\bibinfo {author} {\bibfnamefont {T.}~\bibnamefont
  {Yonehara}}, \bibinfo {author} {\bibfnamefont {K.}~\bibnamefont {Hanasaki}},\
  and\ \bibinfo {author} {\bibfnamefont {K.}~\bibnamefont {Takatsuka}},\
  }\bibfield  {title} {\bibinfo {title} {Fundamental approaches to
  nonadiabaticity: Toward a chemical theory beyond the born--oppenheimer
  paradigm},\ }\href@noop {} {\bibfield  {journal} {\bibinfo  {journal} {Chem.
  Rev.}\ }\textbf {\bibinfo {volume} {112}},\ \bibinfo {pages} {499} (\bibinfo
  {year} {2012})}\BibitemShut {NoStop}%
\bibitem [{\citenamefont {Kitaev}(1995)}]{kitaev1995quantum}%
  \BibitemOpen
  \bibfield  {author} {\bibinfo {author} {\bibfnamefont {A.~Y.}\ \bibnamefont
  {Kitaev}},\ }\bibfield  {title} {\bibinfo {title} {Quantum measurements and
  the abelian stabilizer problem},\ }\href@noop {} {\bibfield  {journal}
  {\bibinfo  {journal} {arXiv:quant-ph/9511026}\ } (\bibinfo {year}
  {1995})}\BibitemShut {NoStop}%
\bibitem [{\citenamefont {Lanyon}\ \emph {et~al.}(2010)\citenamefont {Lanyon},
  \citenamefont {Whitfield}, \citenamefont {Gillett}, \citenamefont {Goggin},
  \citenamefont {Almeida}, \citenamefont {Kassal}, \citenamefont {Biamonte},
  \citenamefont {Mohseni}, \citenamefont {Powell}, \citenamefont {Barbieri}
  \emph {et~al.}}]{lanyon2010towards}%
  \BibitemOpen
  \bibfield  {author} {\bibinfo {author} {\bibfnamefont {B.~P.}\ \bibnamefont
  {Lanyon}}, \bibinfo {author} {\bibfnamefont {J.~D.}\ \bibnamefont
  {Whitfield}}, \bibinfo {author} {\bibfnamefont {G.~G.}\ \bibnamefont
  {Gillett}}, \bibinfo {author} {\bibfnamefont {M.~E.}\ \bibnamefont {Goggin}},
  \bibinfo {author} {\bibfnamefont {M.~P.}\ \bibnamefont {Almeida}}, \bibinfo
  {author} {\bibfnamefont {I.}~\bibnamefont {Kassal}}, \bibinfo {author}
  {\bibfnamefont {J.~D.}\ \bibnamefont {Biamonte}}, \bibinfo {author}
  {\bibfnamefont {M.}~\bibnamefont {Mohseni}}, \bibinfo {author} {\bibfnamefont
  {B.~J.}\ \bibnamefont {Powell}}, \bibinfo {author} {\bibfnamefont
  {M.}~\bibnamefont {Barbieri}}, \emph {et~al.},\ }\bibfield  {title} {\bibinfo
  {title} {Towards quantum chemistry on a quantum computer},\ }\href@noop {}
  {\bibfield  {journal} {\bibinfo  {journal} {Nat. Chem.}\ }\textbf {\bibinfo
  {volume} {2}},\ \bibinfo {pages} {106} (\bibinfo {year} {2010})}\BibitemShut
  {NoStop}%
\bibitem [{\citenamefont {Peruzzo}\ \emph {et~al.}(2014)\citenamefont
  {Peruzzo}, \citenamefont {McClean}, \citenamefont {Shadbolt}, \citenamefont
  {Yung}, \citenamefont {Zhou}, \citenamefont {Love}, \citenamefont
  {Aspuru-Guzik},\ and\ \citenamefont {O'brien}}]{peruzzo2014variational}%
  \BibitemOpen
  \bibfield  {author} {\bibinfo {author} {\bibfnamefont {A.}~\bibnamefont
  {Peruzzo}}, \bibinfo {author} {\bibfnamefont {J.}~\bibnamefont {McClean}},
  \bibinfo {author} {\bibfnamefont {P.}~\bibnamefont {Shadbolt}}, \bibinfo
  {author} {\bibfnamefont {M.-H.}\ \bibnamefont {Yung}}, \bibinfo {author}
  {\bibfnamefont {X.-Q.}\ \bibnamefont {Zhou}}, \bibinfo {author}
  {\bibfnamefont {P.~J.}\ \bibnamefont {Love}}, \bibinfo {author}
  {\bibfnamefont {A.}~\bibnamefont {Aspuru-Guzik}},\ and\ \bibinfo {author}
  {\bibfnamefont {J.~L.}\ \bibnamefont {O'brien}},\ }\bibfield  {title}
  {\bibinfo {title} {A variational eigenvalue solver on a photonic quantum
  processor},\ }\href@noop {} {\bibfield  {journal} {\bibinfo  {journal} {Nat.
  Commun.}\ }\textbf {\bibinfo {volume} {5}},\ \bibinfo {pages} {1} (\bibinfo
  {year} {2014})}\BibitemShut {NoStop}%
\bibitem [{\citenamefont {McClean}\ \emph {et~al.}(2016)\citenamefont
  {McClean}, \citenamefont {Romero}, \citenamefont {Babbush},\ and\
  \citenamefont {Aspuru-Guzik}}]{mcclean2016theory}%
  \BibitemOpen
  \bibfield  {author} {\bibinfo {author} {\bibfnamefont {J.~R.}\ \bibnamefont
  {McClean}}, \bibinfo {author} {\bibfnamefont {J.}~\bibnamefont {Romero}},
  \bibinfo {author} {\bibfnamefont {R.}~\bibnamefont {Babbush}},\ and\ \bibinfo
  {author} {\bibfnamefont {A.}~\bibnamefont {Aspuru-Guzik}},\ }\bibfield
  {title} {\bibinfo {title} {The theory of variational hybrid quantum-classical
  algorithms},\ }\href@noop {} {\bibfield  {journal} {\bibinfo  {journal} {New
  J. Phys.}\ }\textbf {\bibinfo {volume} {18}},\ \bibinfo {pages} {023023}
  (\bibinfo {year} {2016})}\BibitemShut {NoStop}%
\bibitem [{\citenamefont {Tilly}\ \emph {et~al.}(2021)\citenamefont {Tilly},
  \citenamefont {Chen}, \citenamefont {Cao}, \citenamefont {Picozzi},
  \citenamefont {Setia}, \citenamefont {Li}, \citenamefont {Grant},
  \citenamefont {Wossnig}, \citenamefont {Rungger}, \citenamefont {Booth} \emph
  {et~al.}}]{tilly2021variational}%
  \BibitemOpen
  \bibfield  {author} {\bibinfo {author} {\bibfnamefont {J.}~\bibnamefont
  {Tilly}}, \bibinfo {author} {\bibfnamefont {H.}~\bibnamefont {Chen}},
  \bibinfo {author} {\bibfnamefont {S.}~\bibnamefont {Cao}}, \bibinfo {author}
  {\bibfnamefont {D.}~\bibnamefont {Picozzi}}, \bibinfo {author} {\bibfnamefont
  {K.}~\bibnamefont {Setia}}, \bibinfo {author} {\bibfnamefont
  {Y.}~\bibnamefont {Li}}, \bibinfo {author} {\bibfnamefont {E.}~\bibnamefont
  {Grant}}, \bibinfo {author} {\bibfnamefont {L.}~\bibnamefont {Wossnig}},
  \bibinfo {author} {\bibfnamefont {I.}~\bibnamefont {Rungger}}, \bibinfo
  {author} {\bibfnamefont {G.~H.}\ \bibnamefont {Booth}}, \emph {et~al.},\
  }\bibfield  {title} {\bibinfo {title} {The variational quantum eigensolver: a
  review of methods and best practices},\ }\href@noop {} {\bibfield  {journal}
  {\bibinfo  {journal} {arXiv preprint arXiv:2111.05176}\ } (\bibinfo {year}
  {2021})}\BibitemShut {NoStop}%
\bibitem [{\citenamefont {McArdle}\ \emph
  {et~al.}(2019{\natexlab{a}})\citenamefont {McArdle}, \citenamefont {Jones},
  \citenamefont {Endo}, \citenamefont {Li}, \citenamefont {Benjamin},\ and\
  \citenamefont {Yuan}}]{mcardle2019}%
  \BibitemOpen
  \bibfield  {author} {\bibinfo {author} {\bibfnamefont {S.}~\bibnamefont
  {McArdle}}, \bibinfo {author} {\bibfnamefont {T.}~\bibnamefont {Jones}},
  \bibinfo {author} {\bibfnamefont {S.}~\bibnamefont {Endo}}, \bibinfo {author}
  {\bibfnamefont {Y.}~\bibnamefont {Li}}, \bibinfo {author} {\bibfnamefont
  {S.~C.}\ \bibnamefont {Benjamin}},\ and\ \bibinfo {author} {\bibfnamefont
  {X.}~\bibnamefont {Yuan}},\ }\bibfield  {title} {\bibinfo {title}
  {Variational ansatz-based quantum simulation of imaginary time evolution},\
  }\href@noop {} {\bibfield  {journal} {\bibinfo  {journal} {npj Quantum Inf.}\
  }\textbf {\bibinfo {volume} {5}},\ \bibinfo {pages} {1} (\bibinfo {year}
  {2019}{\natexlab{a}})}\BibitemShut {NoStop}%
\bibitem [{\citenamefont {Motta}\ \emph {et~al.}(2020)\citenamefont {Motta},
  \citenamefont {Sun}, \citenamefont {Tan}, \citenamefont {O'Rourke},
  \citenamefont {Ye}, \citenamefont {Minnich}, \citenamefont {Brand{\~a}o},\
  and\ \citenamefont {Chan}}]{motta2020}%
  \BibitemOpen
  \bibfield  {author} {\bibinfo {author} {\bibfnamefont {M.}~\bibnamefont
  {Motta}}, \bibinfo {author} {\bibfnamefont {C.}~\bibnamefont {Sun}}, \bibinfo
  {author} {\bibfnamefont {A.~T.}\ \bibnamefont {Tan}}, \bibinfo {author}
  {\bibfnamefont {M.~J.}\ \bibnamefont {O'Rourke}}, \bibinfo {author}
  {\bibfnamefont {E.}~\bibnamefont {Ye}}, \bibinfo {author} {\bibfnamefont
  {A.~J.}\ \bibnamefont {Minnich}}, \bibinfo {author} {\bibfnamefont {F.~G.}\
  \bibnamefont {Brand{\~a}o}},\ and\ \bibinfo {author} {\bibfnamefont
  {G.~K.-L.}\ \bibnamefont {Chan}},\ }\bibfield  {title} {\bibinfo {title}
  {Determining eigenstates and thermal states on a quantum computer using
  quantum imaginary time evolution},\ }\href@noop {} {\bibfield  {journal}
  {\bibinfo  {journal} {Nat. Phys.}\ }\textbf {\bibinfo {volume} {16}},\
  \bibinfo {pages} {205} (\bibinfo {year} {2020})}\BibitemShut {NoStop}%
\bibitem [{\citenamefont {Kandala}\ \emph {et~al.}(2017)\citenamefont
  {Kandala}, \citenamefont {Mezzacapo}, \citenamefont {Temme}, \citenamefont
  {Takita}, \citenamefont {Brink}, \citenamefont {Chow},\ and\ \citenamefont
  {Gambetta}}]{kandala2017hardware}%
  \BibitemOpen
  \bibfield  {author} {\bibinfo {author} {\bibfnamefont {A.}~\bibnamefont
  {Kandala}}, \bibinfo {author} {\bibfnamefont {A.}~\bibnamefont {Mezzacapo}},
  \bibinfo {author} {\bibfnamefont {K.}~\bibnamefont {Temme}}, \bibinfo
  {author} {\bibfnamefont {M.}~\bibnamefont {Takita}}, \bibinfo {author}
  {\bibfnamefont {M.}~\bibnamefont {Brink}}, \bibinfo {author} {\bibfnamefont
  {J.~M.}\ \bibnamefont {Chow}},\ and\ \bibinfo {author} {\bibfnamefont
  {J.~M.}\ \bibnamefont {Gambetta}},\ }\bibfield  {title} {\bibinfo {title}
  {Hardware-efficient variational quantum eigensolver for small molecules and
  quantum magnets},\ }\href@noop {} {\bibfield  {journal} {\bibinfo  {journal}
  {Nature}\ }\textbf {\bibinfo {volume} {549}},\ \bibinfo {pages} {242}
  (\bibinfo {year} {2017})}\BibitemShut {NoStop}%
\bibitem [{\citenamefont {Hempel}\ \emph {et~al.}(2018)\citenamefont {Hempel},
  \citenamefont {Maier}, \citenamefont {Romero}, \citenamefont {McClean},
  \citenamefont {Monz}, \citenamefont {Shen}, \citenamefont {Jurcevic},
  \citenamefont {Lanyon}, \citenamefont {Love}, \citenamefont {Babbush},
  \citenamefont {Aspuru-Guzik}, \citenamefont {Blatt},\ and\ \citenamefont
  {Roos}}]{hempel2018quantum}%
  \BibitemOpen
  \bibfield  {author} {\bibinfo {author} {\bibfnamefont {C.}~\bibnamefont
  {Hempel}}, \bibinfo {author} {\bibfnamefont {C.}~\bibnamefont {Maier}},
  \bibinfo {author} {\bibfnamefont {J.}~\bibnamefont {Romero}}, \bibinfo
  {author} {\bibfnamefont {J.}~\bibnamefont {McClean}}, \bibinfo {author}
  {\bibfnamefont {T.}~\bibnamefont {Monz}}, \bibinfo {author} {\bibfnamefont
  {H.}~\bibnamefont {Shen}}, \bibinfo {author} {\bibfnamefont {P.}~\bibnamefont
  {Jurcevic}}, \bibinfo {author} {\bibfnamefont {B.~P.}\ \bibnamefont
  {Lanyon}}, \bibinfo {author} {\bibfnamefont {P.}~\bibnamefont {Love}},
  \bibinfo {author} {\bibfnamefont {R.}~\bibnamefont {Babbush}}, \bibinfo
  {author} {\bibfnamefont {A.}~\bibnamefont {Aspuru-Guzik}}, \bibinfo {author}
  {\bibfnamefont {R.}~\bibnamefont {Blatt}},\ and\ \bibinfo {author}
  {\bibfnamefont {C.~F.}\ \bibnamefont {Roos}},\ }\bibfield  {title} {\bibinfo
  {title} {Quantum chemistry calculations on a trapped-ion quantum simulator},\
  }\href {https://doi.org/10.1103/PhysRevX.8.031022} {\bibfield  {journal}
  {\bibinfo  {journal} {Phys. Rev. X}\ }\textbf {\bibinfo {volume} {8}},\
  \bibinfo {pages} {031022} (\bibinfo {year} {2018})}\BibitemShut {NoStop}%
\bibitem [{\citenamefont {Kandala}\ \emph {et~al.}(2019)\citenamefont
  {Kandala}, \citenamefont {Temme}, \citenamefont {C{\'o}rcoles}, \citenamefont
  {Mezzacapo}, \citenamefont {Chow},\ and\ \citenamefont
  {Gambetta}}]{kandala2019error}%
  \BibitemOpen
  \bibfield  {author} {\bibinfo {author} {\bibfnamefont {A.}~\bibnamefont
  {Kandala}}, \bibinfo {author} {\bibfnamefont {K.}~\bibnamefont {Temme}},
  \bibinfo {author} {\bibfnamefont {A.~D.}\ \bibnamefont {C{\'o}rcoles}},
  \bibinfo {author} {\bibfnamefont {A.}~\bibnamefont {Mezzacapo}}, \bibinfo
  {author} {\bibfnamefont {J.~M.}\ \bibnamefont {Chow}},\ and\ \bibinfo
  {author} {\bibfnamefont {J.~M.}\ \bibnamefont {Gambetta}},\ }\bibfield
  {title} {\bibinfo {title} {Error mitigation extends the computational reach
  of a noisy quantum processor},\ }\href@noop {} {\bibfield  {journal}
  {\bibinfo  {journal} {Nature}\ }\textbf {\bibinfo {volume} {567}},\ \bibinfo
  {pages} {491} (\bibinfo {year} {2019})}\BibitemShut {NoStop}%
\bibitem [{\citenamefont {Nam}\ \emph {et~al.}(2020)\citenamefont {Nam},
  \citenamefont {Chen}, \citenamefont {Pisenti}, \citenamefont {Wright},
  \citenamefont {Delaney}, \citenamefont {Maslov}, \citenamefont {Brown},
  \citenamefont {Allen}, \citenamefont {Amini},\ and\ \citenamefont
  {Apisdorf}}]{nam2020ground}%
  \BibitemOpen
  \bibfield  {author} {\bibinfo {author} {\bibfnamefont {Y.}~\bibnamefont
  {Nam}}, \bibinfo {author} {\bibfnamefont {J.-S.}\ \bibnamefont {Chen}},
  \bibinfo {author} {\bibfnamefont {N.~C.}\ \bibnamefont {Pisenti}}, \bibinfo
  {author} {\bibfnamefont {K.}~\bibnamefont {Wright}}, \bibinfo {author}
  {\bibfnamefont {C.}~\bibnamefont {Delaney}}, \bibinfo {author} {\bibfnamefont
  {D.}~\bibnamefont {Maslov}}, \bibinfo {author} {\bibfnamefont {K.~R.}\
  \bibnamefont {Brown}}, \bibinfo {author} {\bibfnamefont {S.}~\bibnamefont
  {Allen}}, \bibinfo {author} {\bibfnamefont {J.~M.}\ \bibnamefont {Amini}},\
  and\ \bibinfo {author} {\bibfnamefont {J.~a.}\ \bibnamefont {Apisdorf}},\
  }\bibfield  {title} {\bibinfo {title} {Ground-state energy estimation of the
  water molecule on a trapped-ion quantum computer},\ }\href@noop {} {\bibfield
   {journal} {\bibinfo  {journal} {npj Quantum Inf.}\ }\textbf {\bibinfo
  {volume} {6}},\ \bibinfo {pages} {1} (\bibinfo {year} {2020})}\BibitemShut
  {NoStop}%
\bibitem [{\citenamefont {Tilly}\ \emph {et~al.}(2020)\citenamefont {Tilly},
  \citenamefont {Jones}, \citenamefont {Chen}, \citenamefont {Wossnig},\ and\
  \citenamefont {Grant}}]{tilly2020computation}%
  \BibitemOpen
  \bibfield  {author} {\bibinfo {author} {\bibfnamefont {J.}~\bibnamefont
  {Tilly}}, \bibinfo {author} {\bibfnamefont {G.}~\bibnamefont {Jones}},
  \bibinfo {author} {\bibfnamefont {H.}~\bibnamefont {Chen}}, \bibinfo {author}
  {\bibfnamefont {L.}~\bibnamefont {Wossnig}},\ and\ \bibinfo {author}
  {\bibfnamefont {E.}~\bibnamefont {Grant}},\ }\bibfield  {title} {\bibinfo
  {title} {Computation of molecular excited states on ibm quantum computers
  using a discriminative variational quantum eigensolver},\ }\href
  {https://doi.org/10.1103/PhysRevA.102.062425} {\bibfield  {journal} {\bibinfo
   {journal} {Phys. Rev. A}\ }\textbf {\bibinfo {volume} {102}},\ \bibinfo
  {pages} {062425} (\bibinfo {year} {2020})}\BibitemShut {NoStop}%
\bibitem [{\citenamefont {Eddins}\ \emph {et~al.}(2021)\citenamefont {Eddins},
  \citenamefont {Motta}, \citenamefont {Gujarati}, \citenamefont {Bravyi},
  \citenamefont {Mezzacapo}, \citenamefont {Hadfield},\ and\ \citenamefont
  {Sheldon}}]{eddins2021doubling}%
  \BibitemOpen
  \bibfield  {author} {\bibinfo {author} {\bibfnamefont {A.}~\bibnamefont
  {Eddins}}, \bibinfo {author} {\bibfnamefont {M.}~\bibnamefont {Motta}},
  \bibinfo {author} {\bibfnamefont {T.~P.}\ \bibnamefont {Gujarati}}, \bibinfo
  {author} {\bibfnamefont {S.}~\bibnamefont {Bravyi}}, \bibinfo {author}
  {\bibfnamefont {A.}~\bibnamefont {Mezzacapo}}, \bibinfo {author}
  {\bibfnamefont {C.}~\bibnamefont {Hadfield}},\ and\ \bibinfo {author}
  {\bibfnamefont {S.}~\bibnamefont {Sheldon}},\ }\bibfield  {title} {\bibinfo
  {title} {Doubling the size of quantum simulators by entanglement forging},\
  }\href@noop {} {\bibfield  {journal} {\bibinfo  {journal} {arXiv:2104.10220}\
  } (\bibinfo {year} {2021})}\BibitemShut {NoStop}%
\bibitem [{\citenamefont {Gao}\ \emph {et~al.}(2021{\natexlab{a}})\citenamefont
  {Gao}, \citenamefont {Nakamura}, \citenamefont {Gujarati}, \citenamefont
  {Jones}, \citenamefont {Rice}, \citenamefont {Wood}, \citenamefont {Pistoia},
  \citenamefont {Garcia},\ and\ \citenamefont
  {Yamamoto}}]{gao2021computational}%
  \BibitemOpen
  \bibfield  {author} {\bibinfo {author} {\bibfnamefont {Q.}~\bibnamefont
  {Gao}}, \bibinfo {author} {\bibfnamefont {H.}~\bibnamefont {Nakamura}},
  \bibinfo {author} {\bibfnamefont {T.~P.}\ \bibnamefont {Gujarati}}, \bibinfo
  {author} {\bibfnamefont {G.~O.}\ \bibnamefont {Jones}}, \bibinfo {author}
  {\bibfnamefont {J.~E.}\ \bibnamefont {Rice}}, \bibinfo {author}
  {\bibfnamefont {S.~P.}\ \bibnamefont {Wood}}, \bibinfo {author}
  {\bibfnamefont {M.}~\bibnamefont {Pistoia}}, \bibinfo {author} {\bibfnamefont
  {J.~M.}\ \bibnamefont {Garcia}},\ and\ \bibinfo {author} {\bibfnamefont
  {N.}~\bibnamefont {Yamamoto}},\ }\bibfield  {title} {\bibinfo {title}
  {Computational investigations of the lithium superoxide dimer rearrangement
  on noisy quantum devices},\ }\href@noop {} {\bibfield  {journal} {\bibinfo
  {journal} {J. Phys. Chem. A}\ }\textbf {\bibinfo {volume} {125}},\ \bibinfo
  {pages} {1827} (\bibinfo {year} {2021}{\natexlab{a}})}\BibitemShut {NoStop}%
\bibitem [{\citenamefont {Gao}\ \emph {et~al.}(2021{\natexlab{b}})\citenamefont
  {Gao}, \citenamefont {Jones}, \citenamefont {Motta}, \citenamefont
  {Sugawara}, \citenamefont {Watanabe}, \citenamefont {Kobayashi},
  \citenamefont {Watanabe}, \citenamefont {Ohnishi}, \citenamefont {Nakamura},\
  and\ \citenamefont {Yamamoto}}]{gao2021applications}%
  \BibitemOpen
  \bibfield  {author} {\bibinfo {author} {\bibfnamefont {Q.}~\bibnamefont
  {Gao}}, \bibinfo {author} {\bibfnamefont {G.~O.}\ \bibnamefont {Jones}},
  \bibinfo {author} {\bibfnamefont {M.}~\bibnamefont {Motta}}, \bibinfo
  {author} {\bibfnamefont {M.}~\bibnamefont {Sugawara}}, \bibinfo {author}
  {\bibfnamefont {H.~C.}\ \bibnamefont {Watanabe}}, \bibinfo {author}
  {\bibfnamefont {T.}~\bibnamefont {Kobayashi}}, \bibinfo {author}
  {\bibfnamefont {E.}~\bibnamefont {Watanabe}}, \bibinfo {author}
  {\bibfnamefont {Y.-y.}\ \bibnamefont {Ohnishi}}, \bibinfo {author}
  {\bibfnamefont {H.}~\bibnamefont {Nakamura}},\ and\ \bibinfo {author}
  {\bibfnamefont {N.}~\bibnamefont {Yamamoto}},\ }\bibfield  {title} {\bibinfo
  {title} {Applications of quantum computing for investigations of electronic
  transitions in phenylsulfonyl-carbazole tadf emitters},\ }\href@noop {}
  {\bibfield  {journal} {\bibinfo  {journal} {npj Comput. Mater.}\ }\textbf
  {\bibinfo {volume} {7}},\ \bibinfo {pages} {1} (\bibinfo {year}
  {2021}{\natexlab{b}})}\BibitemShut {NoStop}%
\bibitem [{\citenamefont {Yeter-Aydeniz}\ \emph {et~al.}(2020)\citenamefont
  {Yeter-Aydeniz}, \citenamefont {Pooser},\ and\ \citenamefont
  {Siopsis}}]{yeter2020practical}%
  \BibitemOpen
  \bibfield  {author} {\bibinfo {author} {\bibfnamefont {K.}~\bibnamefont
  {Yeter-Aydeniz}}, \bibinfo {author} {\bibfnamefont {R.~C.}\ \bibnamefont
  {Pooser}},\ and\ \bibinfo {author} {\bibfnamefont {G.}~\bibnamefont
  {Siopsis}},\ }\bibfield  {title} {\bibinfo {title} {Practical quantum
  computation of chemical and nuclear energy levels using quantum imaginary
  time evolution and lanczos algorithms},\ }\href@noop {} {\bibfield  {journal}
  {\bibinfo  {journal} {npj Quantum Inf.}\ }\textbf {\bibinfo {volume} {6}},\
  \bibinfo {pages} {1} (\bibinfo {year} {2020})}\BibitemShut {NoStop}%
\bibitem [{\citenamefont {Ma}\ \emph {et~al.}(2020)\citenamefont {Ma},
  \citenamefont {Govoni},\ and\ \citenamefont {Galli}}]{ma2020quantum}%
  \BibitemOpen
  \bibfield  {author} {\bibinfo {author} {\bibfnamefont {H.}~\bibnamefont
  {Ma}}, \bibinfo {author} {\bibfnamefont {M.}~\bibnamefont {Govoni}},\ and\
  \bibinfo {author} {\bibfnamefont {G.}~\bibnamefont {Galli}},\ }\bibfield
  {title} {\bibinfo {title} {Quantum simulations of materials on near-term
  quantum computers},\ }\href@noop {} {\bibfield  {journal} {\bibinfo
  {journal} {npj Comput. Mater.}\ }\textbf {\bibinfo {volume} {6}},\ \bibinfo
  {pages} {1} (\bibinfo {year} {2020})}\BibitemShut {NoStop}%
\bibitem [{\citenamefont {Sun}\ \emph {et~al.}(2021)\citenamefont {Sun},
  \citenamefont {Motta}, \citenamefont {Tazhigulov}, \citenamefont {Tan},
  \citenamefont {Chan},\ and\ \citenamefont {Minnich}}]{sun2021quantum}%
  \BibitemOpen
  \bibfield  {author} {\bibinfo {author} {\bibfnamefont {S.-N.}\ \bibnamefont
  {Sun}}, \bibinfo {author} {\bibfnamefont {M.}~\bibnamefont {Motta}}, \bibinfo
  {author} {\bibfnamefont {R.~N.}\ \bibnamefont {Tazhigulov}}, \bibinfo
  {author} {\bibfnamefont {A.~T.}\ \bibnamefont {Tan}}, \bibinfo {author}
  {\bibfnamefont {G.~K.-L.}\ \bibnamefont {Chan}},\ and\ \bibinfo {author}
  {\bibfnamefont {A.~J.}\ \bibnamefont {Minnich}},\ }\bibfield  {title}
  {\bibinfo {title} {Quantum computation of finite-temperature static and
  dynamical properties of spin systems using quantum imaginary time
  evolution},\ }\href {https://doi.org/10.1103/PRXQuantum.2.010317} {\bibfield
  {journal} {\bibinfo  {journal} {PRX Quantum}\ }\textbf {\bibinfo {volume}
  {2}},\ \bibinfo {pages} {010317} (\bibinfo {year} {2021})}\BibitemShut
  {NoStop}%
\bibitem [{\citenamefont {Sim}\ \emph {et~al.}(2019)\citenamefont {Sim},
  \citenamefont {Johnson},\ and\ \citenamefont
  {Aspuru-Guzik}}]{sim2019expressibility}%
  \BibitemOpen
  \bibfield  {author} {\bibinfo {author} {\bibfnamefont {S.}~\bibnamefont
  {Sim}}, \bibinfo {author} {\bibfnamefont {P.~D.}\ \bibnamefont {Johnson}},\
  and\ \bibinfo {author} {\bibfnamefont {A.}~\bibnamefont {Aspuru-Guzik}},\
  }\bibfield  {title} {\bibinfo {title} {Expressibility and entangling
  capability of parameterized quantum circuits for hybrid quantum-classical
  algorithms},\ }\href@noop {} {\bibfield  {journal} {\bibinfo  {journal} {Adv.
  Quantum Technol.}\ }\textbf {\bibinfo {volume} {2}},\ \bibinfo {pages}
  {1900070} (\bibinfo {year} {2019})}\BibitemShut {NoStop}%
\bibitem [{\citenamefont {Cerezo}\ \emph {et~al.}(2021)\citenamefont {Cerezo},
  \citenamefont {Arrasmith}, \citenamefont {Babbush}, \citenamefont {Benjamin},
  \citenamefont {Endo}, \citenamefont {Fujii}, \citenamefont {McClean},
  \citenamefont {Mitarai}, \citenamefont {Yuan}, \citenamefont {Cincio} \emph
  {et~al.}}]{cerezo2021variational}%
  \BibitemOpen
  \bibfield  {author} {\bibinfo {author} {\bibfnamefont {M.}~\bibnamefont
  {Cerezo}}, \bibinfo {author} {\bibfnamefont {A.}~\bibnamefont {Arrasmith}},
  \bibinfo {author} {\bibfnamefont {R.}~\bibnamefont {Babbush}}, \bibinfo
  {author} {\bibfnamefont {S.~C.}\ \bibnamefont {Benjamin}}, \bibinfo {author}
  {\bibfnamefont {S.}~\bibnamefont {Endo}}, \bibinfo {author} {\bibfnamefont
  {K.}~\bibnamefont {Fujii}}, \bibinfo {author} {\bibfnamefont {J.~R.}\
  \bibnamefont {McClean}}, \bibinfo {author} {\bibfnamefont {K.}~\bibnamefont
  {Mitarai}}, \bibinfo {author} {\bibfnamefont {X.}~\bibnamefont {Yuan}},
  \bibinfo {author} {\bibfnamefont {L.}~\bibnamefont {Cincio}}, \emph
  {et~al.},\ }\bibfield  {title} {\bibinfo {title} {Variational quantum
  algorithms},\ }\href@noop {} {\bibfield  {journal} {\bibinfo  {journal} {Nat.
  Rev. Phys.}\ ,\ \bibinfo {pages} {1}} (\bibinfo {year} {2021})}\BibitemShut
  {NoStop}%
\bibitem [{\citenamefont {Bharti}\ \emph {et~al.}(2021)\citenamefont {Bharti},
  \citenamefont {Cervera-Lierta}, \citenamefont {Kyaw}, \citenamefont {Haug},
  \citenamefont {Alperin-Lea}, \citenamefont {Anand}, \citenamefont {Degroote},
  \citenamefont {Heimonen}, \citenamefont {Kottmann}, \citenamefont {Menke}
  \emph {et~al.}}]{bharti2021noisy}%
  \BibitemOpen
  \bibfield  {author} {\bibinfo {author} {\bibfnamefont {K.}~\bibnamefont
  {Bharti}}, \bibinfo {author} {\bibfnamefont {A.}~\bibnamefont
  {Cervera-Lierta}}, \bibinfo {author} {\bibfnamefont {T.~H.}\ \bibnamefont
  {Kyaw}}, \bibinfo {author} {\bibfnamefont {T.}~\bibnamefont {Haug}}, \bibinfo
  {author} {\bibfnamefont {S.}~\bibnamefont {Alperin-Lea}}, \bibinfo {author}
  {\bibfnamefont {A.}~\bibnamefont {Anand}}, \bibinfo {author} {\bibfnamefont
  {M.}~\bibnamefont {Degroote}}, \bibinfo {author} {\bibfnamefont
  {H.}~\bibnamefont {Heimonen}}, \bibinfo {author} {\bibfnamefont {J.~S.}\
  \bibnamefont {Kottmann}}, \bibinfo {author} {\bibfnamefont {T.}~\bibnamefont
  {Menke}}, \emph {et~al.},\ }\bibfield  {title} {\bibinfo {title} {Noisy
  intermediate-scale quantum (nisq) algorithms},\ }\href@noop {} {\bibfield
  {journal} {\bibinfo  {journal} {arXiv:2101.08448}\ } (\bibinfo {year}
  {2021})}\BibitemShut {NoStop}%
\bibitem [{\citenamefont {Nakanishi}\ \emph {et~al.}(2020)\citenamefont
  {Nakanishi}, \citenamefont {Fujii},\ and\ \citenamefont
  {Todo}}]{nakanishi2020sequential}%
  \BibitemOpen
  \bibfield  {author} {\bibinfo {author} {\bibfnamefont {K.~M.}\ \bibnamefont
  {Nakanishi}}, \bibinfo {author} {\bibfnamefont {K.}~\bibnamefont {Fujii}},\
  and\ \bibinfo {author} {\bibfnamefont {S.}~\bibnamefont {Todo}},\ }\bibfield
  {title} {\bibinfo {title} {Sequential minimal optimization for
  quantum-classical hybrid algorithms},\ }\href
  {https://doi.org/10.1103/PhysRevResearch.2.043158} {\bibfield  {journal}
  {\bibinfo  {journal} {Phys. Rev. Research}\ }\textbf {\bibinfo {volume}
  {2}},\ \bibinfo {pages} {043158} (\bibinfo {year} {2020})}\BibitemShut
  {NoStop}%
\bibitem [{\citenamefont {Ostaszewski}\ \emph {et~al.}(2021)\citenamefont
  {Ostaszewski}, \citenamefont {Grant},\ and\ \citenamefont
  {Benedetti}}]{ostaszewski2021structure}%
  \BibitemOpen
  \bibfield  {author} {\bibinfo {author} {\bibfnamefont {M.}~\bibnamefont
  {Ostaszewski}}, \bibinfo {author} {\bibfnamefont {E.}~\bibnamefont {Grant}},\
  and\ \bibinfo {author} {\bibfnamefont {M.}~\bibnamefont {Benedetti}},\
  }\bibfield  {title} {\bibinfo {title} {Structure optimization for
  parameterized quantum circuits},\ }\href@noop {} {\bibfield  {journal}
  {\bibinfo  {journal} {Quantum}\ }\textbf {\bibinfo {volume} {5}},\ \bibinfo
  {pages} {391} (\bibinfo {year} {2021})}\BibitemShut {NoStop}%
\bibitem [{\citenamefont {Watanabe}\ \emph {et~al.}(2021)\citenamefont
  {Watanabe}, \citenamefont {Raymond}, \citenamefont {Ohnishi}, \citenamefont
  {Kaminishi},\ and\ \citenamefont {Sugawara}}]{watanabe2021}%
  \BibitemOpen
  \bibfield  {author} {\bibinfo {author} {\bibfnamefont {H.~C.}\ \bibnamefont
  {Watanabe}}, \bibinfo {author} {\bibfnamefont {R.}~\bibnamefont {Raymond}},
  \bibinfo {author} {\bibfnamefont {Y.-y.}\ \bibnamefont {Ohnishi}}, \bibinfo
  {author} {\bibfnamefont {E.}~\bibnamefont {Kaminishi}},\ and\ \bibinfo
  {author} {\bibfnamefont {M.}~\bibnamefont {Sugawara}},\ }\bibfield  {title}
  {\bibinfo {title} {Optimizing parameterized quantum circuits with free-axis
  selection},\ }\href@noop {} {\bibfield  {journal} {\bibinfo  {journal}
  {arXiv:2104.14875}\ } (\bibinfo {year} {2021})}\BibitemShut {NoStop}%
\bibitem [{\citenamefont {Nelder}\ and\ \citenamefont
  {Mead}(1965)}]{nelder1965simplex}%
  \BibitemOpen
  \bibfield  {author} {\bibinfo {author} {\bibfnamefont {J.~A.}\ \bibnamefont
  {Nelder}}\ and\ \bibinfo {author} {\bibfnamefont {R.}~\bibnamefont {Mead}},\
  }\bibfield  {title} {\bibinfo {title} {A simplex method for function
  minimization},\ }\href@noop {} {\bibfield  {journal} {\bibinfo  {journal}
  {Comput. J.}\ }\textbf {\bibinfo {volume} {7}},\ \bibinfo {pages} {308}
  (\bibinfo {year} {1965})}\BibitemShut {NoStop}%
\bibitem [{\citenamefont {Spall}(1992)}]{spall1992multivariate}%
  \BibitemOpen
  \bibfield  {author} {\bibinfo {author} {\bibfnamefont {J.~C.}\ \bibnamefont
  {Spall}},\ }\bibfield  {title} {\bibinfo {title} {Multivariate stochastic
  approximation using a simultaneous perturbation gradient approximation},\
  }\href@noop {} {\bibfield  {journal} {\bibinfo  {journal} {IEEE Trans.
  Automat. Contr.}\ }\textbf {\bibinfo {volume} {37}},\ \bibinfo {pages} {332}
  (\bibinfo {year} {1992})}\BibitemShut {NoStop}%
\bibitem [{\citenamefont {Yuan}\ \emph {et~al.}(2019)\citenamefont {Yuan},
  \citenamefont {Endo}, \citenamefont {Zhao}, \citenamefont {Li},\ and\
  \citenamefont {Benjamin}}]{yuan2019theory}%
  \BibitemOpen
  \bibfield  {author} {\bibinfo {author} {\bibfnamefont {X.}~\bibnamefont
  {Yuan}}, \bibinfo {author} {\bibfnamefont {S.}~\bibnamefont {Endo}}, \bibinfo
  {author} {\bibfnamefont {Q.}~\bibnamefont {Zhao}}, \bibinfo {author}
  {\bibfnamefont {Y.}~\bibnamefont {Li}},\ and\ \bibinfo {author}
  {\bibfnamefont {S.~C.}\ \bibnamefont {Benjamin}},\ }\bibfield  {title}
  {\bibinfo {title} {Theory of variational quantum simulation},\ }\href@noop {}
  {\bibfield  {journal} {\bibinfo  {journal} {Quantum}\ }\textbf {\bibinfo
  {volume} {3}},\ \bibinfo {pages} {191} (\bibinfo {year} {2019})}\BibitemShut
  {NoStop}%
\bibitem [{\citenamefont {McArdle}\ \emph
  {et~al.}(2019{\natexlab{b}})\citenamefont {McArdle}, \citenamefont {Jones},
  \citenamefont {Endo}, \citenamefont {Li}, \citenamefont {Benjamin},\ and\
  \citenamefont {Yuan}}]{mcardle2019variational}%
  \BibitemOpen
  \bibfield  {author} {\bibinfo {author} {\bibfnamefont {S.}~\bibnamefont
  {McArdle}}, \bibinfo {author} {\bibfnamefont {T.}~\bibnamefont {Jones}},
  \bibinfo {author} {\bibfnamefont {S.}~\bibnamefont {Endo}}, \bibinfo {author}
  {\bibfnamefont {Y.}~\bibnamefont {Li}}, \bibinfo {author} {\bibfnamefont
  {S.~C.}\ \bibnamefont {Benjamin}},\ and\ \bibinfo {author} {\bibfnamefont
  {X.}~\bibnamefont {Yuan}},\ }\bibfield  {title} {\bibinfo {title}
  {Variational ansatz-based quantum simulation of imaginary time evolution},\
  }\href@noop {} {\bibfield  {journal} {\bibinfo  {journal} {npj Quantum
  Information}\ }\textbf {\bibinfo {volume} {5}},\ \bibinfo {pages} {1}
  (\bibinfo {year} {2019}{\natexlab{b}})}\BibitemShut {NoStop}%
\bibitem [{\citenamefont {Yao}\ \emph {et~al.}(2021)\citenamefont {Yao},
  \citenamefont {Gomes}, \citenamefont {Zhang}, \citenamefont {Wang},
  \citenamefont {Ho}, \citenamefont {Iadecola},\ and\ \citenamefont
  {Orth}}]{yao2021adaptive}%
  \BibitemOpen
  \bibfield  {author} {\bibinfo {author} {\bibfnamefont {Y.-X.}\ \bibnamefont
  {Yao}}, \bibinfo {author} {\bibfnamefont {N.}~\bibnamefont {Gomes}}, \bibinfo
  {author} {\bibfnamefont {F.}~\bibnamefont {Zhang}}, \bibinfo {author}
  {\bibfnamefont {C.-Z.}\ \bibnamefont {Wang}}, \bibinfo {author}
  {\bibfnamefont {K.-M.}\ \bibnamefont {Ho}}, \bibinfo {author} {\bibfnamefont
  {T.}~\bibnamefont {Iadecola}},\ and\ \bibinfo {author} {\bibfnamefont
  {P.~P.}\ \bibnamefont {Orth}},\ }\bibfield  {title} {\bibinfo {title}
  {Adaptive variational quantum dynamics simulations},\ }\href@noop {}
  {\bibfield  {journal} {\bibinfo  {journal} {PRX Quantum}\ }\textbf {\bibinfo
  {volume} {2}},\ \bibinfo {pages} {030307} (\bibinfo {year}
  {2021})}\BibitemShut {NoStop}%
\bibitem [{\citenamefont {McLachlan}(1964)}]{mclachlan1964variational}%
  \BibitemOpen
  \bibfield  {author} {\bibinfo {author} {\bibfnamefont {A.}~\bibnamefont
  {McLachlan}},\ }\bibfield  {title} {\bibinfo {title} {A variational solution
  of the time-dependent schrodinger equation},\ }\href@noop {} {\bibfield
  {journal} {\bibinfo  {journal} {Mol. Phys.}\ }\textbf {\bibinfo {volume}
  {8}},\ \bibinfo {pages} {39} (\bibinfo {year} {1964})}\BibitemShut {NoStop}%
\bibitem [{\citenamefont {Nishi}\ \emph {et~al.}(2021)\citenamefont {Nishi},
  \citenamefont {Kosugi},\ and\ \citenamefont
  {Matsushita}}]{nishi2021implementation}%
  \BibitemOpen
  \bibfield  {author} {\bibinfo {author} {\bibfnamefont {H.}~\bibnamefont
  {Nishi}}, \bibinfo {author} {\bibfnamefont {T.}~\bibnamefont {Kosugi}},\ and\
  \bibinfo {author} {\bibfnamefont {Y.}~\bibnamefont {Matsushita}},\ }\bibfield
   {title} {\bibinfo {title} {Implementation of quantum imaginary-time
  evolution method on nisq devices by introducing nonlocal approximation},\
  }\href@noop {} {\bibfield  {journal} {\bibinfo  {journal} {npj Quantum Inf.}\
  }\textbf {\bibinfo {volume} {7}},\ \bibinfo {pages} {1} (\bibinfo {year}
  {2021})}\BibitemShut {NoStop}%
\bibitem [{\citenamefont {Barison}\ \emph {et~al.}(2021)\citenamefont
  {Barison}, \citenamefont {Vicentini},\ and\ \citenamefont
  {Carleo}}]{barison2021efficient}%
  \BibitemOpen
  \bibfield  {author} {\bibinfo {author} {\bibfnamefont {S.}~\bibnamefont
  {Barison}}, \bibinfo {author} {\bibfnamefont {F.}~\bibnamefont {Vicentini}},\
  and\ \bibinfo {author} {\bibfnamefont {G.}~\bibnamefont {Carleo}},\
  }\bibfield  {title} {\bibinfo {title} {An efficient quantum algorithm for the
  time evolution of parameterized circuits},\ }\href
  {https://doi.org/10.22331/q-2021-07-28-512} {\bibfield  {journal} {\bibinfo
  {journal} {{Quantum}}\ }\textbf {\bibinfo {volume} {5}},\ \bibinfo {pages}
  {512} (\bibinfo {year} {2021})}\BibitemShut {NoStop}%
\bibitem [{\citenamefont {Benedetti}\ \emph {et~al.}(2021)\citenamefont
  {Benedetti}, \citenamefont {Fiorentini},\ and\ \citenamefont
  {Lubasch}}]{benedetti2021}%
  \BibitemOpen
  \bibfield  {author} {\bibinfo {author} {\bibfnamefont {M.}~\bibnamefont
  {Benedetti}}, \bibinfo {author} {\bibfnamefont {M.}~\bibnamefont
  {Fiorentini}},\ and\ \bibinfo {author} {\bibfnamefont {M.}~\bibnamefont
  {Lubasch}},\ }\bibfield  {title} {\bibinfo {title} {Hardware-efficient
  variational quantum algorithms for time evolution},\ }\href
  {https://doi.org/10.1103/PhysRevResearch.3.033083} {\bibfield  {journal}
  {\bibinfo  {journal} {Phys. Rev. Research}\ }\textbf {\bibinfo {volume}
  {3}},\ \bibinfo {pages} {033083} (\bibinfo {year} {2021})}\BibitemShut
  {NoStop}%
\bibitem [{\citenamefont {Parrish}\ \emph {et~al.}(2019)\citenamefont
  {Parrish}, \citenamefont {Iosue}, \citenamefont {Ozaeta},\ and\ \citenamefont
  {McMahon}}]{Parrish2019arXiv}%
  \BibitemOpen
  \bibfield  {author} {\bibinfo {author} {\bibfnamefont {R.~M.}\ \bibnamefont
  {Parrish}}, \bibinfo {author} {\bibfnamefont {J.~T.}\ \bibnamefont {Iosue}},
  \bibinfo {author} {\bibfnamefont {A.}~\bibnamefont {Ozaeta}},\ and\ \bibinfo
  {author} {\bibfnamefont {P.~L.}\ \bibnamefont {McMahon}},\ }\bibfield
  {title} {\bibinfo {title} {A {J}acobi {D}iagonalization and {A}nderson
  {A}cceleration {A}lgorithm for {V}ariational {Q}uantum {A}lgorithm
  {P}arameter {O}ptimization},\ }\href@noop {} {\bibfield  {journal} {\bibinfo
  {journal} {arXiv:1904.03206}\ } (\bibinfo {year} {2019})}\BibitemShut
  {NoStop}%
\bibitem [{\citenamefont {Vidal}\ and\ \citenamefont
  {Theis}(2018)}]{vidal2018calculus}%
  \BibitemOpen
  \bibfield  {author} {\bibinfo {author} {\bibfnamefont {J.~G.}\ \bibnamefont
  {Vidal}}\ and\ \bibinfo {author} {\bibfnamefont {D.~O.}\ \bibnamefont
  {Theis}},\ }\bibfield  {title} {\bibinfo {title} {Calculus on parameterized
  quantum circuits},\ }\href@noop {} {\bibfield  {journal} {\bibinfo  {journal}
  {arXiv:1812.06323}\ } (\bibinfo {year} {2018})}\BibitemShut {NoStop}%
\bibitem [{\citenamefont {Mitarai}\ and\ \citenamefont
  {Fujii}(2019)}]{mitarai2019}%
  \BibitemOpen
  \bibfield  {author} {\bibinfo {author} {\bibfnamefont {K.}~\bibnamefont
  {Mitarai}}\ and\ \bibinfo {author} {\bibfnamefont {K.}~\bibnamefont
  {Fujii}},\ }\bibfield  {title} {\bibinfo {title} {Methodology for replacing
  indirect measurements with direct measurements},\ }\href
  {https://doi.org/10.1103/PhysRevResearch.1.013006} {\bibfield  {journal}
  {\bibinfo  {journal} {Phys. Rev. Research}\ }\textbf {\bibinfo {volume}
  {1}},\ \bibinfo {pages} {013006} (\bibinfo {year} {2019})}\BibitemShut
  {NoStop}%
\bibitem [{\citenamefont {Wharton}\ and\ \citenamefont
  {Koch}(2015)}]{wharton2015unit}%
  \BibitemOpen
  \bibfield  {author} {\bibinfo {author} {\bibfnamefont {K.}~\bibnamefont
  {Wharton}}\ and\ \bibinfo {author} {\bibfnamefont {D.}~\bibnamefont {Koch}},\
  }\bibfield  {title} {\bibinfo {title} {Unit quaternions and the bloch
  sphere},\ }\href@noop {} {\bibfield  {journal} {\bibinfo  {journal} {J. Phys.
  A Math.}\ }\textbf {\bibinfo {volume} {48}},\ \bibinfo {pages} {235302}
  (\bibinfo {year} {2015})}\BibitemShut {NoStop}%
\bibitem [{\citenamefont {Foxen}\ \emph {et~al.}(2020)\citenamefont {Foxen},
  \citenamefont {Neill}, \citenamefont {Dunsworth}, \citenamefont {Roushan},
  \citenamefont {Chiaro}, \citenamefont {Megrant}, \citenamefont {Kelly},
  \citenamefont {Chen}, \citenamefont {Satzinger}, \citenamefont {Barends}
  \emph {et~al.}}]{Foxen2020}%
  \BibitemOpen
  \bibfield  {author} {\bibinfo {author} {\bibfnamefont {B.}~\bibnamefont
  {Foxen}}, \bibinfo {author} {\bibfnamefont {C.}~\bibnamefont {Neill}},
  \bibinfo {author} {\bibfnamefont {A.}~\bibnamefont {Dunsworth}}, \bibinfo
  {author} {\bibfnamefont {P.}~\bibnamefont {Roushan}}, \bibinfo {author}
  {\bibfnamefont {B.}~\bibnamefont {Chiaro}}, \bibinfo {author} {\bibfnamefont
  {A.}~\bibnamefont {Megrant}}, \bibinfo {author} {\bibfnamefont
  {J.}~\bibnamefont {Kelly}}, \bibinfo {author} {\bibfnamefont
  {Z.}~\bibnamefont {Chen}}, \bibinfo {author} {\bibfnamefont {K.}~\bibnamefont
  {Satzinger}}, \bibinfo {author} {\bibfnamefont {R.}~\bibnamefont {Barends}},
  \emph {et~al.} (\bibinfo {collaboration} {Google AI Quantum}),\ }\bibfield
  {title} {\bibinfo {title} {Demonstrating a continuous set of two-qubit gates
  for near-term quantum algorithms},\ }\href
  {https://doi.org/10.1103/PhysRevLett.125.120504} {\bibfield  {journal}
  {\bibinfo  {journal} {Phys. Rev. Lett.}\ }\textbf {\bibinfo {volume} {125}},\
  \bibinfo {pages} {120504} (\bibinfo {year} {2020})}\BibitemShut {NoStop}%
\bibitem [{\citenamefont {Kerenidis}\ \emph {et~al.}(2021)\citenamefont
  {Kerenidis}, \citenamefont {Landman},\ and\ \citenamefont
  {Mathur}}]{kerenidis2021}%
  \BibitemOpen
  \bibfield  {author} {\bibinfo {author} {\bibfnamefont {I.}~\bibnamefont
  {Kerenidis}}, \bibinfo {author} {\bibfnamefont {J.}~\bibnamefont {Landman}},\
  and\ \bibinfo {author} {\bibfnamefont {N.}~\bibnamefont {Mathur}},\
  }\bibfield  {title} {\bibinfo {title} {Classical and quantum algorithms for
  orthogonal neural networks},\ }\href@noop {} {\bibfield  {journal} {\bibinfo
  {journal} {arXiv:2106.07198}\ } (\bibinfo {year} {2021})}\BibitemShut
  {NoStop}%
\bibitem [{\citenamefont {ANIS}\ \emph {et~al.}(2021)\citenamefont {ANIS},
  \citenamefont {Abraham}, \citenamefont {AduOffei}, \citenamefont {Agarwal},
  \citenamefont {Agliardi}, \citenamefont {Aharoni}, \citenamefont {Akhalwaya},
  \citenamefont {Aleksandrowicz}, \citenamefont {Alexander}, \citenamefont
  {Amy} \emph {et~al.}}]{Qiskit}%
  \BibitemOpen
  \bibfield  {author} {\bibinfo {author} {\bibfnamefont {M.~S.}\ \bibnamefont
  {ANIS}}, \bibinfo {author} {\bibfnamefont {H.}~\bibnamefont {Abraham}},
  \bibinfo {author} {\bibnamefont {AduOffei}}, \bibinfo {author} {\bibfnamefont
  {R.}~\bibnamefont {Agarwal}}, \bibinfo {author} {\bibfnamefont
  {G.}~\bibnamefont {Agliardi}}, \bibinfo {author} {\bibfnamefont
  {M.}~\bibnamefont {Aharoni}}, \bibinfo {author} {\bibfnamefont {I.~Y.}\
  \bibnamefont {Akhalwaya}}, \bibinfo {author} {\bibfnamefont {G.}~\bibnamefont
  {Aleksandrowicz}}, \bibinfo {author} {\bibfnamefont {T.}~\bibnamefont
  {Alexander}}, \bibinfo {author} {\bibfnamefont {M.}~\bibnamefont {Amy}},
  \emph {et~al.},\ }\href {https://doi.org/10.5281/zenodo.2573505} {\bibinfo
  {title} {Qiskit: An open-source framework for quantum computing}} (\bibinfo
  {year} {2021})\BibitemShut {NoStop}%
\bibitem [{\citenamefont {Gard}\ \emph {et~al.}(2020)\citenamefont {Gard},
  \citenamefont {Zhu}, \citenamefont {Barron}, \citenamefont {Mayhall},
  \citenamefont {Economou},\ and\ \citenamefont {Barnes}}]{gard2020}%
  \BibitemOpen
  \bibfield  {author} {\bibinfo {author} {\bibfnamefont {B.~T.}\ \bibnamefont
  {Gard}}, \bibinfo {author} {\bibfnamefont {L.}~\bibnamefont {Zhu}}, \bibinfo
  {author} {\bibfnamefont {G.~S.}\ \bibnamefont {Barron}}, \bibinfo {author}
  {\bibfnamefont {N.~J.}\ \bibnamefont {Mayhall}}, \bibinfo {author}
  {\bibfnamefont {S.~E.}\ \bibnamefont {Economou}},\ and\ \bibinfo {author}
  {\bibfnamefont {E.}~\bibnamefont {Barnes}},\ }\bibfield  {title} {\bibinfo
  {title} {Efficient symmetry-preserving state preparation circuits for the
  variational quantum eigensolver algorithm},\ }\href@noop {} {\bibfield
  {journal} {\bibinfo  {journal} {npj Quantum Inf.}\ }\textbf {\bibinfo
  {volume} {6}},\ \bibinfo {pages} {1} (\bibinfo {year} {2020})}\BibitemShut
  {NoStop}%
\bibitem [{\citenamefont {Engel}\ \emph {et~al.}(1990)\citenamefont {Engel},
  \citenamefont {Schinke}, \citenamefont {Hennig},\ and\ \citenamefont
  {Metiu}}]{engel1990time}%
  \BibitemOpen
  \bibfield  {author} {\bibinfo {author} {\bibfnamefont {V.}~\bibnamefont
  {Engel}}, \bibinfo {author} {\bibfnamefont {R.}~\bibnamefont {Schinke}},
  \bibinfo {author} {\bibfnamefont {S.}~\bibnamefont {Hennig}},\ and\ \bibinfo
  {author} {\bibfnamefont {H.}~\bibnamefont {Metiu}},\ }\bibfield  {title}
  {\bibinfo {title} {A time-dependent interpretation of the absorption spectrum
  of ch3ono},\ }\href@noop {} {\bibfield  {journal} {\bibinfo  {journal} {The
  Journal of Chemical Physics}\ }\textbf {\bibinfo {volume} {92}},\ \bibinfo
  {pages} {1} (\bibinfo {year} {1990})}\BibitemShut {NoStop}%
\bibitem [{\citenamefont {Zewail}(2000)}]{zewail2000femtochemistry}%
  \BibitemOpen
  \bibfield  {author} {\bibinfo {author} {\bibfnamefont {A.~H.}\ \bibnamefont
  {Zewail}},\ }\bibfield  {title} {\bibinfo {title} {Femtochemistry:
  Atomic-scale dynamics of the chemical bond},\ }\href@noop {} {\bibfield
  {journal} {\bibinfo  {journal} {The Journal of Physical Chemistry A}\
  }\textbf {\bibinfo {volume} {104}},\ \bibinfo {pages} {5660} (\bibinfo {year}
  {2000})}\BibitemShut {NoStop}%
\end{thebibliography}%
